\documentclass[pre,aps,twocolumn]{revtex4}
 \pdfoutput=1
\usepackage{amsfonts}

\usepackage{amssymb}

\usepackage{graphicx}

\usepackage{mathrsfs}

\usepackage{array}

\usepackage{amsmath}

\usepackage{lscape}

\usepackage{ bbold }

\usepackage{ucs}

\begin{document}

\title{Crosstalk and transitions between multiple spatial maps in an attractor neural network model of the hippocampus: Collective motion of the activity}

\author{R. Monasson, S. Rosay}

\affiliation{Laboratoire de Physique Th\'eorique de l'ENS, CNRS \& UPMC, 24 rue Lhomond,75005 Paris, France}

\date{\today}

\begin{abstract}
The dynamics of a neural model for hippocampal place cells storing spatial maps is studied. In the absence of external input, depending on the number of cells and on the values of control parameters (number of environments stored, level of neural noise, average level of activity, connectivity of place cells), a 'clump' of spatially-localized activity can diffuse, or remains pinned due to crosstalk between the environments. In the single-environment case, the macroscopic coefficient of diffusion of the clump and its effective mobility are calculated analytically from first principles, and corroborated by numerical simulations. In the multi-environment case the heights and the widths of the pinning barriers are analytically characterized with the replica method; diffusion within one map is then in competition with transitions between different maps. Possible mechanisms enhancing mobility are proposed and tested. 
\end{abstract}

\maketitle

\section{Introduction}

Since the discovery of place cells in the hippocampus of rodents \cite{OKeefeDostrovsky71}, the hippocampus is believed to support spatial memory and representation. Place cells are neurons that fire specifically when the animal is located at certain positions of space called place fields. Their properties have been extensively studied, revealing striking features. In particular, the memorized places appear to be organized in several discrete 'maps' or 'environments' \cite{Derdikman10}. A given neuron can have place fields in different environments, and these place fields appear randomly allocated, independently of the place cell's location in the neural tissue \cite{OKeefeConway78}. This random reallocation of place fields in each new environment is called 'remapping' \cite{KubieMuller91}. Place fields are also stable in the dark \cite{Quirk90} and after alteration of visual cues \cite{MullerKubie87}, suggesting that their firing is driven in part by  self-motion information ('path integration' \cite{McNaughton96}).

 Many theoretical models have been proposed in order to account for the formation and the firing properties of place cells. An important class of them is formed by attractor neural network models \cite{TsodyksSejnowski95,Samsonovich97,BattagliaTreves98,BrunelTrullier98,Tsodyks99,KaliDayan00}. These models postulate that an environment is memorized when the corresponding neural activities are stable states of the network \cite{Amit89}, such as in the celebrated Hopfied model \cite{Hopfield82}, an assumption motivated here by the high degree of recurrent connectivity in the CA3 area of the hippocampus \cite{Amaral89}. In majority, these studies focus mainly on the static properties of the models, that is the stable states of the network. The conditions of formation of spatially localized attractors, their robustness to noise, the storage capacity of such networks have been investigated in great details. How the network dynamically evolves within one map and between maps remains, however, poorly understood in this framework, leastways analytically. Yet, this dynamical aspect plays a crucial role in most experiments, whether they involve physical motion of the animal \cite{Wilson93,Gothard96,Harris03}, mental trajectory planning \cite{Johnson07}, ``sleep replay'' \cite{Skaggs96} or modification of visual cues \cite{Gothard96,Wills05,Leutgeb05b,Jezek11}.


Attractor neural networks are an important paradigm in the attempt to understand and model the principles of memory. Following their introduction by Hopfield thirty years ago \cite{Hopfield82}, the properties of attractor neural networks have been investigated in detail using tools from statistical mechanics of disordered systems \cite{Amit89}. In the 'basic', most common version, a memorized pattern corresponds to an activity configuration of the network. In the present case of spatial memory, in contrast, a memory item corresponds to a space manifold (a spatial map), {\em i.e.} the whole collection of neural activity configurations obtained when the animal is located in various points of this map. As a consequence attractors are more complex than in the original Hopfield model. As far as dynamics is concerned, again, the present case displays much richer behaviors. Indeed, in the presence of noise in the neural response, the network activity can either jump between maps (as is the case between attractors in the Hopfield model) or evolve continuously within one attractor.  In the latter case, the pattern of activity corresponds successively to positions along a continuous trajectory in one of the maps, as if the neural activity configuration 'moved' in this map.  As a result such an extension of the Hopfield model paves the way for refinements and complexification of the structure of the modelled memory. In this context, the comprehension of its complex dynamics has a theoretical interest in itself.

Furthermore, from the point of view of statistical mechanics, the study of a spatially localized phase as a bump of activity in hippocampal neurons is of great interest. How a 'quasi-particle' emerges from the interactions of microscopic units, and how the dynamics of its location (being considered here as a collective coordinate for the neural activity) can be characterized are non-trivial questions, which highlights the rich connection between statistical mechanics and computational neuroscience.

In a previous article \cite{Monasson13}, we proposed an attractor neural network model for hippocampal place cells encoding one- and two-dimensional spatial maps. We studied the stable states and the phase diagram for varying levels of noise and of memory load. We showed that, under certain conditions, the stable states are 'clumps' (bumps) of activity localized in one of the stored environments, similar to the activity patterns observed in microelectrode single-unit recordings. In the present work, we address the issue of the evolution of the network within one attractor, that is, within one map, when the network is in this clump phase. Its dynamics is studied both analytically and numerically. It appears that the crosstalk between environments has the effect of hindering the motion of the clump, and virtually suppresses motion for a wide range of control parameters. This phenomenon is particularly salient in the one-dimensional case. Neural noise, by itself, may therefore not sufficient to make the clump move, and additional mechanisms have to be proposed to retrieve this motion \cite{Hopfield10}. We show that diffusion within one map is in competition with transitions between maps, corresponding to the sudden disappearance of the localization of the activity at one specific position in the map under consideration, followed by its localization at another position in another map. The detailed study of those transitions and of the distribution of the tunneling positions within the maps will be addressed in a companion publication.

In Section \ref{sec:prelim} we briefly recall the model and summarize the results of \cite{Monasson13} on its stable phases. The main results of the present paper on the dynamics of the activity in one map are reviewed in Section \ref{sec:overview}. In Section \ref{sec:pure} we study the single-environment case, and analytically show that the dynamics can be described by an effective diffusion for the center of the clump; we also characterize the mobility of the clump in response
to an external force. In Section~\ref{sec:activated} we show that the presence of disorder limits drastically the motion of the clump within one environment, and propose additional mechanisms to enhance motion. In Section~\ref{sec:retrieval} we address the retrieval process of the attractor neural network in the presence of input. Finally, in Section \ref{sec:add} we study the effect of other, out-of equilibrium mechanisms on the motion of the clump.

\section{Reminder on the model and its phases}

\label{sec:prelim}

The $N$ place cells are modeled by interacting binary units $\sigma_i$ equal to 0 or 1, and corresponding to, respectively, silent and active neurons. Let us first consider a first environment (that can be either 1 or 2-dimensional). We suppose that, after learning of the environment and random allocation of place fields, each place cell preferentially fires when the animal is located in an environment-specific location in the 1 or 2-dimensional space, defining its place field. For simplicity space is assumed to be a segment of length $N$ in dimension 1, and a square of edge length $\sqrt N$ in dimension 2, with periodic boundary conditions. The $N$ centers of the place fields are located on the nodes of a 1 or 2-dimensional regular grid: two contiguous centers are at unit distance from each other.

Pairs of cells whose place field centers lie within some distance $d_c$ from each other are coupled with an excitatory coupling $J_{ij}^0=\frac 1N$. We choose the cut-off distance $d_c$ such that each cell $i$ is connected to the same number $w\,N$ of other cells $j$, independently of the space dimension: $w(\ll 1)$ is the fraction of the neural population any neuron is coupled to. The $\frac 1N$ scale factor in the coupling $J_{ij}^0$ is such that the total input received by a place cell is finite when the number of cells, $N$, is sent to infinity.

Then, we consider other additional environments. Each time the rodent explores a new environment a remapping of the place fields takes place. We assume that the remapping is represented by a random permutation of the $N$ place-cell indices associated to the place fields on the regular grid. Let $\pi^\ell$ be the permutation corresponding to remapping (environment) number $\ell$, where $\ell=1,\dots, L$ is the index of the new environments.  We assume that all environments contribute equally and additively to the total synaptic matrix, with the result
\begin{equation} \label{rule2}
J_{ij}= J_{ij}^0+\sum\limits_{\ell=1}^L J_{\pi^\ell(i)\pi^\ell(j)}^0 \ .
\end{equation} 
Note that all environments are statistically equivalent. We will look hereafter for the presence of localized activity in the environment 0 (hereafter called reference environment), but this choice is arbitrary.

In addition to pyramidal cells, the network contains long-range, inhibitory interneurons, which maintain the fraction of active place cells at a fixed level, $f$. The probability of a neural activity configuration $\boldsymbol \sigma =(\sigma_1,\sigma_2,\ldots ,\sigma_N)$ is then assumed to be
\begin{equation}\label{dist}
P_J(\boldsymbol\sigma ) = \frac 1{Z_J(T)} \; \exp\big( \sum_{i<j} J_{ij}\, \sigma_i\, \sigma_j/T)\ ,
\end{equation}
where the partition function $Z_J(T)$ is such that the sum of $P_J( \boldsymbol\sigma ) $ over all activity configurations with exactly $f\,N$ active neurons is normalized to unity. Parameter $T$, which plays the role of temperature in statistical mechanics, is indicative of the level of noise in the response of neurons to their inputs (local fields).

In \cite{Monasson13} we have analytically characterized the possible regimes, or phases, of the model in the limit of large size, $N\to \infty$, and at a fixed ratio of the number of environments per neuron, $\alpha\equiv L/ N$, hereafter called load. The phases are defined in terms of the behaviors of the local average of the activity, 
\begin{equation}\label{density_ave}
\rho(x) = \lim _{\epsilon \to 0} \; \lim _{N\to \infty} \;\frac 1{\epsilon N} \sum_{i: |x-\frac iN|< \frac \epsilon 2} \overline{\langle \sigma _i \rangle }\  ,
\end{equation}
and of the Edwards-Anderson overlap  describing the fluctuations of the local activities:
\begin{equation}
q = \frac 1N \sum _{i=1}^N \overline{\langle \sigma_i \rangle ^2} \ .
\end{equation}
The overbar above denotes the average over the random remappings (permutations $\pi^\ell$), while the brackets $\langle\cdot\rangle$ correspond to the average over distribution $P_J$ (\ref{dist}).

\begin{figure}

\centering

\includegraphics[width=\columnwidth]{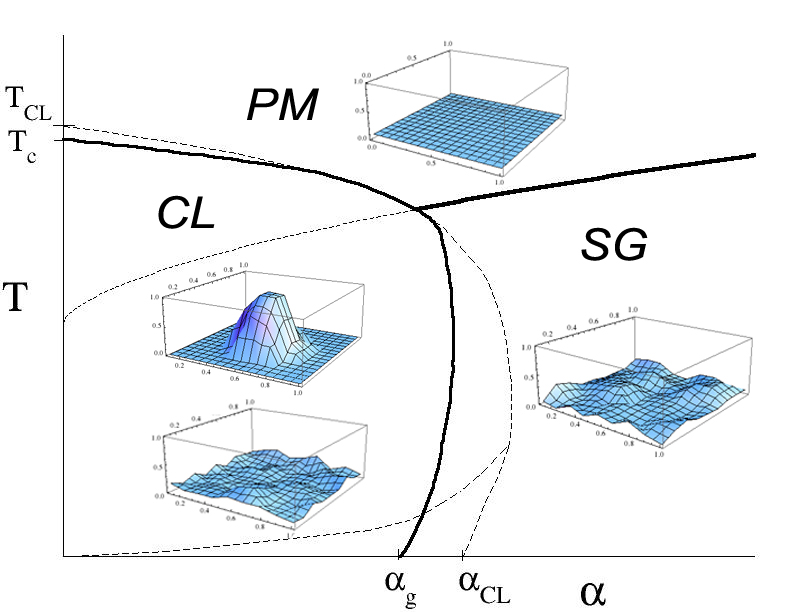}

\caption{Sketch of the phase diagram in the plane of neural noise, $T$, and number of environments per neuron, $\alpha$. Thick solid lines: transitions between phases. Thin dashed lines: stability region of each phase against fluctuations.  Insets show the corresponding activity profiles in the 2D model (averaged over 1 round of Monte Carlo simulations after thermalization). In the clump phase we represent the same activity profile in the retrieved environment (top) and in another stored environment (bottom). See \cite{Monasson13} (Fig. 8) for more quantitative details.}

 \label{phasediag}

\end{figure}

The outcome of the analysis is the phase diagram shown in Fig.~\ref{phasediag}. Three stable phases are found (see \cite{Monasson13} for details): 

\begin{itemize}

\item the paramagnetic phase (PM), corresponding to high levels of noise $T$, in which the average local activity is uniform over space, $\rho(x)=f$, and neurons are essentially uncorrelated, $q=f^2$.  

\item  a glassy phase (SG), corresponding to large loads $\alpha$, in which the local activity $\langle \sigma_i\rangle$ varies from neuron to neuron ($q>f^2$), but does not cluster around any specific location in space in any of the environments ($\rho(x)=f$ after averaging over remappings). In this SG phase the crosstalk between environments is so large that none of them is actually stored in the network activity.

\item a 'clump' phase (CL), for small enough load and noise, where activity depends on space, {\em i.e.} $\rho(x)$ varies with $x$. In the present case, the activity is localized in the first environment (reference environment). This is the consequence of our choice for the reference environment, but in practice the activity could be localized in any environment. Which environment is retrieved may depend on external factors (initial configuration of activity, specific inputs, ...), and may vary with time due to thermal fluctuations. 

\end{itemize}

Unless stated otherwise, we take the parameter values $w=0.05$ and $f=0.1$ in the numerical simulations throughout this work.

\section{Overview of results}

\label{sec:overview}

While the system is in the clump phase, the bump of activity can either move over space in the coherent environment (hence, stay in the same attractor), or switch between environments (transition to another attractor). Transitions from one environment to another have been observed experimentally \cite{Wills05,Jezek11} and will be addressed in a forthcoming publication. In this paper we focus on the dynamics of the neural activity 'within' one map only. We now briefly review our main results.

\begin{figure}

\centering

\includegraphics[width=\columnwidth]{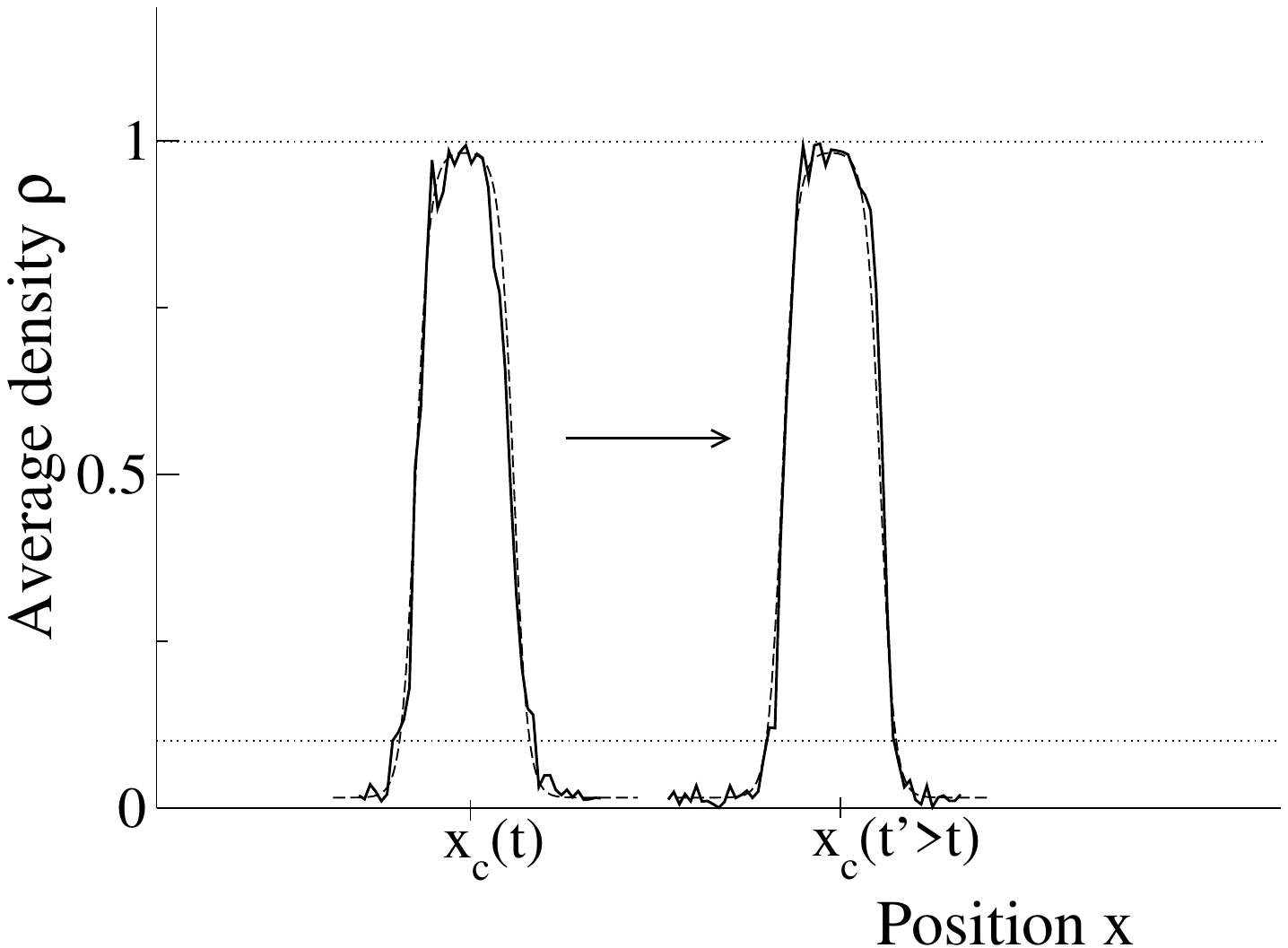}

\caption{Sketches of the clump of neural activity moving in space, shown at two subsequent times (only central parts are shown), in the 1D, single-environment case with $T=0.006$. The dashed lines represent the equilibrium profile $\rho^*(x)$. Full lines correspond to average densities computed at the two times under consideration, which deviate from $\rho^*$ by terms of the order of $N^{-1/2}$. The horizontal dotted lines locate $\rho=f$ and $\rho=1$. Simulations parameters: $N=2000$, activity averaged over short distance (10 spins) and time ($5N$ Monte Carlo steps).}

\label{shape}

\end{figure}

The dynamics we consider defines an evolution for the microscopic configuration of neural activity, that is, the set of all neuron states (silent or active). As we know from the study of equilibrium properties \cite{Monasson13}, the statistics of the activity can be characterized through the average density profile, $\rho ^*(x)$ (the * superscript refers to the equilibrium value). It is a natural question whether such a macroscopic characterization of configurations also exists for dynamics. We show, through a careful study of the single-environment case for which the dynamics can be studied in great analytical details, that the answer is positive. Two main features emerge in the large system size limit, summarized below and in Fig.~\ref{shape}:

\begin{itemize}

\item the position of the center of the clump (center of mass of the activity), $x_c(t)$, plays the role of a collective coordinate for the neural configurations. It undergoes a pure diffusion motion, whose diffusion coefficient is of the order of $\frac 1N$. The clump velocity under an external force satisfies the Einstein relation, with a mobility of the order of $\frac 1N$. The diffusion coefficient and the mobility depend on the exact shape of the equilibrium density profile, as well as some specific details of the microscopic neural evolution. 

\item in addition to the motion of the center of the clump, the activity profile $\rho(x,t)$ shows fluctuations around the equilibrium profile $\rho^*\big(x-x_c(t)\big)$. Those fluctuations are small, of the order of $N^{-1/2}$.

\end{itemize} 

Informally speaking the clump has the status of a quasi-particle. It behaves like a quasi-rigid body, moving in space, and the only time-dependent and relevant variable to consider is the position of its center, as was already observed in simulations of previous models \cite{Samsonovich97}. The properties above and the calculation of the diffusion and mobility coefficients are presented in Section \ref{sec:pure}.

\begin{figure}
\centering
\includegraphics[width=\columnwidth]{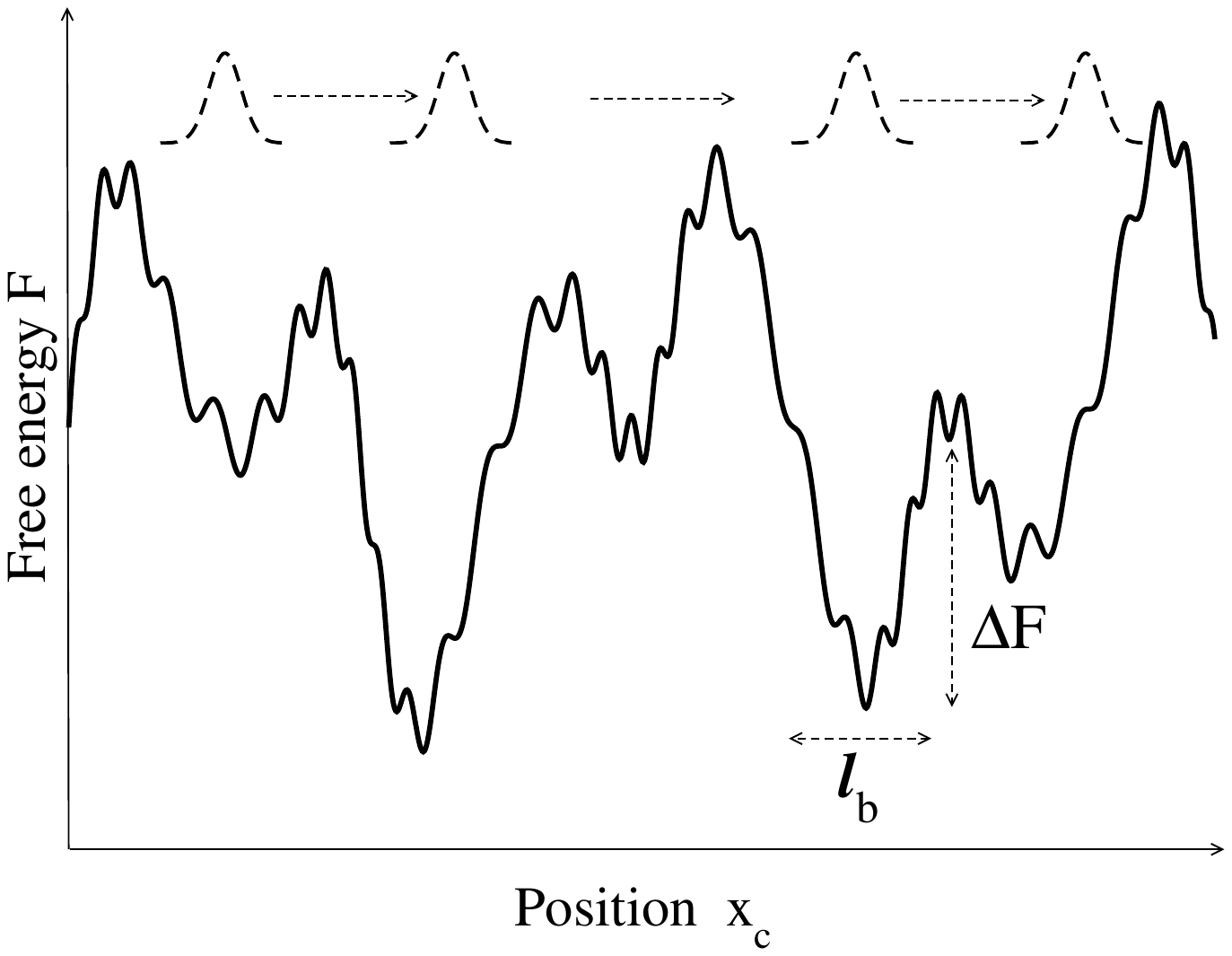}
\caption{Sketch of the free energy landscape probed by the clump of neural activity (dashed curve) moving through space. Fluctuations of the free energy are of the order of $\Delta F$, and are correlated over a space-scale equal to $\ell_b $.}
\label{paysage}
\end{figure}

How does this result extend to the case of multiple environments? We assume that the load and the level of noise are such that the clump is the stable phase of the system. The crosstalk between the environment in which the activity is localized and the other maps encoded in the couplings now hinders the motion of the clump center $x_c$. This effect can be intuitively modeled by the presence of an effective free energy potential acting on the clump, varying with the center position, $x_c$. We expect that this potential will be random and quenched (independent of time). This phenomenon is illustrated in Fig.~\ref{paysage} which sketches the free energy of the clump as a function of $x_c$. Two important features of this free energy landscape are the typical height of free energy fluctuations, $\Delta F$, and the typical space scale over which fluctuations are correlated, $\ell_b$. Those two quantities will be computed in Section \ref{sec:activated}. The barrier height $\Delta F$ is found to increase as the square root of the number $N$ of cells, which makes the diffusion coefficient vanish as the exponential of minus the square root of $N$. Hence, diffusion is strongly activated and the clump may remain trapped for a long time at specific space locations when the size of the neural population exceeds a few tens or hundreds, depending on the values of the control parameters. In practice, therefore, diffusion is possible in a small part of the stability region of the clump phase (close to the small $\alpha$ and large $T$ border) only. As expected the maximal size $N$ for which diffusion is possible increases with the fraction of silent cells in each environment (this fraction ranges from 50 to 80\% according to experiments \cite{ThompsonBest89}).

Diffusion of the clump within one environment coexist with the presence of abrupt transitions from one map to another. In such transitions, the clump of localized activity in the first environment disappear and reform in another environment, where the activity is now localized, and diffusion can resume. We show some examples of transitions in Section \ref{transit}. Disappearance and reformation take place at environment-specific place positions, corresponding to local ressemblances of the environments \cite{Rosay13}. Small values of $N$, which favor diffusion, make transitions more likely to occur, too. Diffusion within maps and transitions between maps are therefore two competing phenomena, both very important for the mobility of the clump.

The results above were obtained in the absence of any external input. In the presence of an external force the clump may however easily move, with a finite velocity. We have investigated the dependence of the velocity on the force value, and on the dimension of the space (1 or 2). However, the force cannot exceed a critical value above which the clump desintegrates, and the neural activity ceases to be localized. We estimate the upper bound on the force in Section \ref{sec:retrieval}. A force can also be used to move the clump towards a specific position in space, to retrieve a particular location. We show in Section \ref{sec:retrieval} that this mechanism can efficiently drive the clump to the desired position, in a time essentially independent of its initial position in the environment. Larger forces make the retrieval time smaller.

Finally we study several biologically inspired mechanisms, including adaptation and theta-related variations of the activity, with numerical simulations in Section \ref{sec:add}, and show how those mechanisms affect the diffusion properties of the neural clump. Adaptation seems to be particularly effective to avoid trapping in local minima of the free energy potential.

\section{Pure diffusion: single-environment case}

\label{sec:pure}

We start with a detailed study of the single-environment case. Since we have considered regularly spaced place fields, neglecting any noise coming from the learning process, there is no disorder in the connections in this case. We first define the dynamics undergone by the microscopic configurations ${\boldsymbol\sigma=\{\sigma_1,\ldots ,\sigma_N\}}$, in terms of transition probabilities between nearby configurations. We then show how the center of the clump emerges as a collective coordinate of the neural population. The dynamics can be described as a diffusion for the clump center, accompanied by low amplitude fluctuations of the clump shape around its equilibrium profile. We then report the results of Monte Carlo simulations, in excellent agreement with the analytical findings.

\subsection{Transition rates for the dynamics of the neural activity configuration $\boldsymbol \sigma$}

\label{sec:transitionrates}

The dynamics is defined as follows. We start from a configuration $\boldsymbol\sigma$ of the neural activity, whose corresponding 'energy' is defined as
\begin{equation}\label{energy}
E= -\sum_{i<j} J_{ij} \, \sigma_i\,\sigma_j\ .
\end{equation}
We then choose (1) a neuron $i$ uniformly at random among the $N(1-f)$ neurons which are silent, {\em i.e.} such that $\sigma_i=0$, and (2) a neuron $j$ uniformly at random among the $Nf$ neurons which are active, {\em i.e.} such that $\sigma_j=1$. Let us define the change in energy, $\Delta E$, when the states of both neurons are flipped, that is, $\sigma_i$ and $\sigma_j$ become, respectively, equal to 1 and 0. A short calculation leads to
\begin{equation}\label{energychange}
\Delta E = -\sum _{k(\ne i,j)} \big(J_{ik}-J_{jk}\big) \sigma_k\ .
\end{equation}
The joint flip of the two spins is accepted with rate (probability per unit of time) $\omega(\Delta E)$, satisfying detailed balance:
\begin{equation}
\frac{\omega(\Delta E)}{\omega(-\Delta E)} = \exp ( -\beta \,\Delta E) \ .
\end{equation}
A possible choice for the rate function is $\omega(\Delta E)=N\,\exp(-\beta \Delta E/2)$, or the Metropolis prescription: $\omega(\Delta E)=N$ if $\Delta E<0$, and $\omega(\Delta E)=N\,\exp(-\beta \Delta E)$ if $\Delta E\ge 0$. The multiplicative $N$ factor in the rate function $\omega$ ensures that the typical time for a round of the dynamical procedure ($N$ joint flip attempts) is independent of the system size, and equal to unity in the infinite size limit.

Note that the joint flip allows us to keep the global activity unchanged. The procedure is then iterated (choice of a new couple of spins, acceptance or rejection of the joint flip, and so on). As a consequence of detailed balance and of the obvious irreducibility of the Markov chain the system reaches equilibrium at long times.

\subsection{The clump is an emergent collective 'coordinate' of the neural activity}

\subsubsection{Transition rates for the dynamics of the density $\boldsymbol \rho$}

The previous dynamics over neurons defines an effective dynamics for the average density profile over space, $\boldsymbol \rho=\{\rho(x)\}$. Let us denote by $a=i/N$ and $b=j/N$ the reduced positions of the two spins we attempt to flip. Let also $J_w(u)=1$ if $|u|<\frac w2$, and 0 otherwise.  Observe first that the change in energy resulting from a joint flip is, according to (\ref{energychange}),
\begin{equation}\label{energychange2}
\Delta E = -\int dx \, \big(J_w(a-x)-J_w(b-x)\big)\, \rho (x)\ ,
\end{equation}
up to corrections of the order of $N^{-1/2}$ (the contributions coming from the spins $i$ and $j$, which are discarded in (\ref{energychange}), are of the order of $N^{-1}$). In the formula above $\boldsymbol \rho$ denotes the activity density associated to the configuration $\boldsymbol\sigma$. A rigorous procedure would require to bin the activity into boxes of width $W$, with $1\ll W \ll N$, and send $N\to \infty$ first, $W \to \infty$ next. To lighten notations we omit this binning procedure here.

The joint flip results in a change $\Delta \boldsymbol \rho$ of the activity density equal to 
\begin{equation}\label{activitychange}
\Delta \rho(x) =\frac 1N \delta(x-a) - \frac 1N \delta(x-b) \ ,
\end{equation}
and in a change of the free-energy (see Eq.~(11) in \cite{Monasson13}) given by

\begin{eqnarray}\label{fechange}
\Delta F &=&N\, {\cal F} [\boldsymbol \rho +  \Delta \boldsymbol \rho] -N\, {\cal F} [\boldsymbol \rho ]\nonumber\\
&  =& \frac{\delta {\cal F}}{\delta \rho(a)} -
\frac{\delta {\cal F}}{\delta \rho(b)} \nonumber \\
&=&  -\int dx \, \big(J_w(a-x)-J_w(b-x)\big)\, \rho (x)\nonumber\\
&& + T\log\left[ \frac{\rho(a)}{(1-\rho(a))}\right] - T\log\left[ \frac{\rho(b)}{(1-\rho(b))}\right]
\ ,
\end{eqnarray}
when $N$ is sent to infinity.

As the probability of choosing a silent spin at reduced position $a$ and an active spin at reduced position $b$ is equal to $ \frac{(1-\rho(a))\rho(b)}{f(1-f)}$ we may write the rate for the small change $\boldsymbol \rho \to \boldsymbol \rho +  \Delta \boldsymbol\rho$, 
\begin{eqnarray}\label{rate17}
\hat \omega(\boldsymbol\rho;a , b) &=& \frac{(1-\rho(a))\rho(b)}{f(1-f)}\;\omega(\Delta E) \nonumber\\
&=& \frac{(1-\rho(a))\rho(b)}{f(1-f)}\times\\
&&\omega\left( -\int dx \, \big(J_w(a-x)-J_w(b-x)\big)\, \rho (x) \right) \ .\nonumber
\end{eqnarray}
It is a simple check from equation (\ref{fechange}) that the ratio of the forward and backward rates is equal to
\begin{equation}\label{db2}
\frac{\hat \omega(\boldsymbol\rho;a , b)}{ \hat \omega(\boldsymbol\rho+\Delta \boldsymbol\rho; b,a)} =\exp ( -\beta \,\Delta F[\boldsymbol\rho]) \ .
\end{equation}
Hence detailed balance is obeyed at the level of activity density profiles $\boldsymbol\rho$.

\subsubsection{Fokker-Planck equation for the activity density $\boldsymbol \rho$}

Let us call ${\cal P}[\boldsymbol\rho, t]$ the probability density that the average density profile is equal to $\boldsymbol\rho$ at time $t$. Detailed balance condition (\ref{db2}) ensures that, at long times, equilibrium is reached and the activity density converges to its equilibrium value $\boldsymbol\rho^*$, as the infinite-size limit suppresses fluctuations.  We now propose a heuristic derivation of the Fokker-Planck equation satisfied by ${\cal P}$ at finite times $t$. For simplicity we will restrict to a simplified version of this equation, describing the evolution around the equilibrium profile $\boldsymbol\rho^*$ only.

The essential components of the Fokker-Planck equation are the diffusion tensor, the effective force as a function of the activity density, and the mobility tensor. The diffusion tensor is given by
\begin{eqnarray}
D( x,y)&=& \langle \Delta \rho(x) \;  \Delta \rho(y) \rangle \nonumber \\
&=& \frac{\delta(x-y)}{N\,f(1-f)} \bigg[ (1-\rho^*(x))\int db\,\rho^*(b)\, \omega ^*(x,b)\nonumber \\
&& +  \rho^*(x)\int da\,(1-\rho^*(a))\, \omega ^*(a,x)\bigg] \nonumber \\
&-& \frac 1{N\,f(1-f)} \bigg[ (1-\rho^*(x))\rho^*(y) \, \omega ^*(x,y)\nonumber \\
&&+ \rho^*(x)(1-\rho^*(y)) \, \omega ^*(y,x)\bigg]\ , 
\end{eqnarray}
where the average is taken over the joint flips $a,b$ with rate $\hat \omega$ (\ref{rate17}), and
\begin{equation}
\omega^*(x,y) \equiv\frac1N\, \omega \left( -\int dz \, \big(J_w(x-z)-J_w(y-z)\big)\, \rho^* (z) \right) \ .
\end{equation}
Note that $\omega^*$  is of the order of 1 as $\omega$ is of the order of $N$. We have $\langle \Delta \rho(x)\rangle =0$ for all positions $x$ since fluctuations cancel on average around the equilibrium density $\boldsymbol\rho^*$. It is easy to check that $\bf D$ is a real-valued, symmetric, and semi-definite positive operator:
\begin{eqnarray}
&&N\int dx \,dy \,\Phi(x)D(x,y)\Phi(y)  = \\
&&\int dx dy \frac{(1-\rho^*(x))\rho^*(y)}{f(1-f)} \, \omega ^*(x,y)\, \big( \Phi(x)-\Phi(y)\big)^2 \ge 0 \ .\nonumber
\end{eqnarray}
The only zero mode of $\bf D$ is uniform over space: ${\Phi(x)=\Phi_0}$. 

Under the action of diffusion a current of probability ${\bf J}^{dif}[\boldsymbol\rho,t]$ is produced, proportional to the gradient of ${\cal P}[\boldsymbol\rho, t]$ over the density space, and to the diffusion tensor. This current is an infinite-dimensional vector whose component $x$ is given by 
\begin{equation}
J^{dif}[\boldsymbol \rho,t] (x) = - \frac 12 \int dy\, D(x,y)\,  \frac{\delta  {\cal P}[\boldsymbol \rho,t]}{\delta \rho(y)} \ .
\end{equation}

We now turn to the force acting on the activity density, which we denote by ${\bf A}$. The force includes thermodynamic contributions, proportional to minus the gradient of the free-energy function $N\,{\cal F}$, and external input contributions (to be made more precise in Section \ref{sec:retrieval}). Under the action of this effective force a velocity ${\bf v}$ in the activity density space is produced, whose component $x$ at 'point' $\boldsymbol\rho$ is
\begin{equation}\label{mobility}
v[\boldsymbol \rho,t](x) = \int dy\, \mu(x,y)\, A[\boldsymbol \rho ,t] (y) \ ,
\end{equation}
where $ {\boldsymbol \mu}$ is the mobility tensor, and ${\bf A}[\rho ,t]$ is the force at 'point' $\boldsymbol\rho$ and time $t$. The components of the current of probability ${\bf J}^{force}[\boldsymbol\rho,t]$ resulting from the action of the force are
\begin{equation}
J^{force}[\boldsymbol\rho,t](x) =   {\cal P}[\boldsymbol \rho,t]\, v[\boldsymbol \rho,t](x) \ .
\end{equation}

The corresponding Fokker-Planck equation for ${\cal P}[\boldsymbol\rho, t]$ reads
\begin{equation}
\label{eq:FP}
\frac{\partial {\cal P}[\boldsymbol\rho, t]}{\partial t} = - \int dx \,\frac{\delta}{\delta \rho(x)}\big[
J^{dif}[\boldsymbol\rho,t](x)  + J^{force}[\boldsymbol\rho,t](x)\big]\ . 
\end{equation}
We see that ${\cal P}[\boldsymbol\rho]\propto \exp(-N\beta {\cal F}[\boldsymbol\rho])$ is a stationary solution of the Fokker-Planck equation above with the force given by $A(y)=\delta (-N {\cal F}[\boldsymbol \rho])/ \delta \rho(y)$,  if the mobility tensor is chosen to be
\begin{equation}\label{einstein}
 \mu (x,y)=\frac{\beta}{2}\, D(x,y)\ ,
\end{equation}
which is the celebrated Einstein identity.

\subsubsection{Quasiparticle description around the equilibrium density $\boldsymbol \rho^*$ and effective diffusion coefficient}\label{sec:langevin}

We are now able to write the Langevin equation for the activity density equivalent to the previous Fokker-Planck equation, with the result
\begin{eqnarray}\label{la1}
\frac{\partial \rho(x,t)}{\partial t} &=& -\int dy \, \mu(x,y)\,  \frac{\delta N {\cal F}[\boldsymbol \rho]}{ \delta \rho(y)}\nonumber\\
&&+ \int dy \, D^{1/2} (x,y)\, \eta(y,t)\ ,
\end{eqnarray}
where $\eta$ is a white noise process, uncorrelated in space and in time:
\begin{equation}
\langle \eta(y,t)\rangle=0 \ , \quad \langle \eta(y,t)\, \eta(y',t')\rangle = \delta(y-y') \, \delta (t-t')\ ,
\end{equation}
and ${\bf D}^{1/2}$ is the square root of $\bf D$ (in operator terms):
\begin{equation}
D(x,y) = \int dz\,  D^{1/2} (x,z)\,  D^{1/2} (z,y)\ . 
\end{equation}
Note that the drift term in (\ref{la1}) is of the order of 1 as ${N\gg 1}$, while the effective noise term is of the order of $N^{-1/2}$. We stress that the Langevin equation (\ref{la1}) is expected to be valid for $\boldsymbol\rho$ close to $\boldsymbol\rho^*$; far away from $\boldsymbol\rho^*$ the diffusion tensor would have a different value, as one would need to compute the connected 2-point correlation of the activity density fluctuations.

Let us write now $\boldsymbol\rho=\boldsymbol\rho^*+\boldsymbol\epsilon$, with $\boldsymbol\epsilon$ 'small'. Then
\begin{equation}
\frac{\delta \beta {\cal F}[\boldsymbol \rho]}{ \delta \rho(y)}= \int dy' \, H(y,y') \,\epsilon (y')\ ,
\end{equation} 
where
\begin{align}
H(x,y) &= \left.\frac{\delta ^2 \beta {\cal F}}{\delta \rho(x) \delta \rho(y)} \right|_{\boldsymbol\rho^*} \nonumber
\\ &= -\beta \, J_w(x-y) + \frac{\delta (x-y)}{\rho^*(x)(1-\rho^*(x))}\ .
\end{align}
Langevin equation (\ref{la1}) reduces to a Ornstein-Uhlenbeck process for $\boldsymbol\epsilon$, described by 
\begin{eqnarray}\label{la2}
\frac{\partial \epsilon(x,t)}{\partial t} &=& - \frac N2 \int dy\,dz\,D(x,y)\, H(y,z)\,\epsilon(z,t) \nonumber 
\\ &+&\int dy\,D^{1/2}(x,y)\,\eta(y,t)\ .
\end{eqnarray}
The integral of the right hand side member above over $x$ vanishes since the constant function 1 is an eigenmode of ${\bf D}$ and ${\bf D}^{1/2}$ with zero eigenvalue. So $\int dx\, \epsilon (x,t)$ is independent of time, and equal to zero according to the initial condition at time $t=0$: the activity is constant, as was expected from the use of joint flips for the elementary moves of the dynamics.

Let us denote by $u_m (x)$ and $\lambda_m$ the eigenmodes and the (real-valued) eigenvalues of the operator $\frac N2{\bf D}\cdot {\bf H}$. Then
\begin{equation}
\frac{d \epsilon_m}{dt}(t)=-\lambda_m \, \epsilon_m(t) + \xi_m(t)\ , 
\end{equation}
 where  $\xi_m(t)$ and $\epsilon_m(t)$ denote the components on $\boldsymbol u_m$ of $\boldsymbol \eta(t)$ and $\boldsymbol\epsilon(t)$ respectively. Note that all eigenvalues are positive as the equilibrium profile of the clump is a minimum of the free energy. We find that:
\begin{itemize}
\item {For the modes $m$ with $\lambda _m>0$}: 
\begin{equation}
\epsilon_m(t)=\epsilon_m(0)e^{-\lambda_m t}+\int_0^t\mathrm{d}s\,\xi_m(s)e^{-\lambda_m(t-s)}\ ,
\end{equation}
These modes reach equilibrium at long times. More precisely the equilibrium distribution of the coefficient $\epsilon_m$ is asymptotically Gaussian with a variance proportional to the variance of the noise term and to the inverse of $\lambda_m$. Loosely speaking, those modes are thermalized at very low temperature (of the order of $1/N$) and describe very weak fluctuations around the equilibrum clump shape $\boldsymbol\rho ^*$.
\item {For the zero mode (associated to $\lambda_{0}=0$)}: 
\begin{equation}
\epsilon_{0}(t)=\epsilon_{0}(0)+\int_0^t\mathrm{d}s\,\xi_{0}(s)\ .
\end{equation}
 This mode  freely diffuses with a small diffusion coefficient of the order of $1/N$.
\end{itemize}

 It is easy to convince oneself that the only zero mode of $\bf H$, denoted by $u_0$, is proportional to the derivative of the equilibrium clump shape, 
\begin{equation}\label{v0}
u_0(x) = \frac 1{\sqrt{\int dy \left(\frac{d\rho^*(y)}{dy} \right)^2}}  \frac{d\rho^*(x)}{dx}\ .
\end{equation}
Indeed, a global translation of the clump by $\delta x$ does not affect the free energy. As ${\rho^*(x+\delta x) \simeq \rho^*(x)+\delta x\,  \frac{d\rho^*(x)}{dx}}$ we conclude that (\ref{v0}) is the normalized zero mode of $\bf H$. Note that, in more than one dimension, the derivative of $\rho^*(x)$ in (\ref{v0}) must be replaced by the gradient vector with respect to the space coordinates.

Hence, the effective diffusion coefficient characterizing the diffusive motion of the center of the clump is  given by
\begin{align}
\label{eq:D0}
D_0 &=  \langle u_0| D | u_0 \rangle= \\
& \frac 1{N}\int dx dy  \frac{(1-\rho^*(x))\rho^*(y)}{f(1-f)} \omega ^*(x,y) \big( u_0(x)-u_0(y)\big)^2\nonumber\ . 
\end{align}
This prediction is in very good agreement with simulations, as detailed in Section \ref{sec:numsim}.

\subsubsection{Effective mobility of the quasiparticle}

The velocity $\bf v$ of the density profile in the $\boldsymbol\rho$-space in response to an external force $\bf A$ is controlled by the mobility tensor $\boldsymbol\mu$, see (\ref{mobility}) and (\ref{einstein}). Here we derive an explicit expression for the effective mobility velocity of the center of the clump, hereafter denoted by $V_0$, as a function of the applied force. We assume that the clump behaves as a quasiparticle, \emph{i.e.} that the temperature and the applied force are not too large.

The velocity $v(x)$ in (\ref{mobility}) can be decomposed as a linear combination of the different eigenmodes $u_m(x)$, see Section \ref{sec:langevin}. According to the results above all projections on the modes $m\ne 0$ will decay exponentially fast to zero. The projection along $u_0(x)$ is simply related to the velocity $V_0$ of the center of the clump. Indeed, consider the displacement of the clump during the time $\delta t$, from the activity profile $\rho(x,0)=\rho^*(x)$ to $\rho(x,\delta t) =\rho^*(x-V_0\,\delta t)$. The velocity of the profile in the $\boldsymbol\rho$-space is
\begin{eqnarray}\label{gh8}
v(x) &=&\frac{\rho^*(x-V_0\,\delta t) -\rho^*(x)}{\delta t} \\
&=& - V_0 \, \frac{d\rho^*(x)}{dx} = -V_0\, \sqrt{\int dy \left(\frac{d\rho^*(y)}{dy} \right)^2}\, u_0(x)\ . \nonumber
\end{eqnarray}
Comparing expressions (\ref{mobility}), (\ref{einstein}), and (\ref{gh8}) we deduce the following expression for the effective velocity of the center of the clump:
\begin{equation}
\label{eq:mob}
V_0 = \int dx\,  \mu_0 (x)\; A(x) \ ,
\end{equation}
where $A(x)$ is the force acting on position $x$ of the clump, and the component $\mu_0(x)$ of the effective mobility is
\begin{equation}
\label{eq:mob2}
 \mu_0 (x) =-\beta\;\frac{\int dy\;D(x,y)\;u_0(y)}{2\sqrt{\int dy \left(\frac{d\rho^*(y)}{dy} \right)^2}}\ .
\end{equation}
Note that the effective mobility is, as the effective diffusion coefficient, of the order of $1/N$. This theoretical prediction will be shown to be in very good agreement with simulations in Section \ref{sec:retrieval}.

\subsection{Numerical simulations}

\label{sec:numsim}

We now report Monte Carlo simulations done with the Metropolis prescription above, and in the region of stability of the clump phase. In this Section we consider only the motion in the absence of an external force; the case of an input is considered in Section \ref{sec:retrieval}.

\begin{figure}

\centering

\includegraphics[width=\columnwidth]{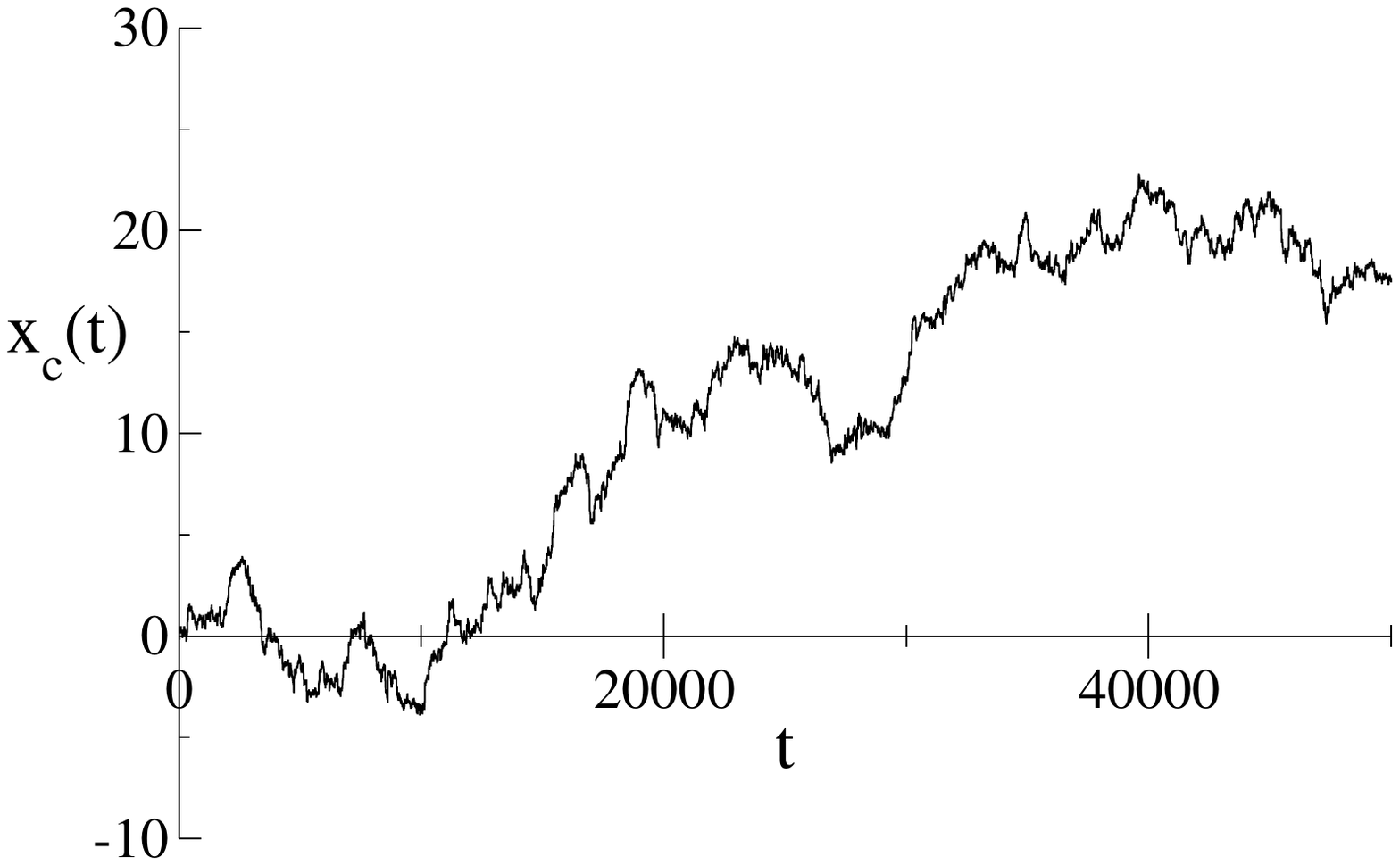}

\caption{Position $x_c$ vs. time $t$ of a freely diffusing clump in dimension 1, for $\alpha=0$ and $50000$ rounds of Monte Carlo simulation with $N=333$ neurons, and noise $T=0.006$. Time unit = 1 round of $20N$ steps.}

 \label{fig:traj1d}

\end{figure}

\begin{figure}

\centering

\includegraphics[width=\columnwidth]{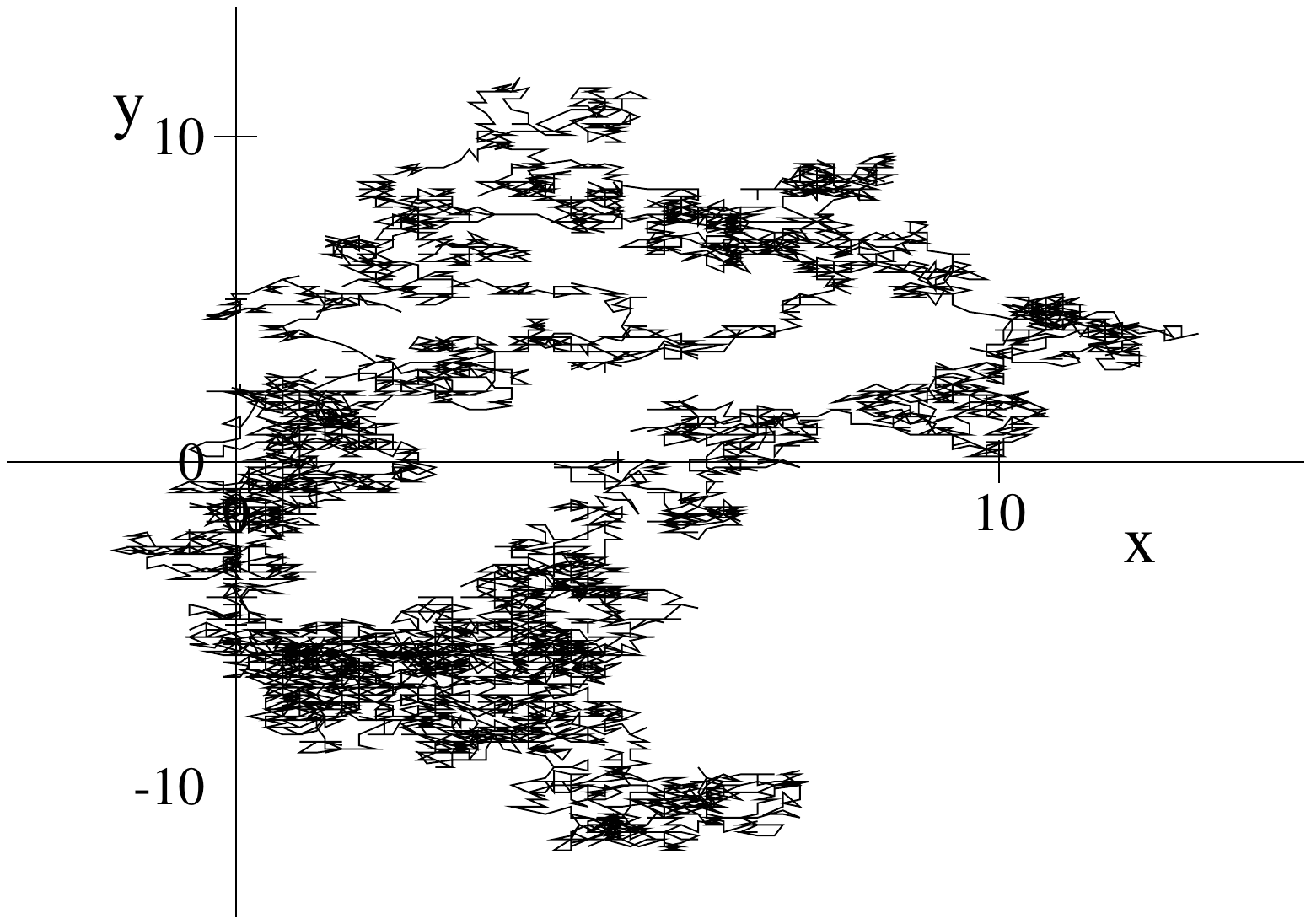}

\caption{Trajectory of a freely diffusing clump in dimension 2, for $\alpha=0$ and $50000$ rounds of Monte Carlo simulation with $N=32\times32$ neurons, and noise $T=0.005$.  Time unit = 1 round of $20N$ steps.}

 \label{fig:traj2d}

\end{figure}

We observe that the stochastic evolution of neural units at the microscopic level results in a macroscopic erratic motion of the clump, both in one and two dimensions. To characterize this motion we compute the position of the clump center from the coarse-grained activity of the network. Space is binned into boxes of size approximatively equal to the clump width. We look for the box where the activity is maximal at time $t$, and compare it to the box of maximal activity at time $t-1$, taking into account periodic boundary conditions. This provides us with the displacement of the clump between times $t-1$ and $t$. The position of the clump is obtained by adding those displacements over time. Two examples of trajectories are shown in Figs.~\ref{fig:traj1d} and \ref{fig:traj2d}.

\subsubsection{Method for estimating the diffusion coefficient}

We assume that the trajectories of the clump correspond to realizations of a diffusion process with diffusion constant $D$. We want to infer $D$ from the $t_M$ measured displacements $\{\Delta x_t\}_{t=1,\ldots,t_M}$. Bayes' formula gives the posterior distribution for $D$:
\begin{equation}
\label{eq:bayes}
P(D|\{\Delta x_t\})=\frac{P(\{\Delta x_t\}|D)\;P_0(D)}{P(\{\Delta x_t\})}\ .
\end{equation}
We choose a flat prior over the diffusion coefficients: $P_0(D)=\Theta(D)$ (Heaviside step function). The likelihood of the trajectories given $D$ is 
\begin{equation}
 P(\{\Delta x_t\}|D)=\prod\limits_{t=1}^{t_M}\frac{1}{\sqrt{2\pi D}}\exp\left(-\frac{\Delta x_t^2}{2D}\right) \ ,
\end{equation}
where we have fixed the time interval between two successive measured positions to unity. The denominator in (\ref{eq:bayes}) is a normalization factor.

Maximization of $P(D|\{\Delta x_t\})$ with respect to $D$ in (\ref{eq:bayes}) gives the most likely value for $D$, here denoted $D^*$:
\begin{equation}
D^*=\frac1{t_M}\sum\limits_{t=1}^{t_M} \Delta x_t^2 \ ,
\end{equation}
and the standard deviation of $D$ with posterior distribution (\ref{eq:bayes}) is about $\delta=D^*/\sqrt{t_M}$.

\subsubsection{Corrections of systematic errors due to binning}

 The exact position of the center of the clump of activity is not well defined in simulations. As explained above, we therefore bin space into boxes of length $a$ roughly equal to the width of the clump, and estimate the diffusion coefficient through
\begin{equation}\label{dmes}
D^\text{mes}\equiv\frac1{t_M}\sum\limits_{t=1}^{t_M} (a\;\Delta_t)^2 \ ,
\end{equation}
with $\Delta _t=0,\pm 1, \pm 2, ...$ denotes the change in the box number between times $t-1$ and $t$.

We now want to estimate the error on the estimate of the diffusion coefficient due to binning.
Let us consider a pure diffusion process with coefficient $D$ in one-dimensional continuous space $x$. The trajectory is observed during $t_M$ steps, and $D^\text{mes}$ is estimated according to (\ref{dmes}). During a unit time interval $t\to t+1$ the continuous walker has moved by a quantity $z_t$, which is a Gaussian random variable with zero mean, and standard deviation equal to $\sqrt D$. We generically note $k$ the integer part of the ratio of $z_t$ over $a$, and $u$ the remainder of the division, {\em i.e.} $z_t=k\,a+u$.  We need to relate $\Delta_t$ to $z_t$, that is, to $k$ and $u$. 








For simplicity, we consider that, up to time ${t=t_1\equiv\frac{a^2}{4D}}$ (diffusion time in a box), the displacement is counted from the middle of a box, while, for larger times $t$, the clump position is uniform at random in the box. (This approximation is not valid when $D$ is too small, typically $D\lesssim10^{-5}$: in simulations, we therefore have to adapt the length of one round in order to avoid low $D$ effects when applying the correction.)  It is then easy to show that, for $t>t_1$, $\Delta _t =k$ with probability $1-\frac ua$ and $\Delta _t=k+1$ with probability $\frac ua$. 





We conclude that the estimate of the diffusion coefficient is on average
\begin{eqnarray}
\langle D^\text{mes}\rangle&=&\frac{a^2}{t_M}\bigg( t_1 \sum\limits_{k=-\infty}^{+\infty}\int\limits_{-\frac a2}^{\frac a2}\frac{\mathrm{d}u}{\sqrt{2\pi D}}e^{-(ka+u)^2/(2D)}\, k^2\nonumber \\
&+& (t_M-t_1) \sum\limits_{k=-\infty}^{+\infty}\int\limits_{0}^{a}\frac{\mathrm{d}u}{\sqrt{2\pi D}}\,e^{-(ka+u)^2/(2D)}\nonumber \\
&\times & \left[ k^2 \left(1-\frac ua\right)+(k+1)^2 \left(\frac ua\right) \right]\bigg)\ .
\end{eqnarray}
The formula above gives the estimated $D^\text{mes}$ as a function of the 'true' diffusion coefficient $D$. In practice, for each $D^*$ estimated according to (\ref{dmes}) we numerically solve $D^\text{mes}(D)=D^*$. 

The same reasoning in two dimensions leads to a similar result (with a multiplicative factor 2 because we bin both the $x$ and the $y$ axes).

\subsubsection{Statistical error bars}

Once the individual values ${\{D^*_n\}_{n=1...N_\text{sim}}}$ measured in $N_\text{sim}$ simulations have thus been corrected, we estimate the diffusion coefficient $D$ as their average:
\begin{equation}
D = \frac 1{N_{sim}} \sum_{n=1}^{N_{sim}}D^*_n \ .
\end{equation}
The error bars on the inferred $D$ must take into account two sources of uncertainty: the width $\delta_n=D^*_n/\sqrt{(t_M)_n}$ of the distribution of each $D^*_n$ due to the randomness in the Monte Carlo process, and the standard deviation $\tilde \delta$ of the diffusion coefficients due to the random realization of the maps in each simulation. In practice, for the long MC runs, we consider that the former error is negligible compared to the latter. We therefore estimate the error bar on $D$ through
\begin{equation}
\tilde\delta  = \frac 1{N_{sim}}\sqrt{ \sum_{n=1}^{N_{sim}} \left(\big(D^*_n\big)^2 - D^2\right) }\ .
\end{equation}

\begin{figure}
\centering
 \includegraphics[width=\columnwidth]{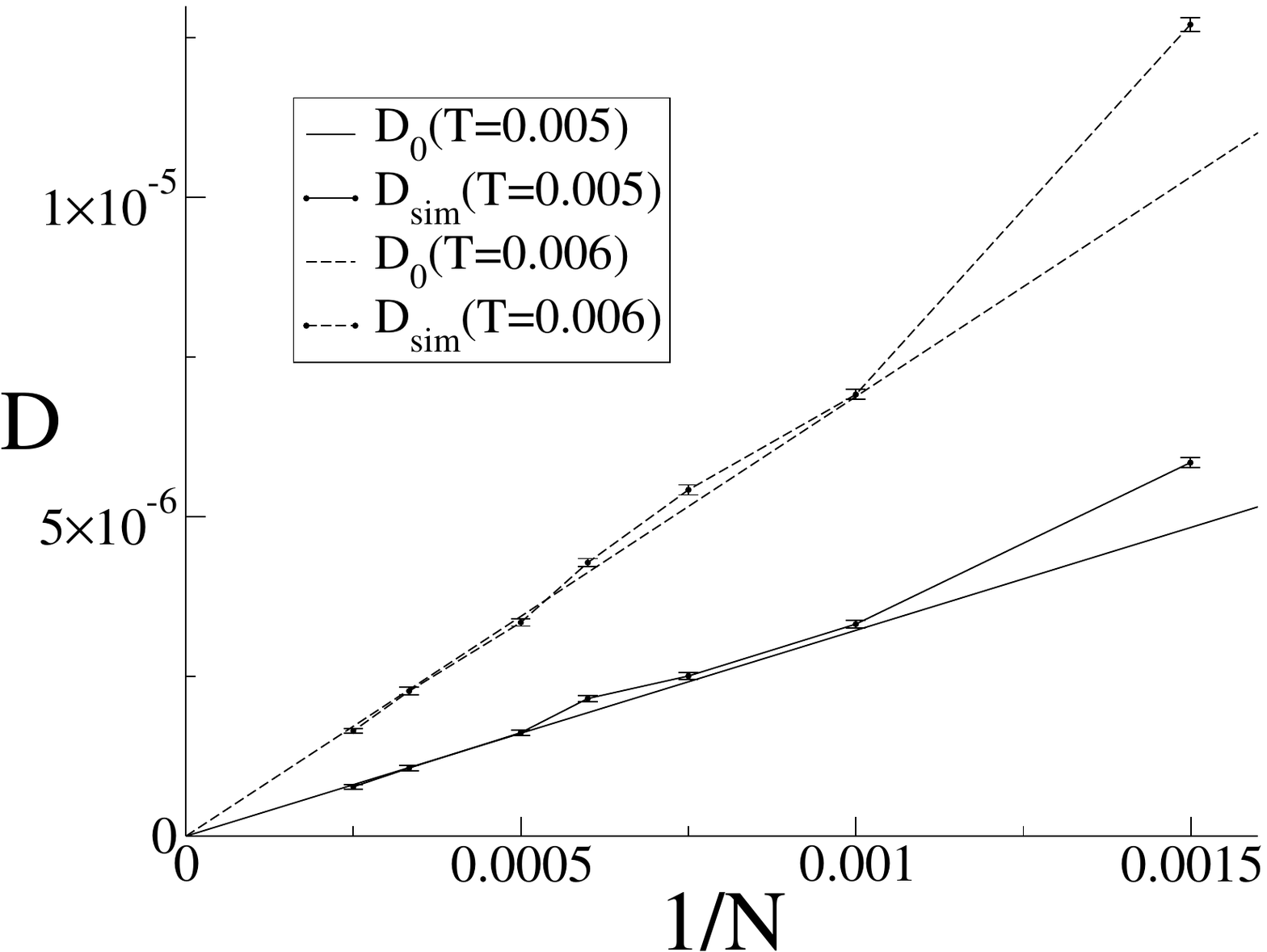}
\caption{Diffusion of the clump in the single environment case ($\alpha=0$) and 1-dimensional space. The theoretical prediction for the diffusion constant, $D_0$, given by Eq.(\ref{eq:D0}), is plotted as a function of $\frac1N$ for $T=0.005$ (dashed lines) and $T=0.006$ (full lines) and compared to the results  of Monte Carlo simulations $D_{\text{sim}}$ (after correction of the binning effect). The agreement with the analytical prediction (done in the $N\to\infty$ limit) improves as $N$
increases. This also explains why the discrepancy is larger than error bars for smaller $N$. Therefore, simulations corroborate well the theoretical analysis and the diffusion properties of the clump can be understood analytically in the single environment case. Simulation time: 1000 rounds of $100N$ steps. Depending on the computational cost, each point is averaged over a number of simulations ranging from 5 (for large $N$) to 100. }
 \label{fig:dif0}
\end{figure}

We compare the value of $D$ to the theoretical prediction $D_0$ given by (\ref{eq:D0}). The results in dimension 1 are plotted in Fig.~\ref{fig:dif0}, and show that the agreement is very good. The prediction gets better and better as $N$ increases: indeed, it is valid in the large $N$ limit.

\section{Activated diffusion: multiple-environment case}

\label{sec:activated}

In the presence of multiple environments the motion of the clump within the retrieved environment is not purely diffusive any longer. The crosstalk between the stored maps indeed creates an effective (free energy) potential for the clump, which is not uniform over the space, as sketched in Fig.~\ref{paysage}. In this section we first compute the typical height $\Delta F$ of the barriers in this potential, and their typical width $\ell_b$. We then show results of simulations, and address the issue of partial activity of place cells.

\subsection{Characterization of free energy barriers}

\label{sec:F1}

\subsubsection{Barrier heights}

In the presence of disorder, the distribution of the free energy $F_J=- T \log Z_J(T)$ is centered around its typical value, with a non-zero width for finite size $N$. To compute this width, we use the replica method. Expanding the $n^{th}$ moment of the partition function, $ \overline{Z_J(T)^n}$, in cumulants of $F_J$ we write
\begin{eqnarray}
\overline{Z_J(T)^n}&=&\overline{\exp\big(-n\,\beta\,F_J\big)}\\
&=&\exp \big( -n\,\beta\,\overline{F_J}+\frac 12 n^2\beta^2(\overline{F_J^2}-\overline{F_J}^2)+\cdots \big)\ ,\nonumber
\end{eqnarray}
Hence, the variance of $F_J$ can be computed from the knowledge of the second derivative of  $ \overline{Z_J(T)^n}$ in $n=0$:
\begin{equation}
\overline{F_J^2}-\overline{F_J}^2 = \left. \frac{\partial^2 }{\partial n^2}\right|_{n\to 0} T^2\log  \overline{Z_J(T)^n}
\ .
\end{equation}
The calculation of this second derivative is reported in Appendix B, with the result:
\begin{equation}
\label{eq:deltaF}
\overline{F_J^2}-\overline{F_J}^2 = V(\alpha,T)\;N
\ ,
\end{equation}
where
\begin{eqnarray}
\label{eq:V}
V(\alpha,T) &=&-\alpha \,r\, q+\alpha T^2(q-f^2)^2\,\varphi(q,T)\\
&&+T^2\int\mathrm{d}x\int\mathrm{D}z\log^2(1+e^{\beta z\sqrt{\alpha r}+\beta\mu(x)})\nonumber\\
&&-T^2\int\mathrm{d}x\left(\int\mathrm{D}z\log\big(1+e^{\beta z\sqrt{\alpha r}+\beta\mu(x)}\big)\right)^2\ .\nonumber
\end{eqnarray}
In the formula above, $\mu(x)$ is the field conjugated to the average density $\rho(x)$ (not to be confused with the mobility tensor $\boldsymbol\mu$ introduced above), and $r$ is the conjugated force to $q$, see Appendix A; $Dz=\frac{dz}{\sqrt{2\pi}}\exp(-z^2/2)$ denotes the Gaussian measure. The function $\varphi(q,T)$ is given by
\begin{equation}
\label{eq:varphi1d}
\varphi^{1D}=\sum\limits_{k\geq1}\bigg(\frac{T\,\pi k}{\sin(\pi k w)}+q-f\bigg)^{-2}
\end{equation}
in dimension 1, and by
\begin{align}
\label{eq:varphi2d}
\varphi^{2D}=2\sum_{\underset{\neq(0,0)}{(k_1,k_2)}}\bigg(\frac{T\,\pi^2\,k_1\,k_2}{\sin(\pi\ k_1\sqrt{w})\sin(\pi\ k_2\sqrt{w})} +q-f\bigg)^{-2}
\end{align}
in dimension 2.

\begin{figure}

\centering

\includegraphics[width=\columnwidth]{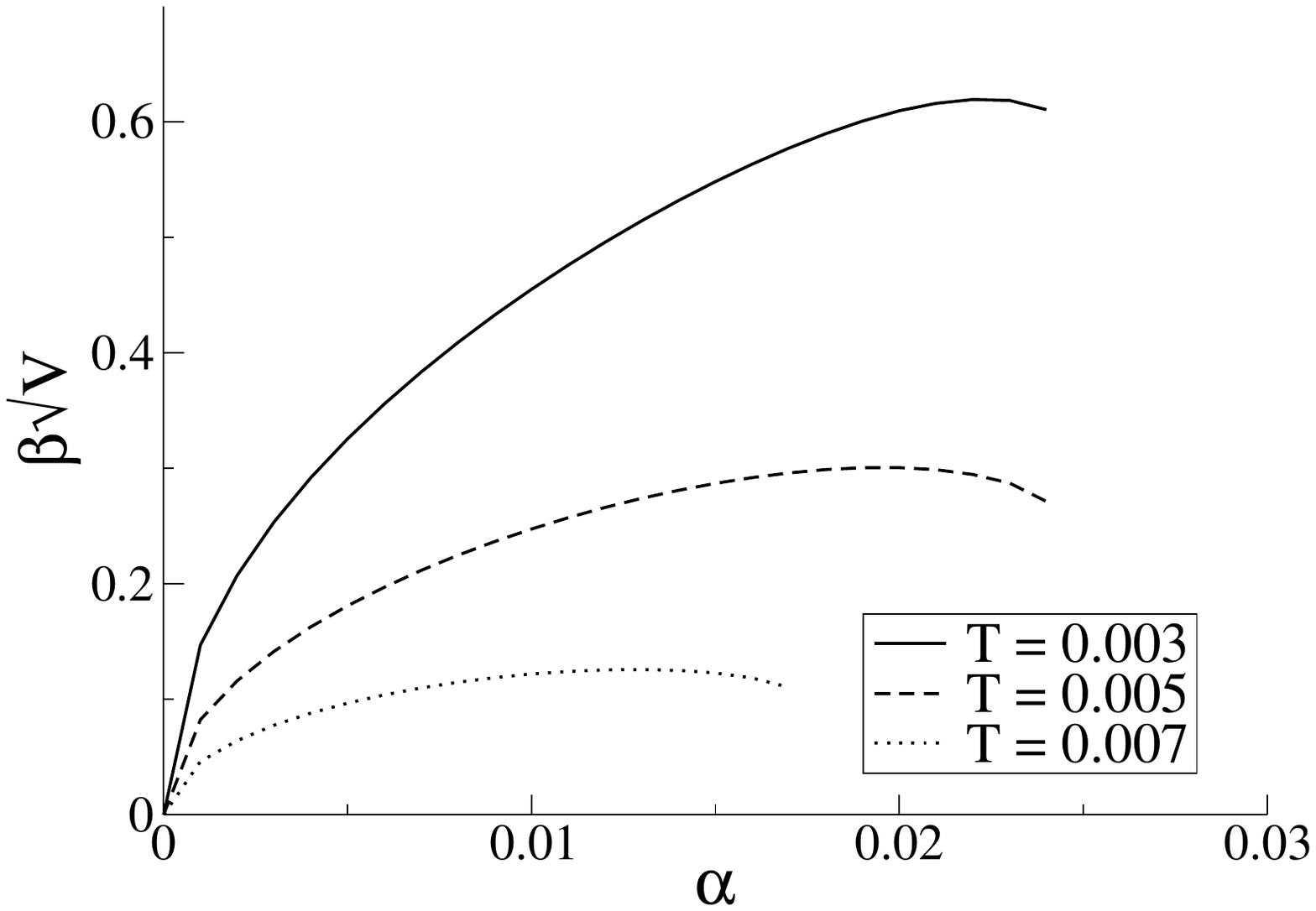}

\caption{Standard deviation $\beta\sqrt{V}$ of the free energy (in units of the temperature and divided by $\sqrt N$) as a function of the load $\alpha$ for fixed temperature $T$. Lines end at the clump instability limit.}

 \label{fig:F1}

\end{figure}

The typical barrier height, $\Delta F$, is given by the standard deviation of the free energy: $\Delta F= \sqrt N\,\sqrt{V}$ from Eq.~(\ref{eq:deltaF}). We have computed $V$ for different values of $\alpha,T$ and verified that it is a definite positive quantity.  We plot in Fig.~\ref{fig:F1} the barrier height $\Delta F$, after division by $\sqrt{N}$, as a function of the load $\alpha$. We see that $\Delta F$ increases very quickly with the load for small $\alpha$, and reaches a maximal value close to the stability boundary of the clump phase.

To gain some intuition on the barriers heights  we look for a simple estimate of the standard deviation $\Delta E$ of the energy $E=-\sum_{i<j} J_{ij}\sigma_i\sigma_j$. To do so, we keep the spin configuration fixed, and compute the variations due to the stochastic coupling matrix $J$, with the result 
\begin{equation}
 \Delta E \sim f(1-f)\sqrt{\frac{\alpha \;w(1-w)}{2}}\sqrt N \ ,
\end{equation}
to dominant order in $N$. Numerically, we find that $\Delta E$ in the formula above takes values close to $\Delta F$. Hence, the much simpler formula for $\Delta E$ offers some insight on the order of magnitude of the barriers, as well as on their dependence on the model parameters.

\begin{figure}

\centering

 \includegraphics[width=\columnwidth]{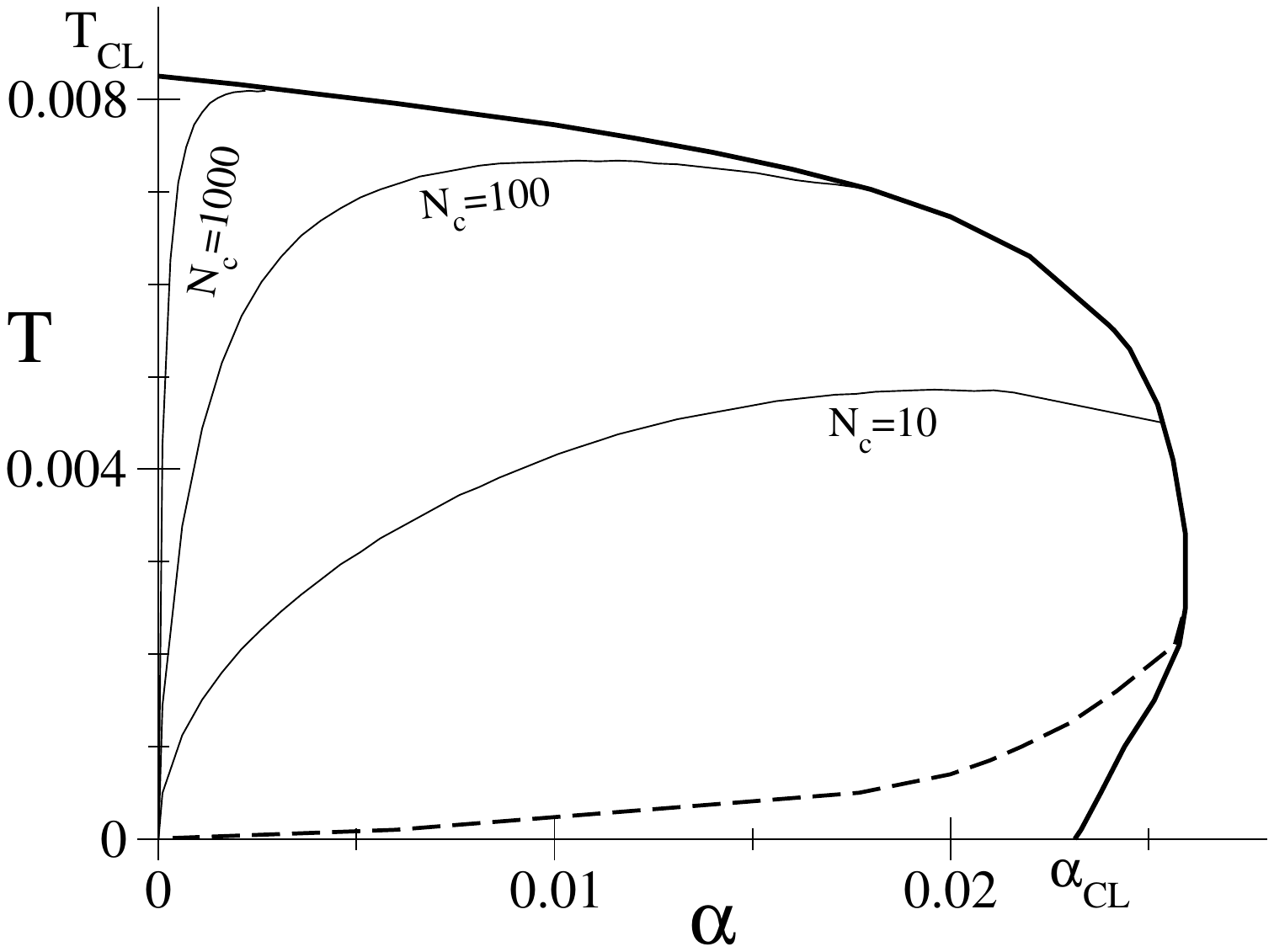}

\includegraphics[width=\columnwidth]{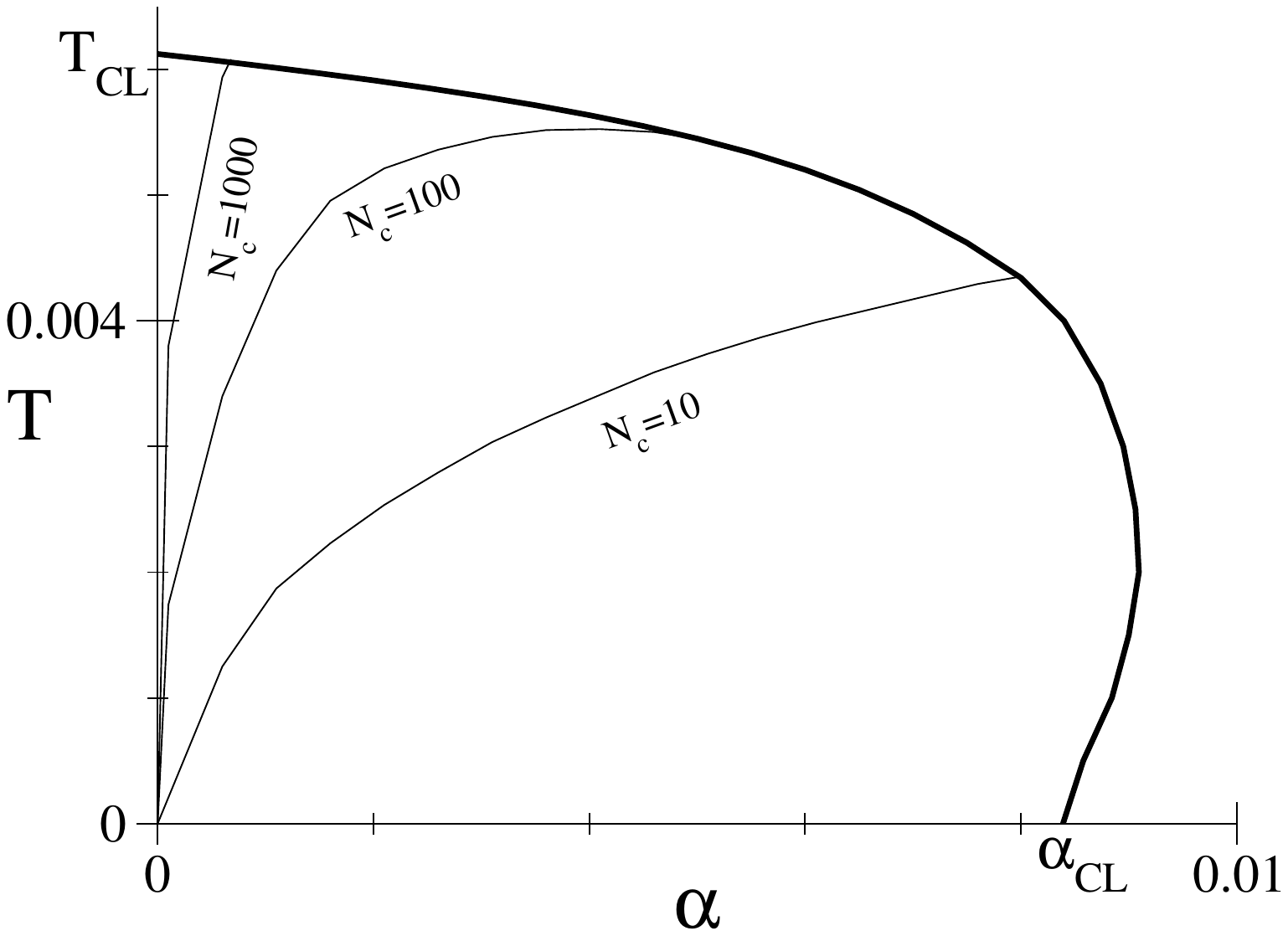}

\caption{Contour lines of constant $N_c$ in the phase diagrams of the one- (top) and two-dimensional (bottom) models. In one dimension, for a given $N_c$, the area of the diagram above the contourline corresponds to free diffusion, while in the area below the diffusion is activated. In two dimensions, this distinction is less clear due to the possible by-passing of free energy barriers (see text).}

 \label{fig:ln}

\end{figure}

As the barrier heights against diffusion scale as $\sqrt{N}$ we can plot in the phase diagram the contour lines of different cross-over sizes $N_c$, corresponding to barrier heights such that ${\beta\Delta F=1}$. The cross-over size $N_c$ is thus defined through
\begin{equation}\label{nc}
N_c= \frac 1{\beta^2 V(\alpha,T)}\ . 
\end{equation}
The outcome is shown in Fig.~\ref{fig:ln}. In dimension 1 we can estimate that diffusion will be approximatively free for $N<N_c$. For $N>N_c$ barriers cannot be neglected, and diffusion is activated. We see that, except in a narrow region of the phase diagram, the clump cannot freely diffuse for realistic values of $N$ (of the order of thousands). In dimension 2, this argument is not true anymore because barriers can be bypassed. Nevertheless, simulations show that diffusion is quite limited also in that case, albeit to a lesser extent (see Section \ref{sec:retrieval}). Furthermore, in both 1 and 2 dimensions, in the low $\alpha$ - high $T$ region where diffusion can occur, we observe in simulations that this process is in competition with transitions between environments (see Section \ref{transit}).

\subsubsection{Barrier widths}

\label{sec:width}

In order to estimate the typical width $\ell_b$ of the barriers depicted in Fig.~\ref{paysage}, we calculate the correlation between the free energies (denoted $F_J(x)$ and $F_J(y)$) of the clump centered respectively on two positions $x$ and $y$ of space, that is 
\begin{equation}
\text{cov}\big(F_J(x),F_J(y)\big)\equiv\overline{ F_J (x) F_J (y)}-\overline{F_J (x)}\ \overline{F_J (y)}\ .
\end{equation}
This quantity can be derived using the replica method. We split the $n$ replicas in two groups: the first $\frac n2$ replicas have an activity profile centered in $x$, while the remaining $\frac n2$  replicas have an activity profile centered in $y$. All $n$ replicas share the same interaction matrix $J$, and are coupled once these quenched couplings are averaged out. The resulting partition function for the $n$-replica system reads 
\begin{equation}
Z(n,x,y)=\overline{\exp\left(-\frac n2 \,\beta\,\big(F_J (x)+F_J (y)\big)\right)} \ .
\end{equation}
Similarly to the  calculation above, by expanding in cumulants and taking the second derivative of  $Z(n,x,y)$ in $n=0$, 
\begin{equation}
\label{eq:VW}
\left. \frac{\partial^2 }{\partial n^2}\right|_{n\to 0} T^2\log  Z(n,x,y)=\frac{N}{2}\left(V+W(x,y)\right)\ ,
\end{equation}
where 
\begin{equation}
W(x,y) \equiv \frac 1N \, \text{cov}\big(F_J(x),F_J(y)\big)\ .
\end{equation}
$V$ was defined in (\ref{eq:deltaF},\ref{eq:V}) and we use that, by translational invariance, the average of $F_J(x)$ over $J$ does not depend on the position $x$. By translational invariance again, $W(x,y)$ only depends on the distance $x-y$, and is equal to $W(x-y)$.

The calculation of $Z(n,x,y)$  is detailed in Appendix \ref{app:spatialcorr}. We denote $q_{12}$ the overlap between two replicas respectively belonging to the group with a clump in $x$ and the group with a clump in $y$. The outcome is 
\begin{align}
\label{eq:W}
W(x-y)&=-\alpha \,r_{12}\,q_{12}+\alpha T^2(q_{12}-f^2)^2\,\varphi(q,T)\nonumber\\
&-T^2\int\mathrm{d}x'\bigg[\int\mathrm{D}u\log \left(1+e^{\beta\sqrt{\alpha r}u+\beta\mu(x')}\right)\nonumber\\
&\cdot\int\mathrm{D}v\log \left(1+e^{\beta\sqrt{\alpha r}v+\beta\mu_1(x'-x+y)}\right)\nonumber\\
&-\int\mathrm{D}u\mathrm{D}v\ \kappa(u,v)\nonumber\\
& \cdot\log \left(1+e^{\beta\sqrt{\alpha (r-r_{12})}u+\beta\mu(x')}\right)\nonumber\\
&\cdot\log \left(1+e^{\beta\sqrt{\alpha (r+r_{12})}v+\beta\mu(x'-x+y)}\right)\bigg] \,
\end{align}
where
\begin{equation}\label{kappa}
\kappa(u,v) = \exp\left(\frac{r_{12}}{2}\left(\frac{u^2}{r+r_{12}}-\frac{v^2}{r-r_{12}}+\frac{2uv}{\sqrt{r^2-r_{12}^2}}\right)\right)\ ,
\end{equation}
and
\begin{align}
q_{12}=\int\mathrm{d}x'&\int\mathrm{D}u\mathrm{D}v\,\kappa(u,v)/ \big[1+e^{-\beta u\sqrt{\alpha (r-r_{12})}-\beta\mu(x')}\big]\nonumber\\
&/ \big[1+e^{-\beta v\sqrt{\alpha (r+r_{12})}-\beta\mu(x'-x+y)}\big]\ .
\end{align}
The conjugated parameter is $r_{12}=2T^2(q_{12}-f^2)\varphi(q,T)$. Parameters $q,r,\mu(x)$ are found from the extremization of the free energy given in Appendix A.

\begin{figure}

\centering

 \includegraphics[width=\columnwidth]{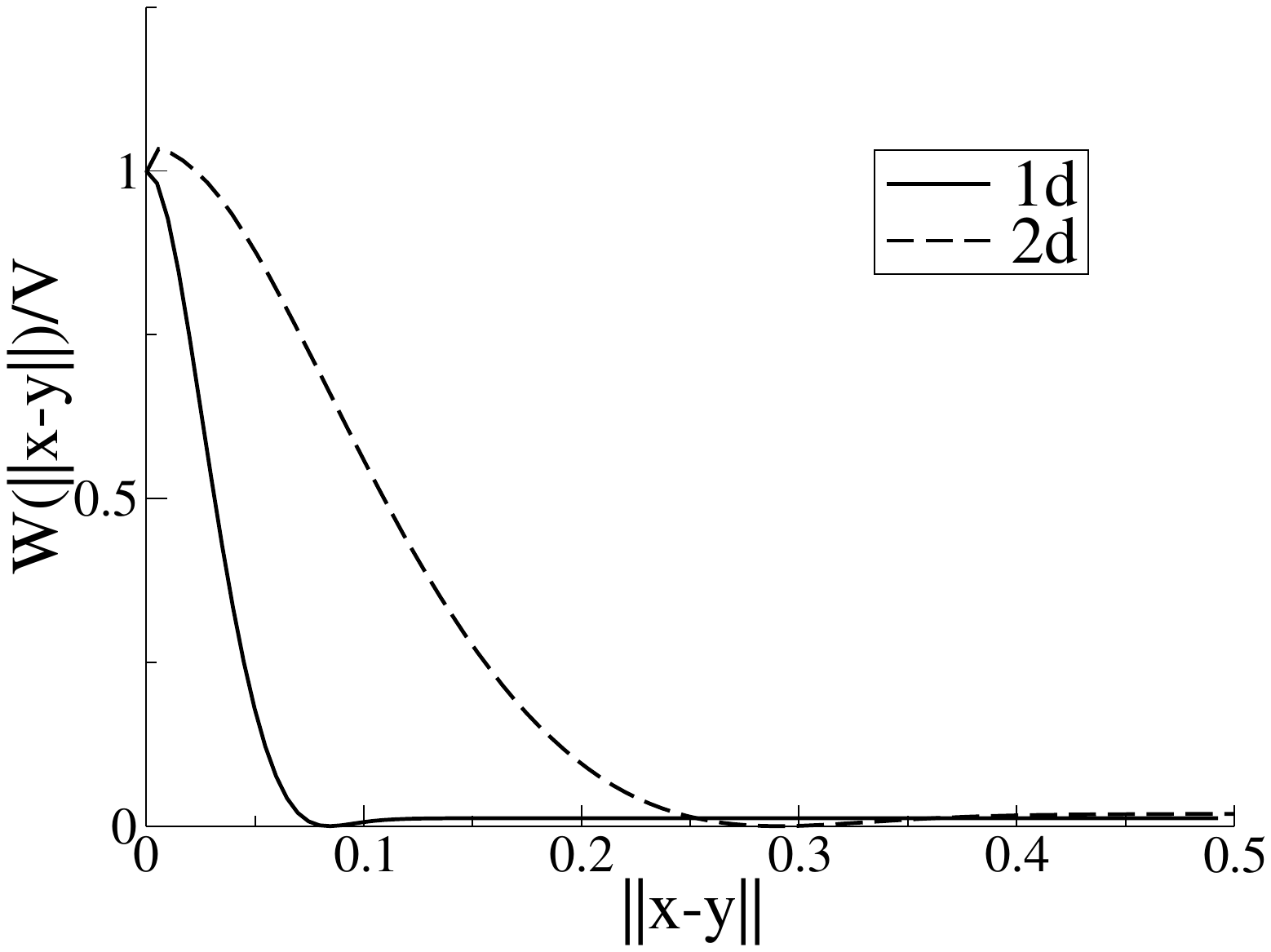}

\caption{Covariance $W(x-y)$ of the free energies of the clump centered on positions $x$ and $y$, normalized by $V$. Results are shown for dimension 1, with $T=0.006$, $\alpha=0.01$ (full line) and in dimension 2 with $T=0.004$, $\alpha=0.002$ (dashed line). }

 \label{fig:W}

\end{figure}

 We observe that $W(x-y)$ is of the order of $V$ on a distance $x-y$ equal to the typical size of the clump, and sharply decreases at larger distances (Fig.~\ref{fig:W}). Therefore, the typical width of the barriers $\ell_b$ is comparable to the size of the clump. A more quantitative comparison is obtained from the following quantities (computed for the parameters of Fig.~\ref{fig:W}): ${\int\mathrm{d}x\,x\, W(x)/\int\mathrm{d}x\,W(x)=0.057}$ and ${\int\mathrm{d}x\,x\,\rho(x)/\int\mathrm{d}x\,\rho(x)=0.082}$ in one dimension, ${\int\mathrm{d}x\,x\,W(x)/\int\mathrm{d}x\,W(x)=0.088}$ and ${\int\mathrm{d}x\,x\,\rho(x)/\int\mathrm{d}x\,\rho(x)=0.097}$ in two dimensions. The overlap $q_{12}$ decreases on a similar typical distance, see Fig.~\ref{fig:q12} in Appendix \ref{app:spatialcorr}.

\subsection{Numerical simulations}

\subsubsection{Activated diffusion}

\begin{figure}
\centering
\includegraphics[width=\columnwidth]{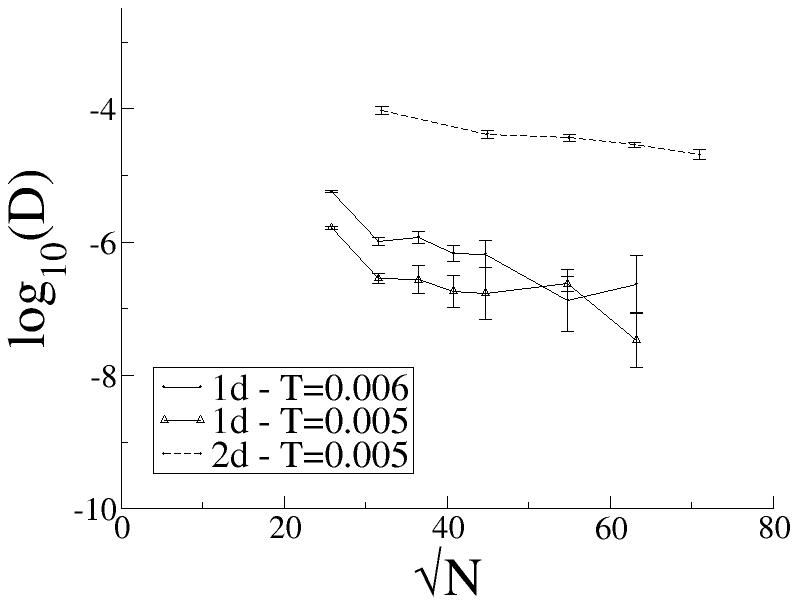}
\caption{Logarithm of the diffusion constant $D$ as a function of $\sqrt{N}$ with constant $L+1=2$, measured in Monte Carlo simulations in both dimensions 1 and 2. For sufficiently large $N$, $\log_{10}(D)$ seems to decrease linearly with $\sqrt N$. The simulations length depends on the frequency of transitions: typically, of the order of ${10-10^2}$ rounds for $\sqrt N=18$ and 1000 rounds for $\sqrt N > 35$. Depending on the computational cost, each point is averaged over a number of simulations ranging from 5 (for large $N$) to 100.}
\label{fig:logd}
\end{figure}

\begin{figure}
\centering
\includegraphics[width=\columnwidth]{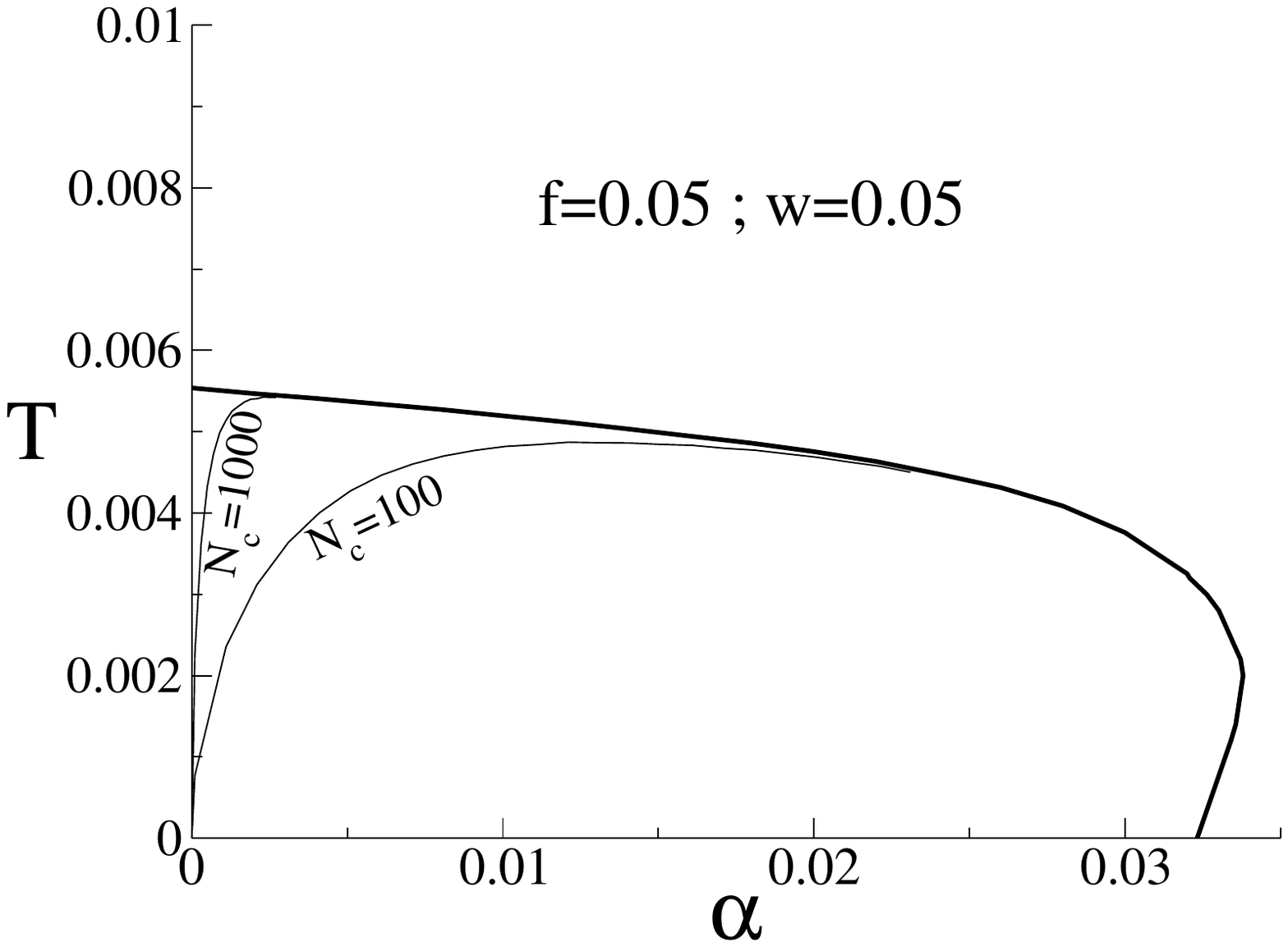}
\includegraphics[width=\columnwidth]{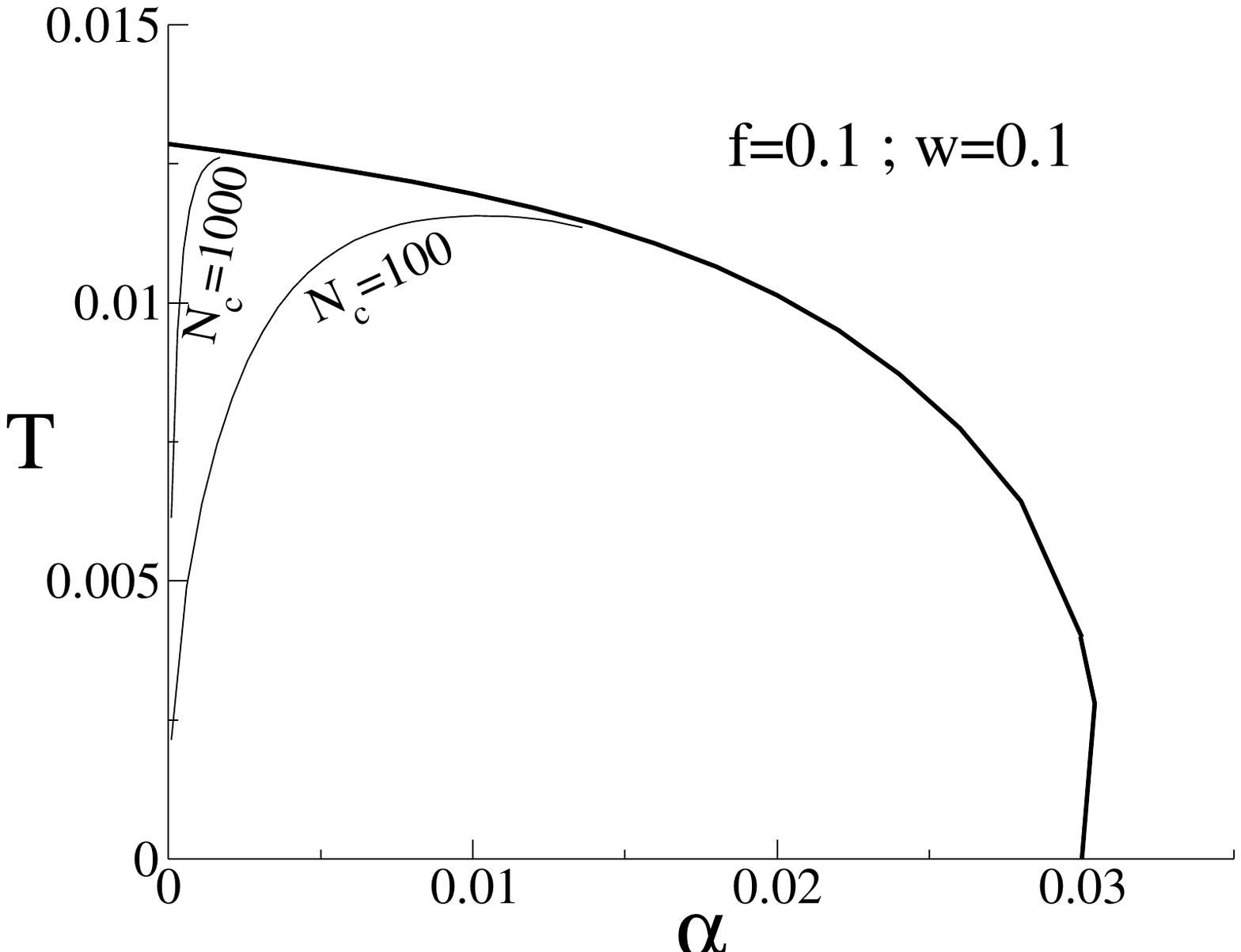}
\caption{Contour lines of constant $N_c$ in the 1-dimensional phase diagram for different values of $w$, $f$. Note the quantitative change in the $T$ axis. The qualitative aspect is remarkably preserved.}
\label{fig:lnf5}
\end{figure}

We ran Monte Carlo simulations of the model with multiple environments and measured the quantity $D$ defined above (Sec.~\ref{sec:numsim}). Results are plotted in Fig.~\ref{fig:logd}. In agreement with the predictions above, we observe that the clump is trapped as soon as $N$ exceeds a few hundreds or when $T$ is too low or $\alpha$ too high. We nevertheless note that $D$ is in general higher in 2d than in 1d: this effect will be discussed later (see Section \ref{sec:retrieval}). Interestingly, the crossover size $N_c$ (\ref{nc})  is very robust to changes in parameters. Figure~\ref{fig:lnf5} shows that the constant-$N_c$ lines remain qualitatively unchanged with respect to the clump stability region as $f$ and $w$ vary, while the absolute location of the stability region in the $(\alpha,T)$ plane varies, see \cite{Monasson13}. 

Estimating the diffusion coefficient would require simulations long enough to allow the clump to move on distances larger than the environment size. The occurrence of transitions to other environments forbid such long simulation times for most parameter values (Fig.~\ref{fig:logd}). As a consequence, the displacement of the clump during our simulations is generally smaller than the environment size. The values of $D$ we measure are therefore indicative of the motion of the clump on a limited time scale, and allow us to study the influence of parameters, e.g. the size $N$ in Fig.~\ref{fig:logd}, on this motion. Note that, in two dimensions, diffusion is easier, and the simulation times required to explore the environment are smaller. 

It is interesting to notice that, due to disorder effects, the diffusion constant for a same set of stored environments varies with the environment the clump of activity is coherent with. In other words, in each attractor (stored map), the clump phase has a different diffusion dynamics. For some maps diffusion is relatively 'easy', while the clump will remain trapped for very long times and hardly diffuse in other maps. This phenomenon is illustrated in Table~\ref{tab:Denvenv}.

\begin{table}
\begin{tabular}{|c|c|c|c|c|}
\hline
$\ell$ & 0 & 1 & 2 & 3\\
\hline
$D^{(\ell)}$ & $1.1\cdot10^{-5}$ & $1.1\cdot10^{-5}$ & $5.6\cdot10^{-6}$ & $5.7\cdot10^{-6}$ \\
&$\pm2.0\cdot10^{-6}$&$\pm1.8\cdot10^{-6}$&$\pm5.1\cdot10^{-7}$&$\pm7.8\cdot10^{-7}$\\
\hline
\end{tabular}
\caption{The diffusion constant $D^{(\ell)}$ differs significantly from an environment $\ell$ to another within a same given system (set of couplings created from the $L+1$ environments). The table shows the results obtained for one set of simulations with $N=1000$ neurons, $L+1=4$ randomly drawn environments, and $T=0.005$. Each value is averaged over 100 simulations of 1000 rounds, initialized at different positions of space. The variations of $D^{(\ell)}$ from environment to environment is larger than error bars.}
\label{tab:Denvenv}
\end{table}

\subsubsection{Transitions to other environments}\label{transit}

\begin{figure}
\centering
\includegraphics[width=\columnwidth]{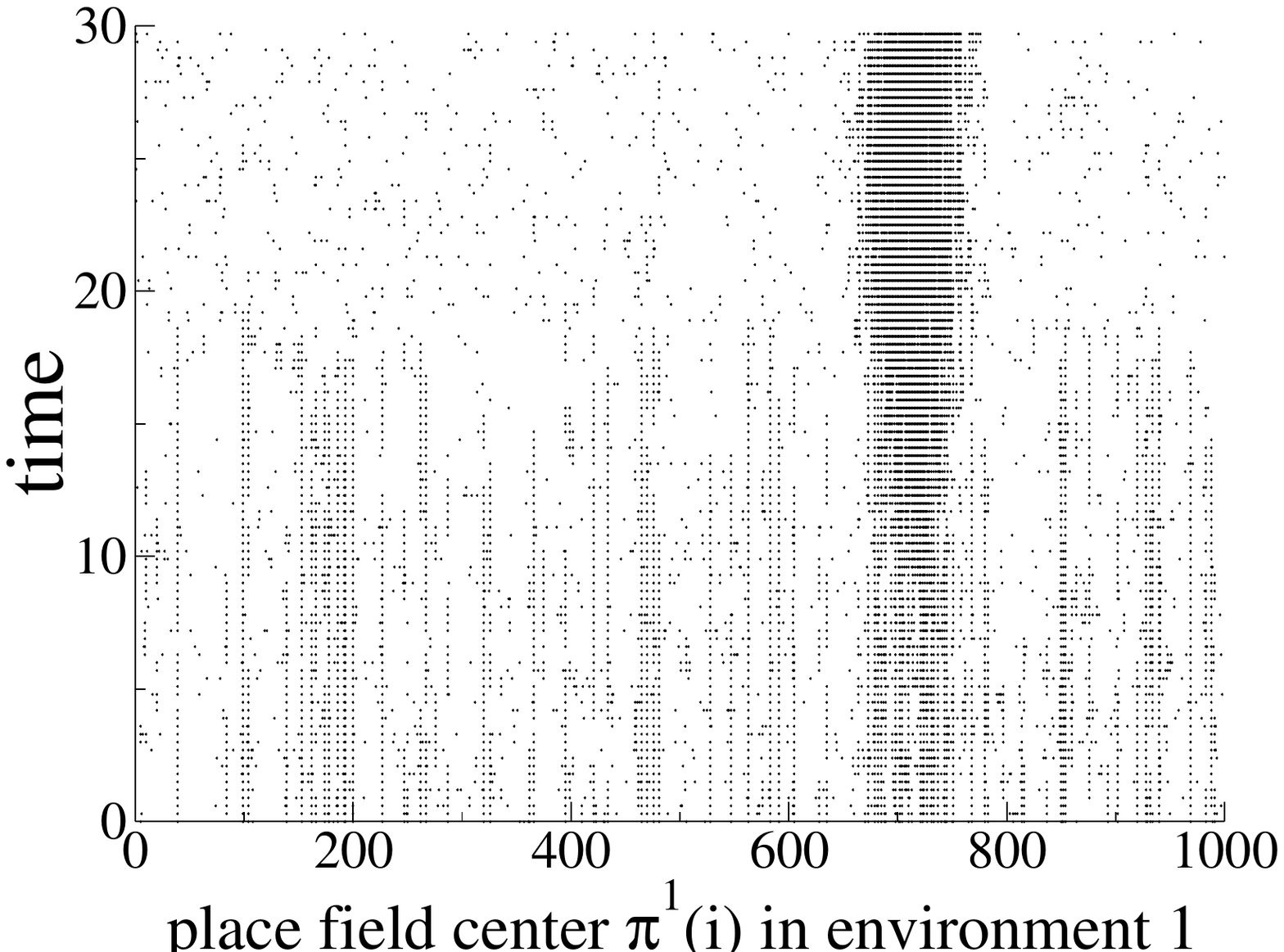}
\includegraphics[width=\columnwidth]{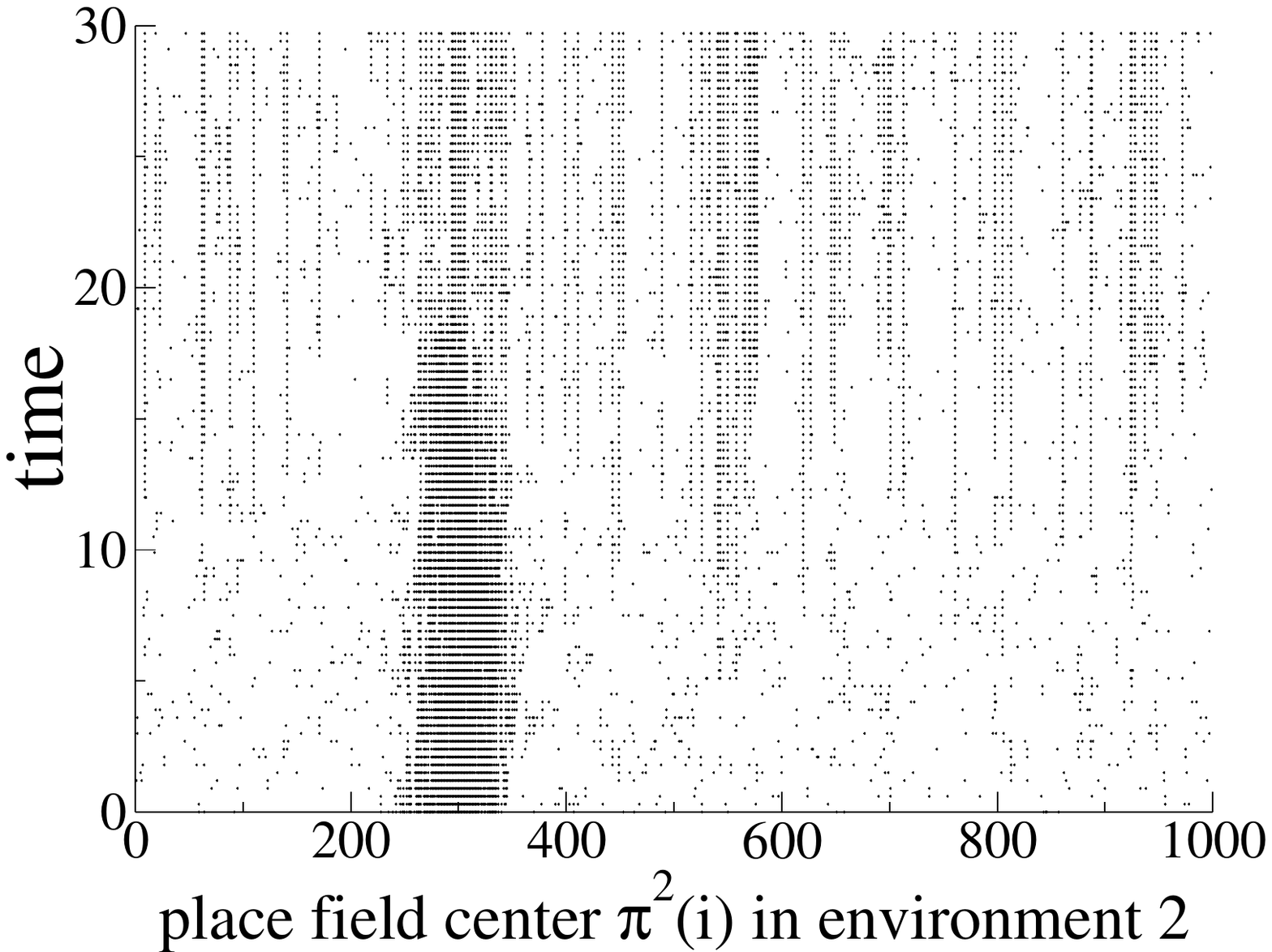}
\caption{Example of a transition observed in a Monte Carlo simulation with $N=1000$ neurons, $L+1=2$ environments and $T=0.006$. Neural  configurations $\boldsymbol\sigma$ are shown at different times (black dots correspond to active neurons). Both panels show the same data, with the difference that neurons are ordered according to their place field centers $\pi^{1}(i)$ in environment 1 (top) and  $\pi^{2}(i)$ in  environment 2 (bottom). The transition takes place around time $t\simeq 15$ (time unit: 1 round of $N$ steps). }
\label{fig:mecatransi}
\end{figure}

\begin{figure}
\centering
\includegraphics[width=\columnwidth]{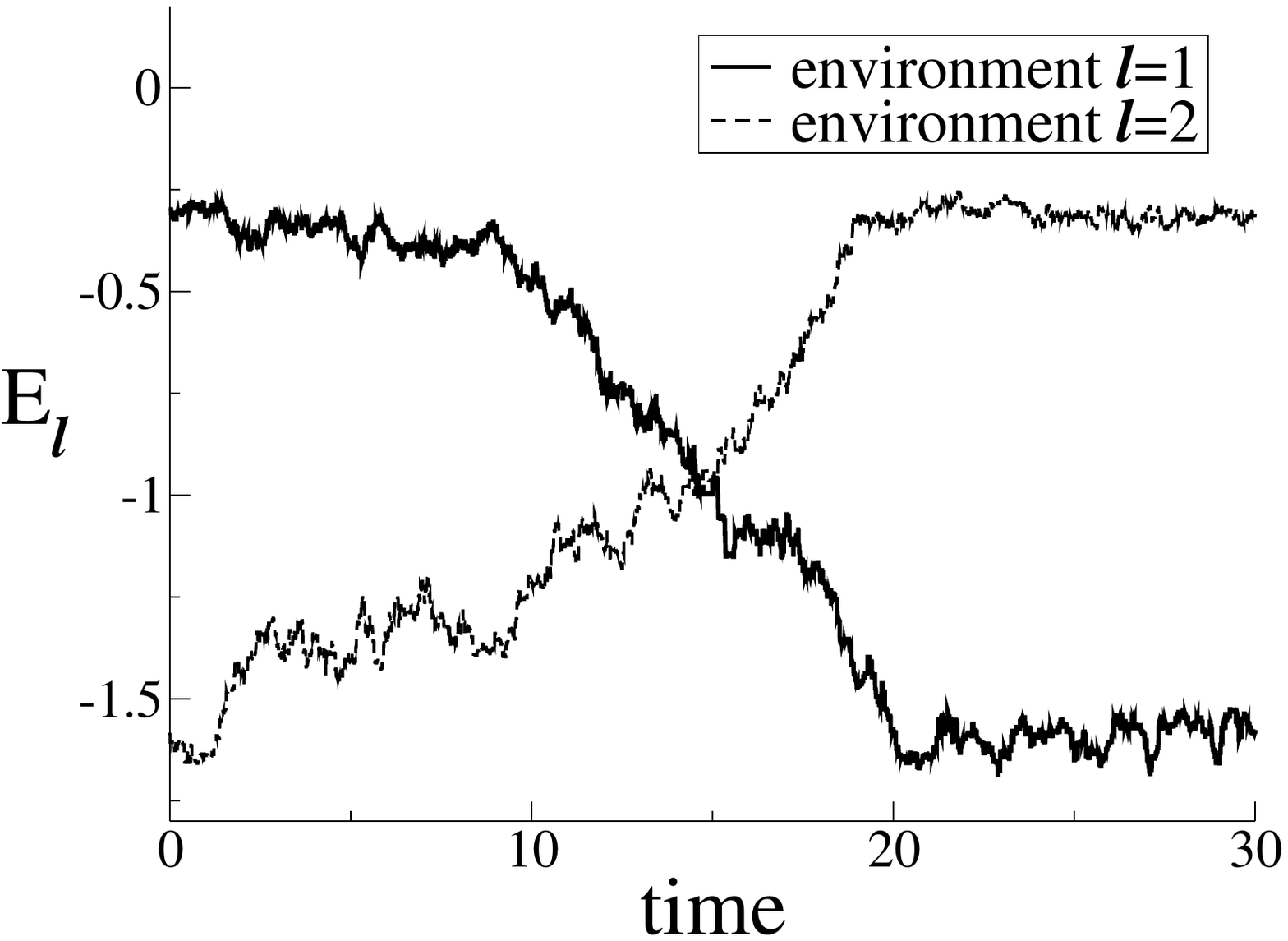}
\caption{Evolution of ${E_{\ell}\equiv\sum\limits_{i<j}J_{ij}^\ell\sigma_i\sigma_j}$, for the same transition event as in Fig.~\ref{fig:mecatransi}. $E_{\ell}$ is the contribution of environment $\ell$ to the logarithm of the probability of the neural configuration $\boldsymbol\sigma$, see (\ref{dist}). The crossing of $E_1$ and $E_2$ defines the transition between the two maps, as well as the intermediary state, where the activity is weakly localized in both maps. }
\label{fig:mecatransi2}
\end{figure}

Abrupt jumps between maps are often observed in Monte Carlo simulations with several environments. A detailed study of those transitions is postponed to a companion paper; hereafter we limit ourselves to briefly report the salient features of transitions, which are of interest to the dynamics of activity within one map studied in the present paper. An example of transitions is shown in Fig.~\ref{fig:mecatransi}. We observe that the activity configuration goes from being localized in the first environment (clump state) to being localized in the second environment, through an intermediary state which weakly localized in both environments. This can be seen directly on the microscopic configuration $\boldsymbol\sigma$, or, alternatively, by looking at the contributions of both environments to the log. probability $P_J(\boldsymbol\sigma)$ of the neural configuration, as shown in Fig.~\ref{fig:mecatransi2}.

\begin{figure}
\centering
\includegraphics[width=6cm,angle=90]{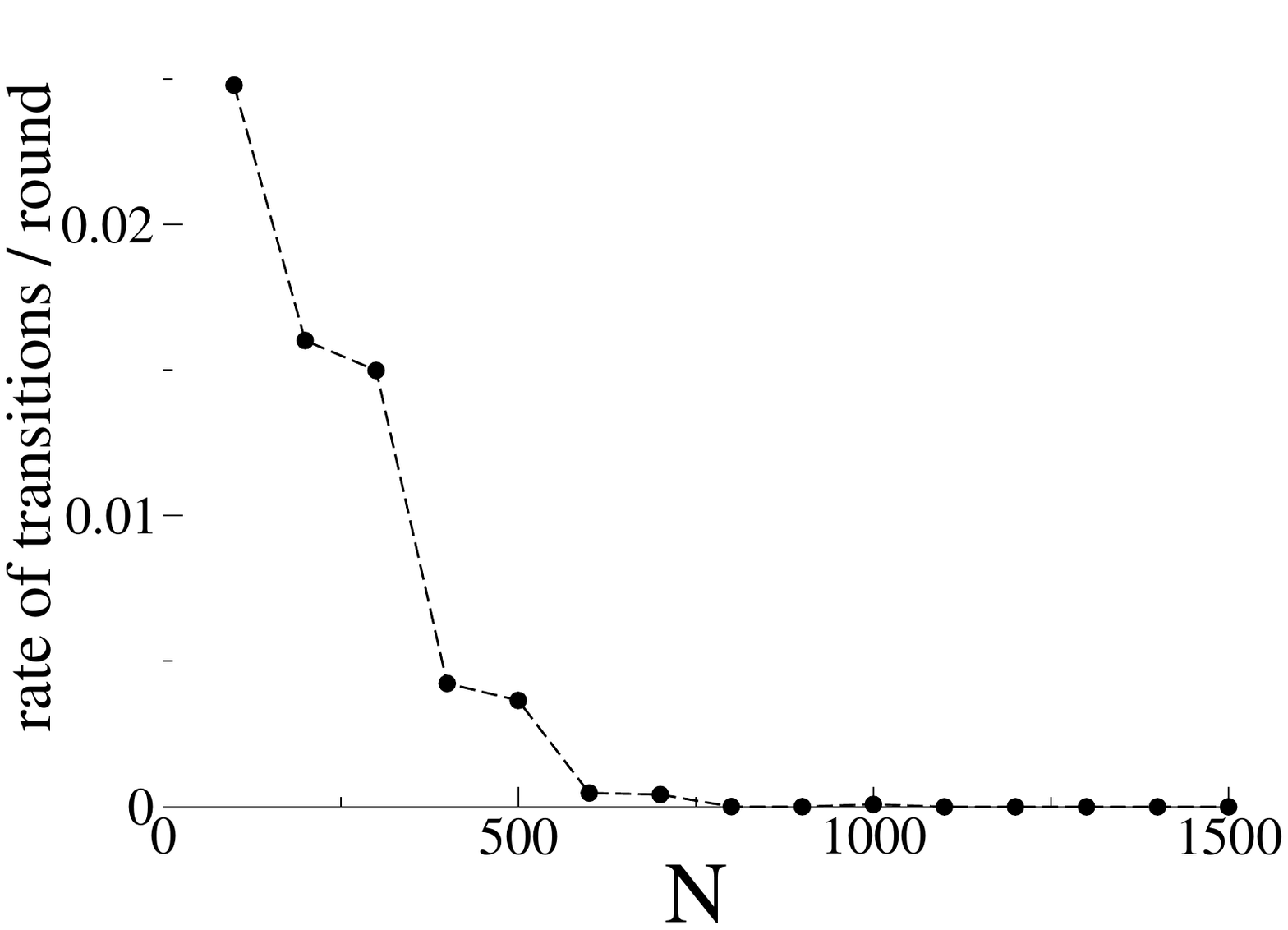}
\caption{Rate of transitions to other environments as a function of $N$ for one realization of $L+1=2$ one-dimensional environments and $T=0.006$. Each point is averaged over 10 simulations of 1000 MC rounds. Time unit: 1 round of $N$ steps. The decay of the rate is consistent with an exponentially decreasing function of $N$, hence with Arrhenius' law and the existence of free energy barriers proportional to $N$.}
\label{fig:transiden}
\end{figure}

Transitions are less and less frequent as $N$ increases. The decrease of rate of transitions with $N$ is shown in Fig.~\ref{fig:transiden}. An important consequence is that the presence of transitions is in competition with diffusion. As $N$ decreases the motion of the clump is facilitated, but so are transitions to other environments. 
We observe the existence of preferred 'tunelling' locations, where map-to-map transitions are likely to take place. As transitions are made possible by the existence of intermediary activity configurations where the activity is partially localized in both maps, it is natural to expect that those preferred positions correspond to sites of local ressemblance between the random permutations defining the maps. Such a similarity in the permutations at places where transitions happen most often is indeed observed \cite{Rosay13}. A detailed study of those properties will be reported in a forthcoming publication.

\subsection{Effects of partial activity}
\label{partial}

The study above can be repeated under the more realistic assumption that there exist many 'silent' place cells, in the sense that only a fraction $c\, (<1)$ of the neurons have place fields in a given environment. For instance, in one dimension, the variance $V$ of the free energy, given by (\ref{eq:V}) in the case $c=1$, becomes (see \cite{Monasson13} for details about the $c<1$ calculations)
\begin{eqnarray}
V_{c}&=&-\alpha rq+\alpha T^2 c^2(q-f^2)^2\varphi_c(q,T)\\
&&+T^2c \int\mathrm{d}x\mathrm{D}z\log^2(1+e^{\beta z\sqrt{\alpha r}+\beta\mu(x)})\nonumber\\
&&-T^2c\int\mathrm{d}x\left(\int\mathrm{D}z\log(1+e^{\beta z\sqrt{\alpha r}+\beta\mu(x)})\right)^2\nonumber\\
&&+T^2(1-c)\int\mathrm{D}z\log^2(1+e^{\beta z\sqrt{\alpha r}+\beta\mu_2})\nonumber\\
&&-T^2(1-c)\left(\int\mathrm{D}z\log(1+e^{\beta z\sqrt{\alpha r}+\beta\mu_2})\right)^2\ ,\nonumber
\end{eqnarray}
where
\begin{eqnarray}
\varphi_c(q,T)
&=&\sum\limits_{k\geq1}\bigg(\frac{T\,\pi k}{\sin(\pi k w)}+c(q-f)\bigg)^{-2}\ ,
\end{eqnarray}
and $\mu_2$ is such that $\int\mathrm{D}z[1+e^{-\beta z\sqrt{\alpha r}-\beta\mu_2}]^{-1}=f$.

\begin{figure}
\centering
\includegraphics[width=\columnwidth]{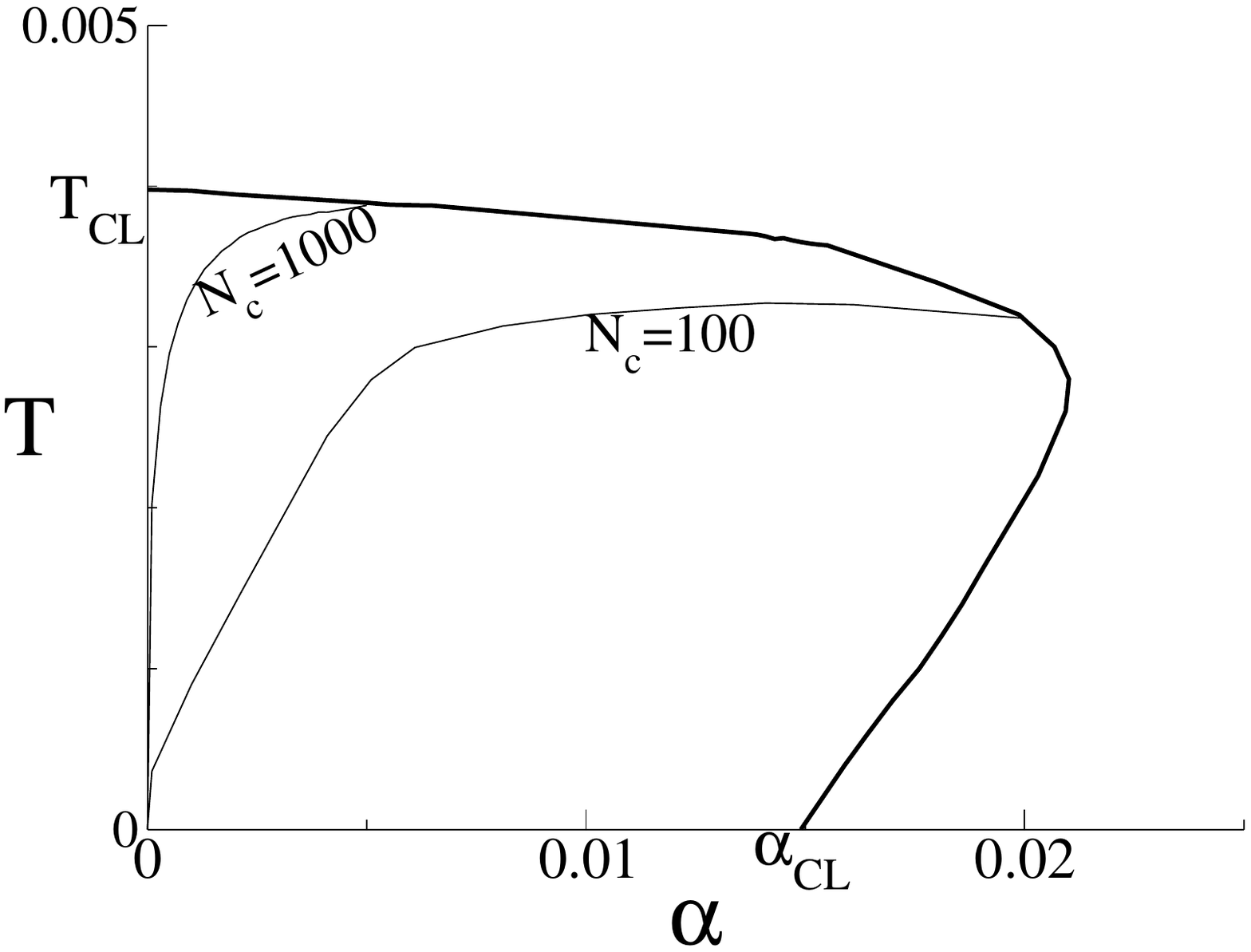}
\caption{Contour lines in the $(\alpha,T)$-plane corresponding to a fixed crossover size, $N_c$, for the 1-dimensional case with $c=0.5$.}
\label{fig:lnc5}
\end{figure}

Having $c<1$ quantitatively changes the stability region of the clump phase, but does not have any qualitative effect on the static properties of the system \cite{Monasson13}. Here we look at the effect of partial activity on the diffusion. Interestingly, it turns out that again the location of the contour lines for $N_c$ with respect to the stability domain of the clump phase remains essentially unchanged with $c$. This robustness phenomenon is illustrated in Fig.~\ref{fig:lnc5}.

\begin{figure}
\centering
 \includegraphics[width=\columnwidth]{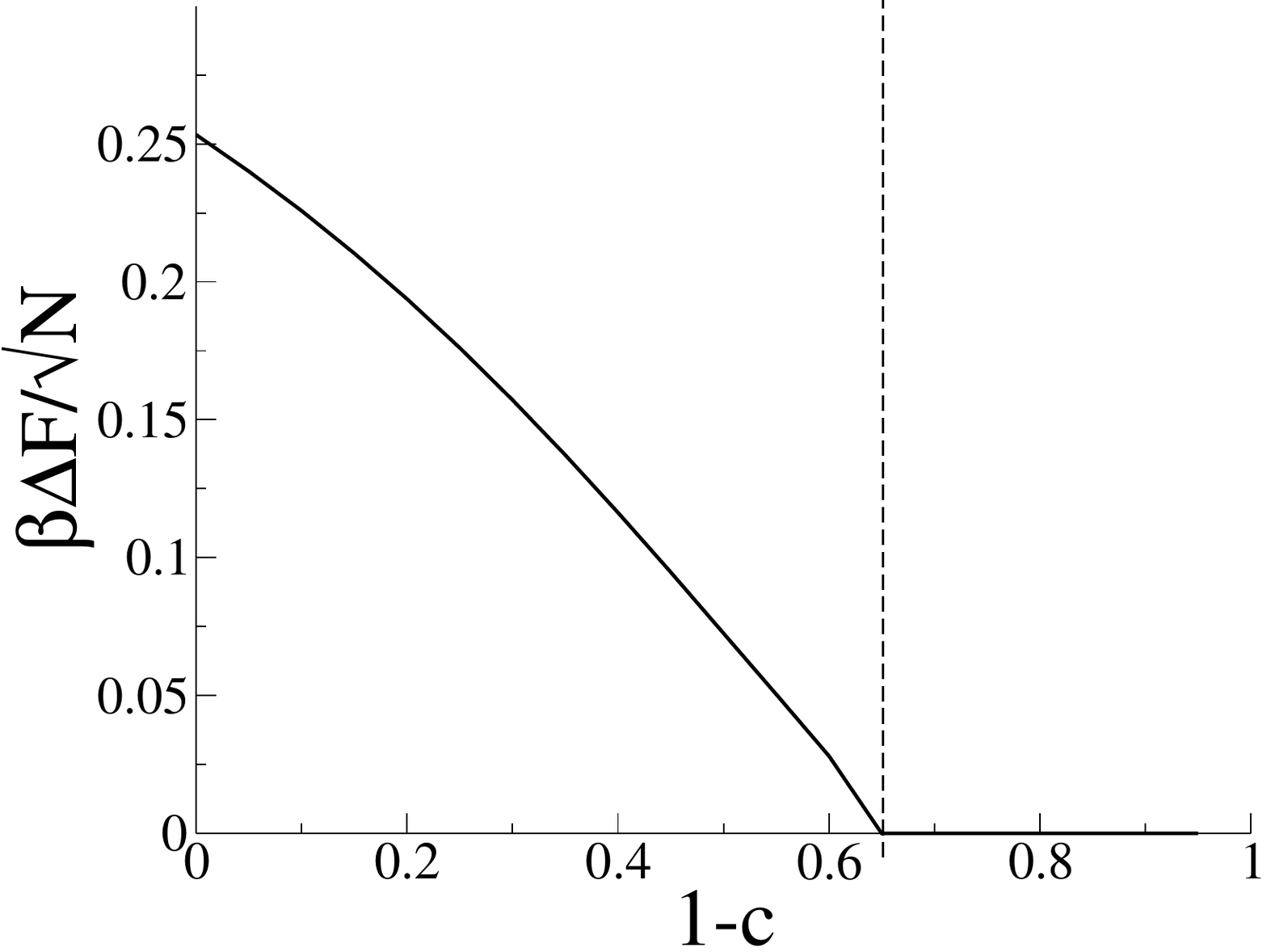}
\includegraphics[width=\columnwidth]{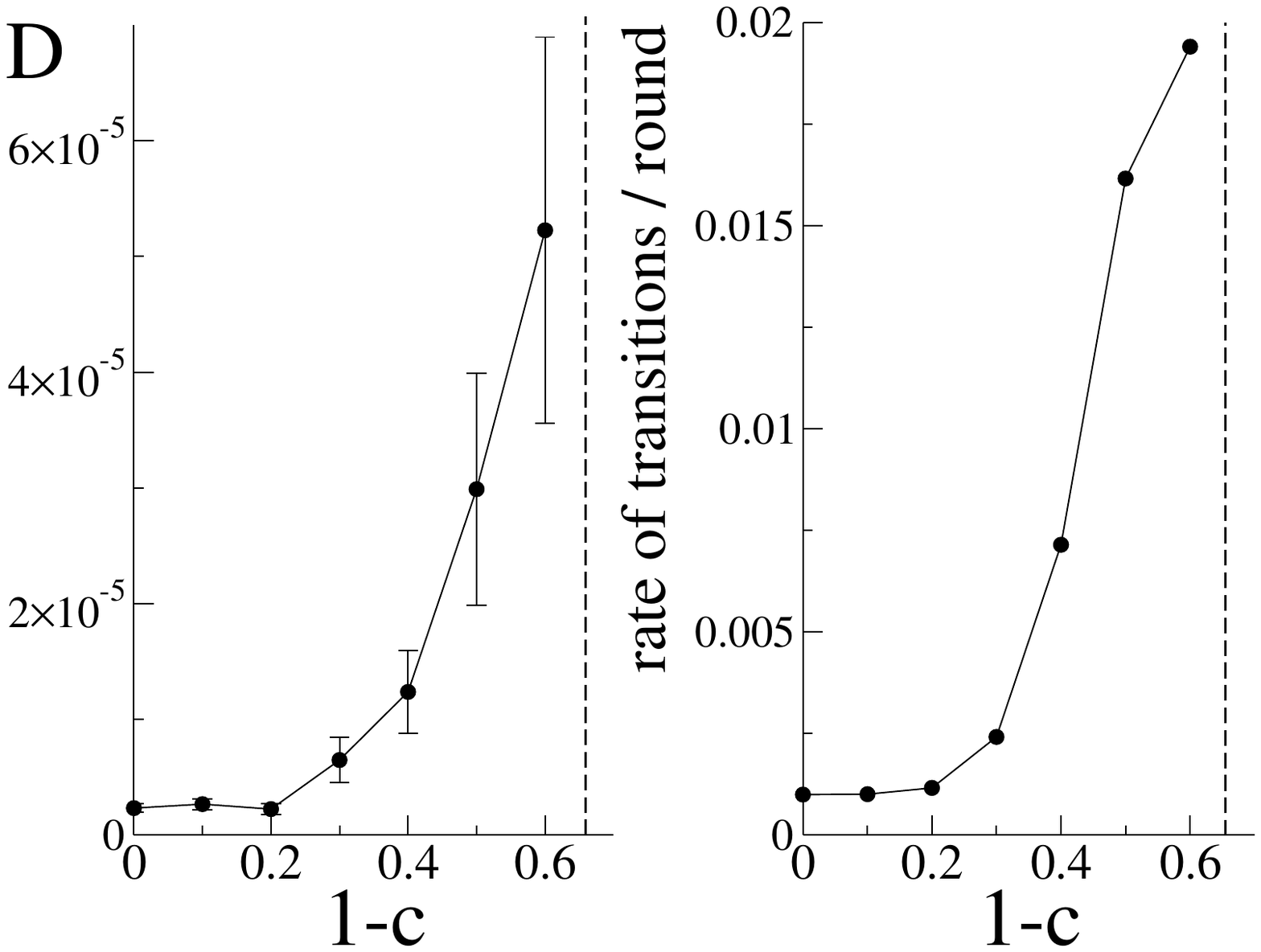}
\caption{Effect of partial activity on the theoretical free energy barriers $\beta\sqrt{V}$ (top), on the diffusion constant $D$ (bottom, left) and on the rate of transitions per round (bottom,right). Results correspond to the 1-dimensional case, $T=0.003$, $\alpha=0.003$, $N=1000$. The dashed line indicates the limit of stability of the clump. The simulations length depends on the frequency of transitions: typically, 1000 rounds for $1-c=0$ and of the order of ${10^2}$ rounds for $1-c=0.6$. 1 round = $100N$ steps. Each point is averaged over 100 simulations.}
 \label{fig:F1c}
\end{figure}

\begin{figure}
\centering
 \includegraphics[width=\columnwidth]{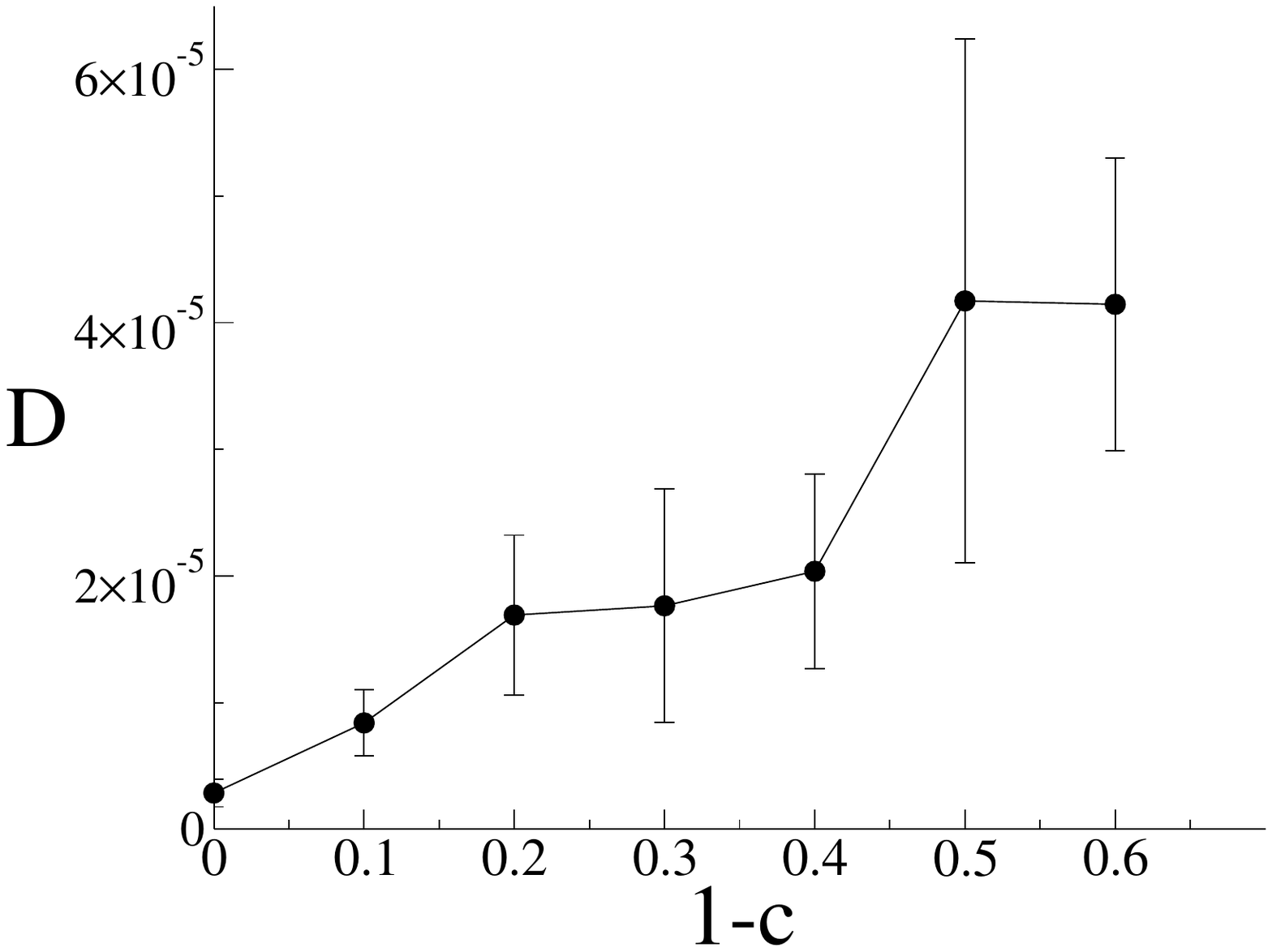}
\caption{Effect of partial activity on the diffusion constant $D$ in the 2-dimensional case, with $T=0.002$, $\alpha=0.001$, $N=45\times45$ units. The clump phase is not stable anymore when $1-c$ exceeds $\simeq 0.6$. The simulations length depends on the frequency of transitions: typically, of the order of $10^3$ rounds for $1-c=0$ and of the order of ${10^2}$ rounds for $1-c=0.6$. 1 round = $100N$ steps. Each point is averaged over 30 simulations.}
 \label{fig:Ddec2d}
\end{figure}

As a consequence, for given $\alpha,T$, decreasing $c$, {\em i.e.} increasing the sparsity of the representation will have the effect of increasing the diffusion constant, mostly because the neural noise is relatively more important. The rate of transitions to other environments increases, too. When $c$ becomes too low, the clump is not stable anymore, and disappears. Simulations are in good agreement with this prediction, as shown in Fig.~\ref{fig:F1c}. In dimension 2 the behavior with decreasing $c$ is the same, see Fig.~\ref{fig:Ddec2d}.

\section{Motion under an external force}
\label{sec:retrieval}

We now investigate the motion of the clump under an external input. 
\subsection{Drift under an external force}
\label{sec:drift}

We consider the behavior of the model when the environment is 'tilted', {\em i.e.} when a force is applied to make the clump move in a given direction. In the absence of disorder in the interactions (single-environment case) the force is expected to move the clump with a positive, and constant velocity. In the presence of disorder, the wrinkled energy landscape combined to the tilt will pin the activity. The motion will be strongly activated, with the clump trapped in minima most of the time, until the force exceeds some critical threshold, above which the clump will acquire a positive velocity.

This scenario is corroborated by simulations. We model the presence of a force through an increase of the probability of the two-neuron flip ${\sigma_i=1,\sigma_j=0\to \sigma_i=0,\sigma_j=1}$ with respect to ${\sigma_i=0,\sigma_j=1 \to \sigma_i=1,\sigma_j=0}$, for $i<j$ (1-dimensional case).  This creates a bias in favor of motion to the right. More precisely, the Metropolis rate defined in Section \ref{sec:transitionrates} is modified as follows:
\begin{align}
\omega(\Delta E)&=N e^{-\beta(\Delta E-A\Delta x_c)}\quad \text{ if }\quad \Delta E\ge A\,\Delta x_c\, \nonumber \\
&=N\quad \text{ if }\quad \Delta E< A\, \Delta x_c\ ,
\end{align}
where 
\begin{equation}\label{dx}
\Delta x_c=\frac{j-i+\epsilon(i,j)N}{f\,N^2}\ ,
\end{equation} 
is the displacement of the center of gravity of the clump when neuron $i$ goes from being active to silent, and neuron $j$ goes from being silent to active; ${\epsilon(i,j)\in\{-1,0,+1\}}$ enforces periodic boundary condition. Parameter $A$ denotes the intensity of the applied force.

\subsubsection{Critical values of the force}

Using the estimates $\Delta F$ and $\ell_b$ for, respectively, the height and the width of the free energy barriers derived in Section \ref{sec:F1}, we evaluate the critical intensity $A_{depin}$ of the force above which the clump can overcome barriers. A rough estimate of this depinning force is obtained by imposing that the work of the force in moving the clump through the barrier, $A_{depin}\times\ell_b$, compensates the barrier height, $\Delta F=\sqrt{V(\alpha,T)N}$ (\ref{eq:deltaF}). We obtain the typical value
\begin{equation}
\label{eq:Ac}
A_{depin}^{typ}\simeq\frac{\sqrt{V(\alpha,T)N}}{\ell_b}\ .
\end{equation}
Drift is mostly hindered by the highest barriers. The maximal height can be estimated by considering that barriers heights are Gaussian variables, drawn independently and at random for each one of the $1/\ell_b$ segments of length $\ell_b$. Hence, according to extreme value theory, the maximal barrier heights is about $\sqrt{2\log (1/\ell_b)}$ times the typical value computed above, 
\begin{equation}
\label{eq:Ac2}
A_{depin}^{max}\simeq\sqrt{2\log\left(\frac1{\ell_b}\right)}\;\frac{\sqrt{V(\alpha,T)N}}{\ell_b}\ .
\end{equation}

As the force is applied at the microscopic level on the neuron states, and not at the macroscopic scale on the clump itself, taking $A$ too large will make the clump desintegrate. This will happen if the work of the force exceeds the cohesion energy of the clump. We estimate the critical intensity $A_{break}$ based on the following reasoning. Silencing a neuron within the clump and activating another neuron outside the clump costs on average (for the 1-dimensional case)
\begin{align}
\Delta E&\simeq \langle\mu\rangle_{\text{inside}}-\langle\mu\rangle_{\text{outside}}\\
& = \frac 1{\ell_b}\,\int\limits _{|x-x_c| < \ell_b/2} dx\, \mu(x) -\frac1 {1 -\ell_b} \int\limits _{|x-x_c| > \ell_b/2} dx\, \mu(x)\ . \nonumber
\end{align}

This energy cost is decreased by the work of the force, $A\, \Delta x_c$, where $\Delta x_c$ is the change in the average position of the clump following a microscopic flip of two neuron states, see (\ref{dx}). The most favorable case, corresponding to the largest shift of the clump center, is $\Delta x_\text{max}=1/(2fN)$.  We conclude that the cost decreases linearly with $A$ (and can even become negative at large $A$), leading to the breaking apart of the localized collective activity. An estimate of the critical force at which this happens can be obtained from the comparison of the cost with the temperature of stability of the clump at zero force, $T_{CL}$, see Section \ref{sec:prelim} and \cite{Monasson13}. We expect
\begin{equation}
\frac{\Delta E - A_{break} \,\Delta x_{max} }{T}\simeq \frac{\Delta E}{T_{CL}}\ ,
\end{equation}
or, equivalently,
\begin{equation}
A_{break}\simeq2\,f\,N\,\Delta E\, \bigg(1-\frac T{T_{CL}}\bigg)\ .
\end{equation}

\subsubsection{Simulations}

\begin{figure}

\centering

 \includegraphics[width=\columnwidth]{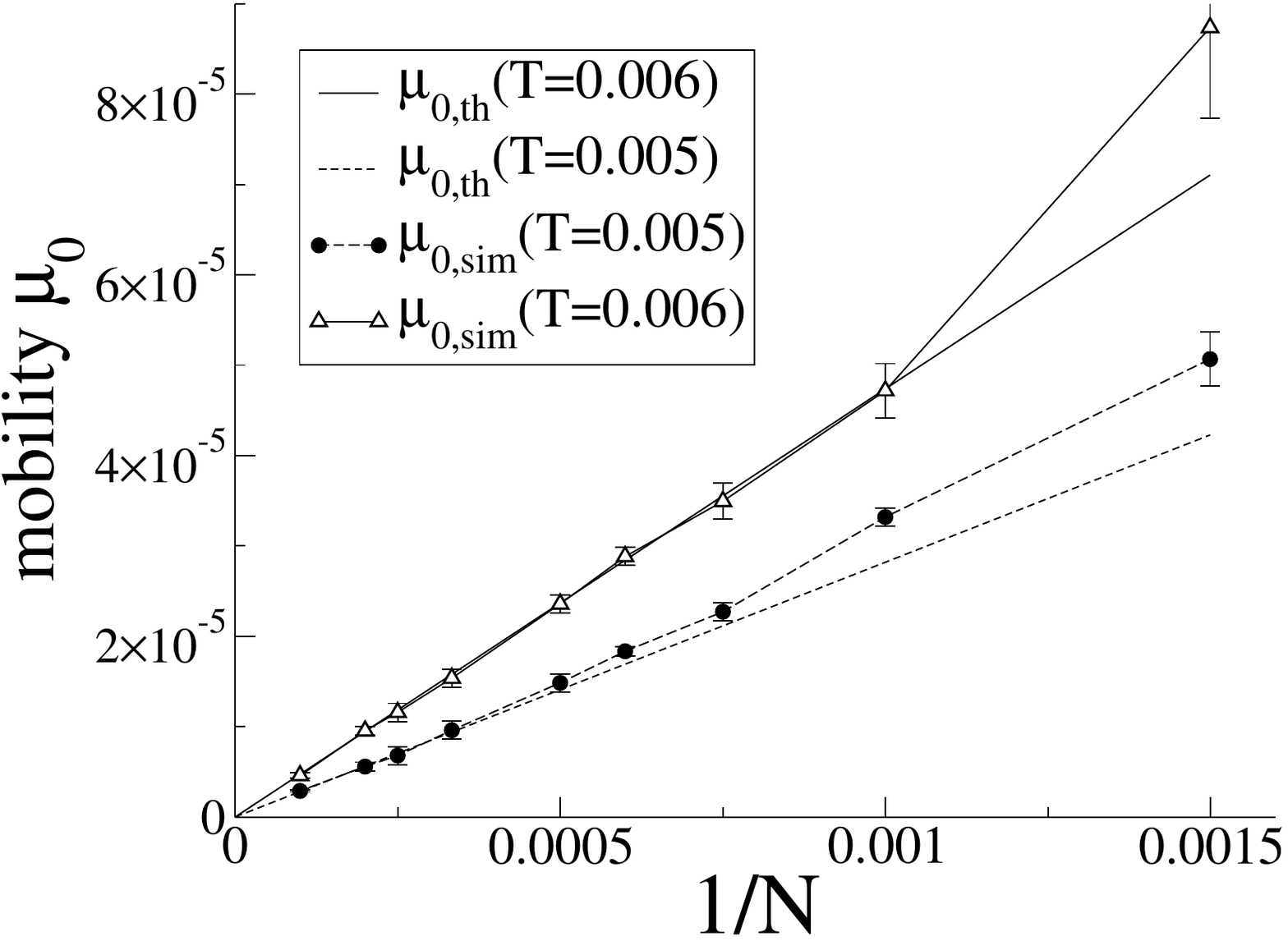}

\caption{Mobility of the clump in response to an external force, in the single-environment ($\alpha=0$), one-dimensional case. The theoretical prediction for the effective mobility, $\mu_{0,\text{th}}$, computed from Eq.~(\ref{eq:mob3}), is plotted as a function of $\frac1N$ for $T=0.005$ (dashed lines) and $T=0.006$ (full lines) and compared to the results  of Monte Carlo simulations $\mu_{0\text{sim}}$. The agreement with the analytical prediction (which neglects terms smaller than $O(\frac 1N)$) improves as $N$ increases. Each point is averaged over 10 simulations, in which the clump, initially at location $x=0$, had moved over 4 space-bins (the environment is covered by 11 bins). }

 \label{fig:mob}

\end{figure}

First, we tested the theoretical prediction (\ref{eq:mob2}) for the effective mobility of the quasiparticle in the one-dimensional, single environment case. We ran simulations for different values of $N$ and $A$ and measured the velocity of the center of the clump. Taking $A(x)=-A\,x$ in Eq.~(\ref{eq:mob}) gives
\begin{equation}
 V_0=\mu_{0,th}\,A\ ,
\end{equation}
where
\begin{equation}
 \mu_{0,th}\equiv\beta\;\frac{\int dx\; dy\;x\; D(x,y)\;u_0(y)}{2\sqrt{\int dy \left(\frac{d\rho^*(y)}{dy} \right)^2}}
\label{eq:mob3}
\end{equation}
is the predicted mobility of the clump.
As expected, for a fixed number $N$ of cells, the velocity increases linearly with $A$ (up to $A_{break}$).  The slope of this curve is our numerical estimate $\mu_{0,sim}$ for the mobility of the clump.  This measure of the mobility is in very good agreement with theory, as shown in Fig.~\ref{fig:mob}.

\begin{figure}

\centering

 \includegraphics[width=\columnwidth]{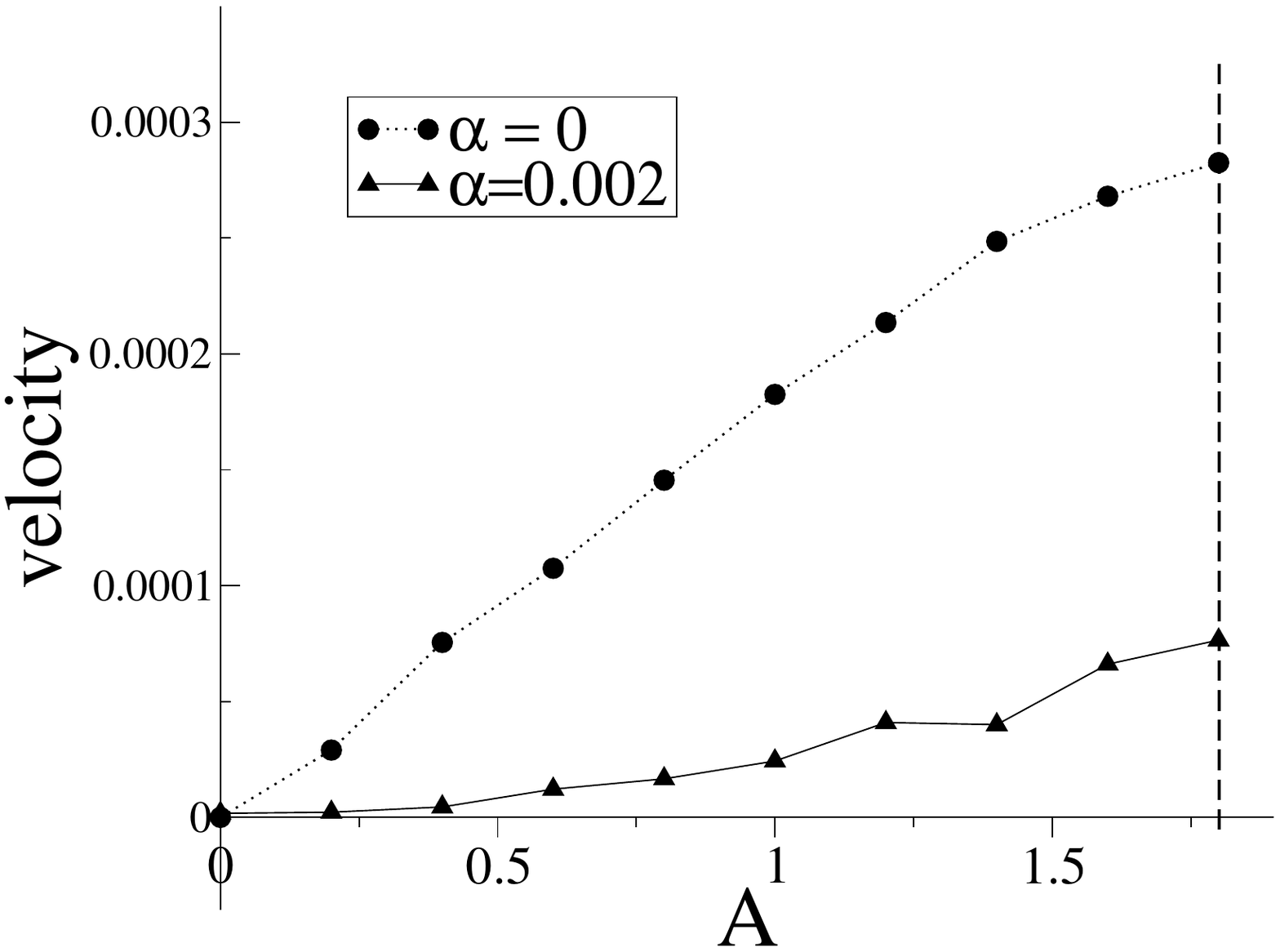}

 \includegraphics[width=\columnwidth]{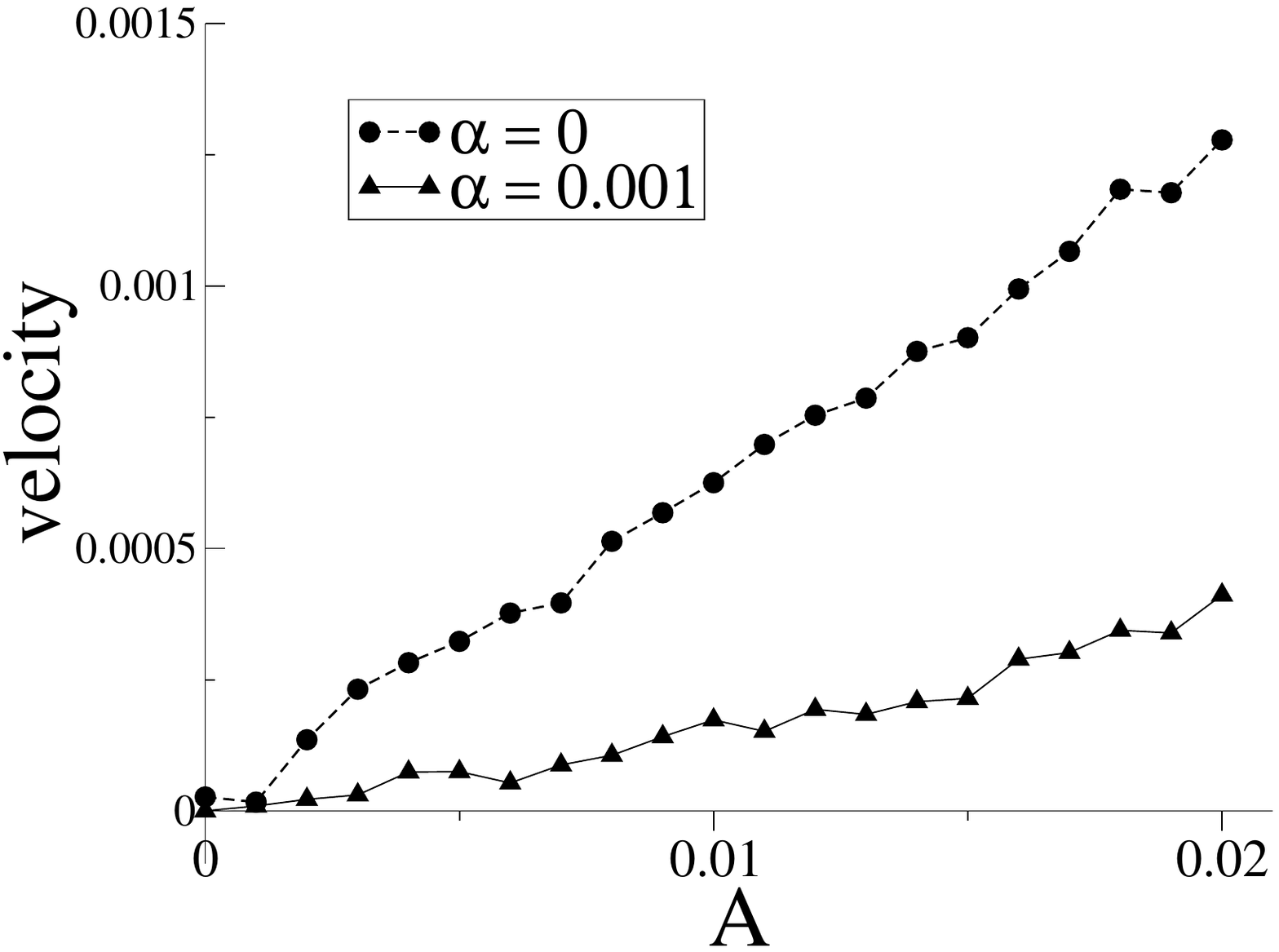}

\caption{Velocity of the clump under a force $A$. Top: dimension one, $T=0.006$, $N=1000$ (the clump is not stable for larger $A$ as indicated by the dashed line). Bottom: dimension two, $T=0.005$, $N=32\times32$. Simulation time: 1000 rounds for $\alpha=0$; around $10^3-10^4$ rounds for $\alpha>0$. Each point is averaged over 10 simulations. 1 round = $20N$ steps.}

 \label{fig:drift}

\end{figure}

In one dimension, the pinning effect due to the  environments other than the one in which the activity is localized is observed in simulations. An example is shown in Fig.~\ref{fig:drift} for one realization of the disorder.  For the parameters values of the simulation of Fig.~\ref{fig:drift}, we find, according to (\ref{eq:Ac},\ref{eq:Ac2}), $A_{depin}^{typ}\simeq0.16$ and $A_{depin}^{max}\simeq0.23$. In the simulation we observe $A_{depin}^{\text{sim}}\simeq0.25$, in good agreement with $A_{depin}^{max}$ as expected. In addition, note that the depinning force, $A_{depin}^{\text{sim}}$, is found to fluctuate from realization to realization, while our theoretical estimate is sample independent.

We also estimate the force at which the clump disintegrates, under Fig.~\ref{fig:drift} simulation conditions. We find $A_{break}\simeq 1.82$, in excellent agreement with the results of simulations, $A_{break}^{\text{sim}}\simeq1.8$.

\begin{figure}
\centering
 \includegraphics[width=\columnwidth]{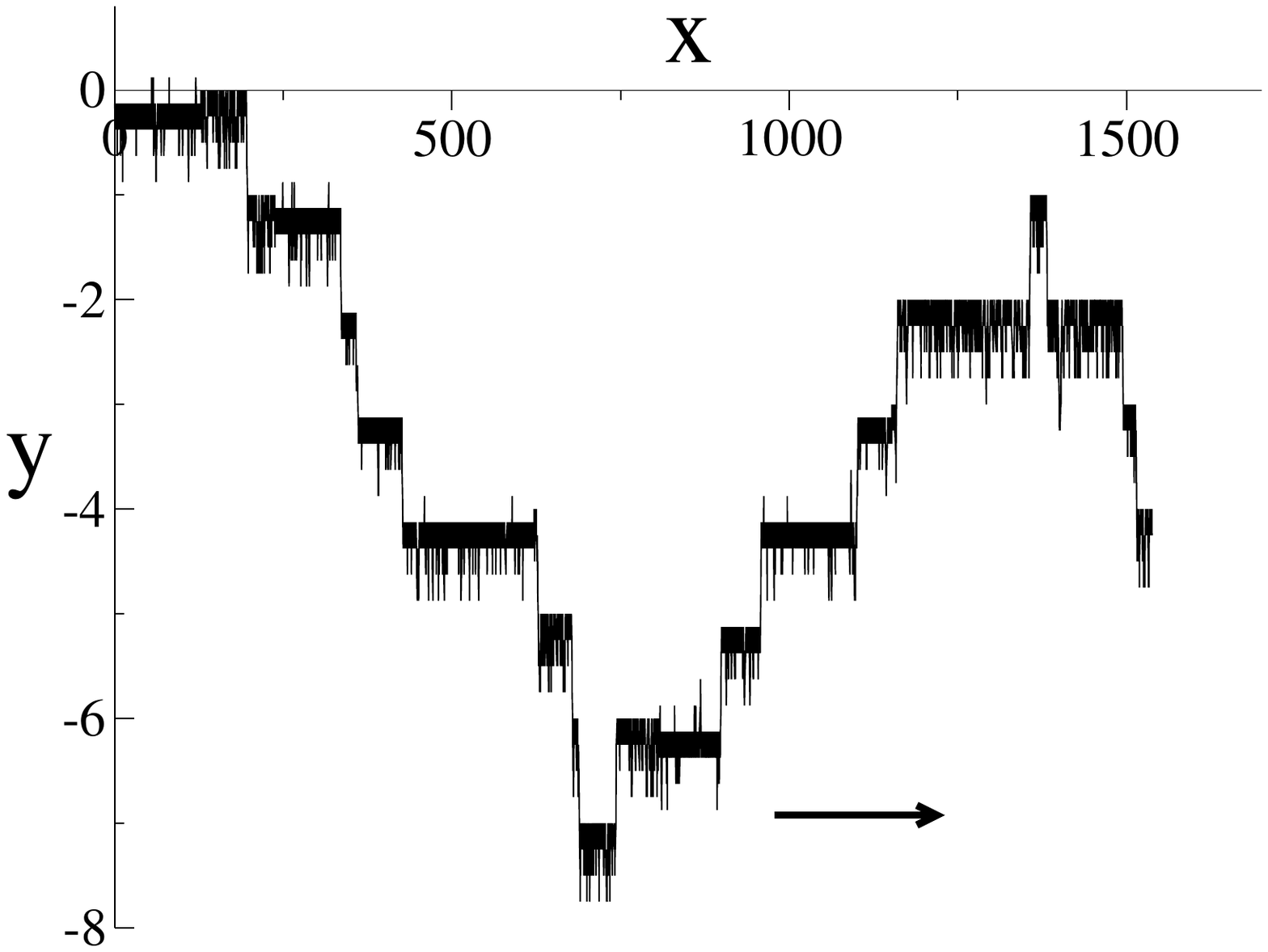}
\includegraphics[width=6cm]{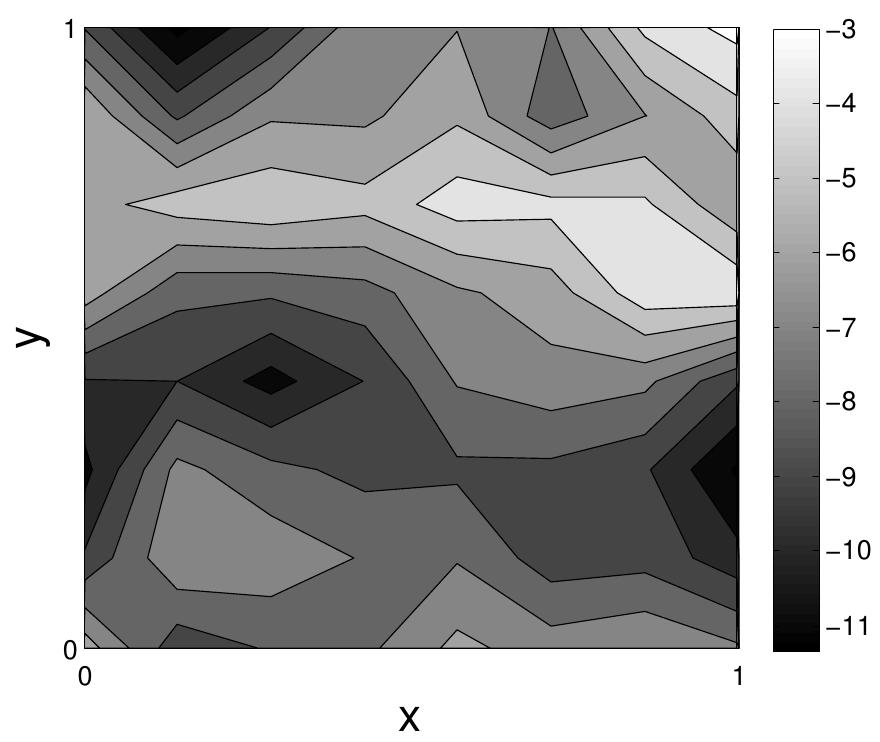}
\caption{Bypassing of barriers in 2d. Top: Trajectory of the clump in the $(x,y)$ plane under the effect of a force oriented rightward along the $x$ axis, indicated by an arrow. Parameters are $T=0.004$, $\alpha=0.001$, $N=32\times32$, $A=0.02$, simulation time = {$3.5\cdot10^5$ rounds} of $20N$ steps. Bottom: resulting contour plot of ${\hat{F}\equiv-\log \tau_\text{tot}(x,y)}$, where $\tau_\text{tot}(x,y)$ is the total time spent in position $(x,y)$ after incorporation of periodic boundary conditions. $\hat{F}$ is an estimate of the free energy landscape: deep local minima, hills and valleys appear, see grey-level scale on the right side.}
 \label{fig:trajdrift}
\end{figure}

In two dimensions, contrary to the one-dimensional case, free energy barriers can be bypassed. Drift can occur even with forces that are not strong enough to cross the barriers, and the value of $A_{depin}$ given above is not relevant. Simulations indeed show that the pinning of the clump is much weaker than in one dimension, see Fig.~\ref{fig:drift}. This is an important point, which shows that the dynamics of the clump within one map strongly differ in the one- and two-dimensional cases. In two dimensions, contrary to one dimension, barriers can be bypassed by the clump trajectories. This phenomenon could explain the fact that the diffusion constants measured in 2-dimensional simulations are in general larger than their 1-dimensional counterparts (Fig.~\ref{fig:logd}). We checked the existence of this bypassing mechanism by looking at trajectories of the clump in the $(x,y)$ plane when an external force is applied along the $x$ axis (Fig.~\ref{fig:trajdrift}, top). We observe displacements along the $y$ axis, with preferred values for $y$, indicating that the overall rightward motion is the result of the clump motion around the barriers, instead of crossing them. We looked at the time spent in each position of the unit square (Fig.~\ref{fig:trajdrift}, bottom). Favored positions clearly appear, where the total time spent is several orders of magnitude greater than in other positions. The opposite of the logarithm of these residence times is an estimate of the free-energy landscape probed by the moving clump. 

In experiments, place fields have been studied in both one- and two-dimensional environments ('one' referring to a linear track whose width is small compared to the length), but the two-dimensional case is obviously of particular importance for natural environments.

\subsection{Retrieval}

In Hopfield's original model for attractor neural networks (ANN), a memory item corresponds to one activity configuration of the network. The retrieval phase consists in stabilizing the network activity in this configuration, starting from a different initial configuration. In contrast, in our ANN model for the hippocampus, a memory item corresponds to a map, \emph{i.e.} a whole set of activity configurations corresponding to clumps centered around positions along the map. What does retrieval mean in this case? Two views are possible. First, it is of course possible to retrieve (in Hopfield's model sense) one particular activity pattern starting from a similar configuration, that is, a clump centered on one particular position in one particular environment. This retrieval mechanism, requiring a specific input, will be addressed in Section \ref{sec:loc}. Secondly, one can focus on the broader issue of map retrieval. In this case one map would be retrieved, if the activity is coherent (localized) in the map,  while the clump is free to wander in the environment, see Section \ref{sec:global}.

\subsubsection{Retrieval of one position in a given environment}

\label{sec:loc}

We investigate the dynamics of the model when one given position in a given environment is selected by a local field. The pattern to be retrieved is an activity configuration $\boldsymbol{\xi}$ corresponding to a clump centered on, say, position $x_0$ in environment $0$. A local field $h_i$ is applied on the spins: 
\begin{equation}
h_i=\left\{\begin{array}{l}
h\ \text{if}\ \big|\frac{i}{N}-x_0\big|\leq d_0\ ,\\
0\ \text{otherwise}\ .
\end{array}\right.
\end{equation}

\begin{figure}

\centering

\includegraphics[width=\columnwidth]{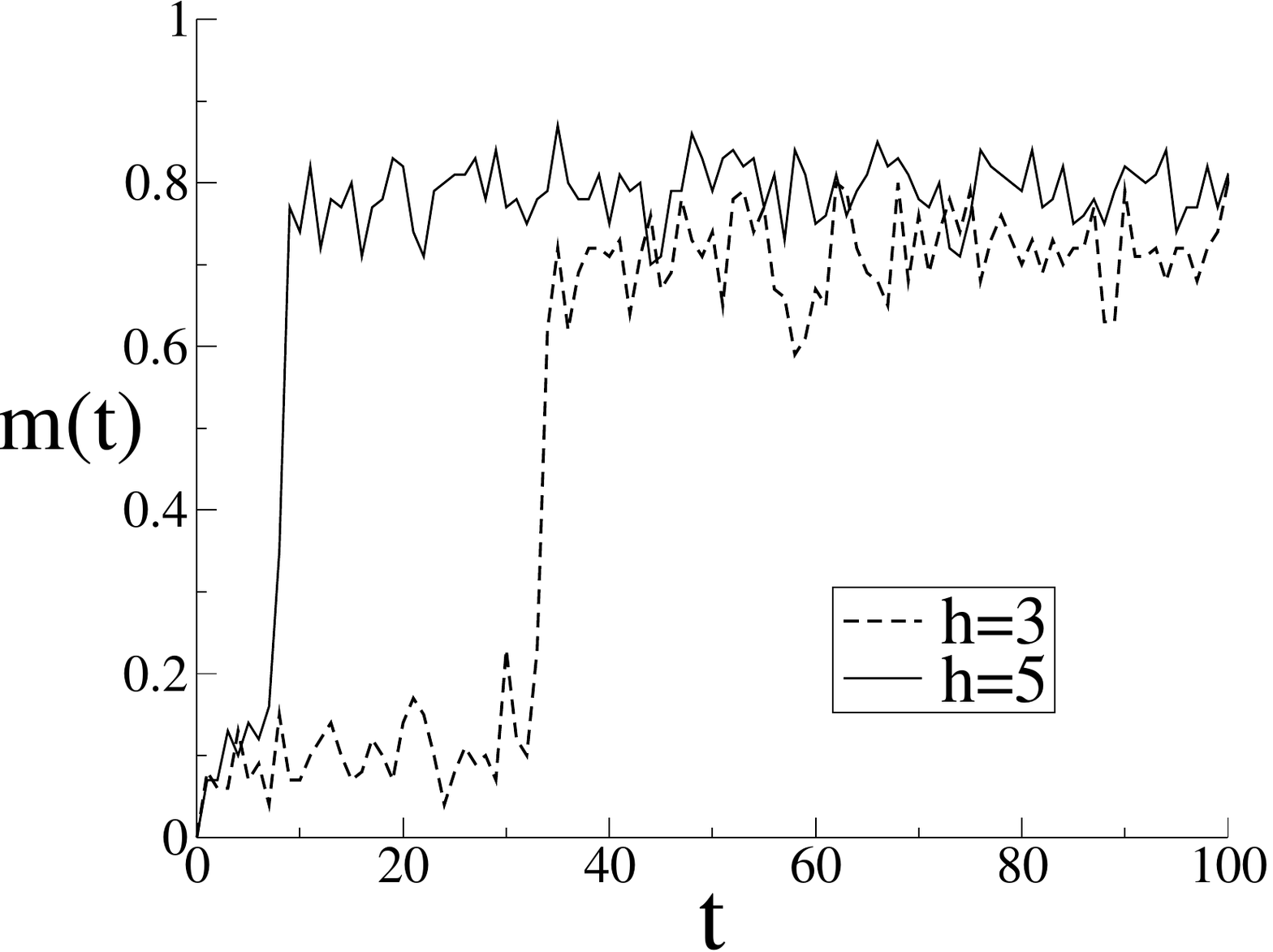}

\caption{Evolution of the overlap with the retrieved pattern as a function of time, during two Monte Carlo simulations initialized in the clump phase in the same environment, at a position different from $x_0$. $N=1000$, $T=0.006$, $\alpha=0.01$, $d_0=0.05$, time unit = 1 round of $20N$ steps.}

\label{fig:retrieval1}

\end{figure}

 Retrieval is detected by the measure of the overlap 
\begin{equation}
 m\equiv\frac1{fN}\sum\limits_i\sigma_i\xi_i\ .
\end{equation}
An example of the retrieval process is given in Fig.~\ref{fig:retrieval1}: it occurs abruptly, as a global switching of the network activity to a configuration close to $\boldsymbol{\xi}$.

\begin{figure}

\centering

\includegraphics[width=\columnwidth]{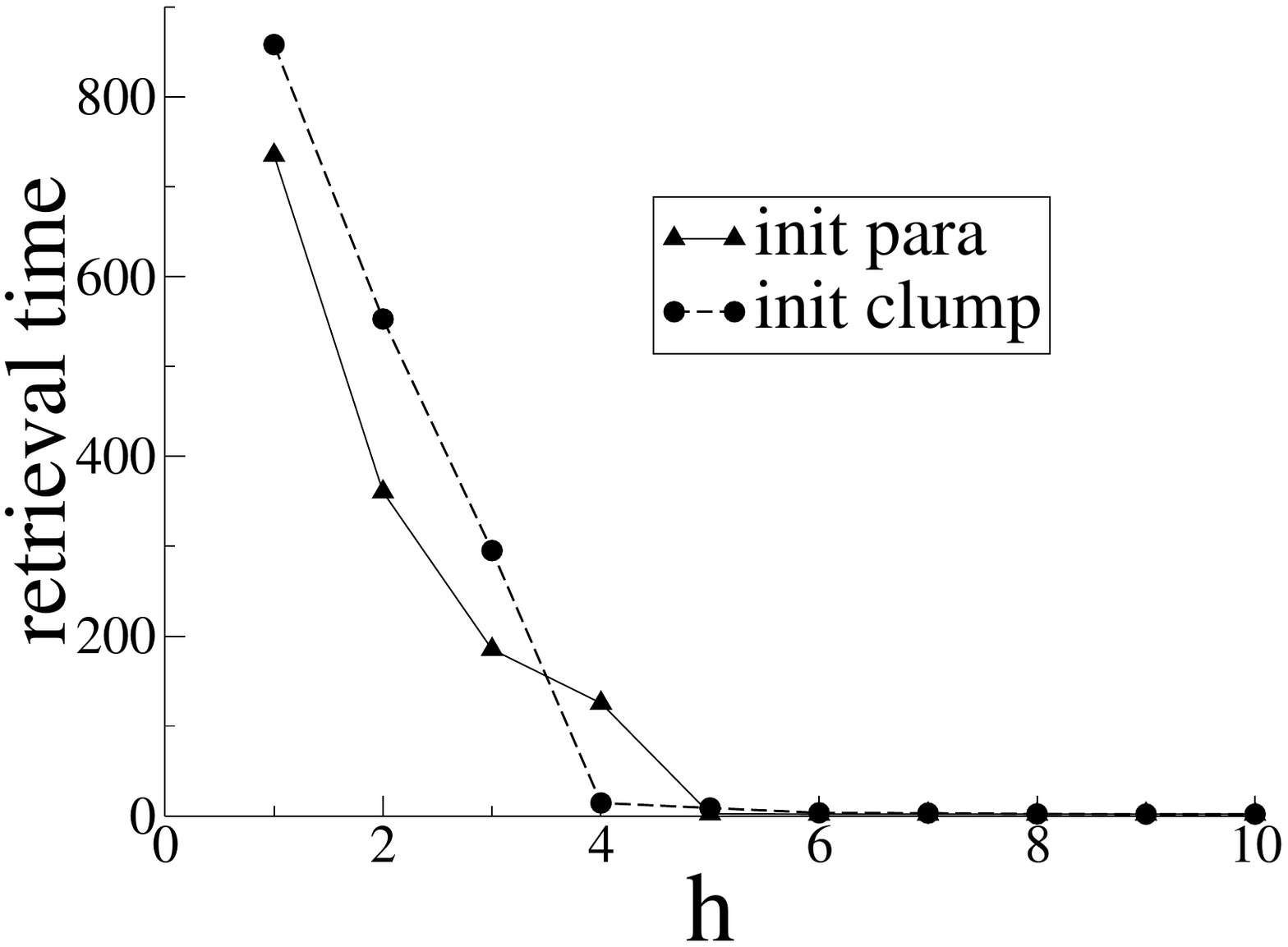}

\includegraphics[width=\columnwidth]{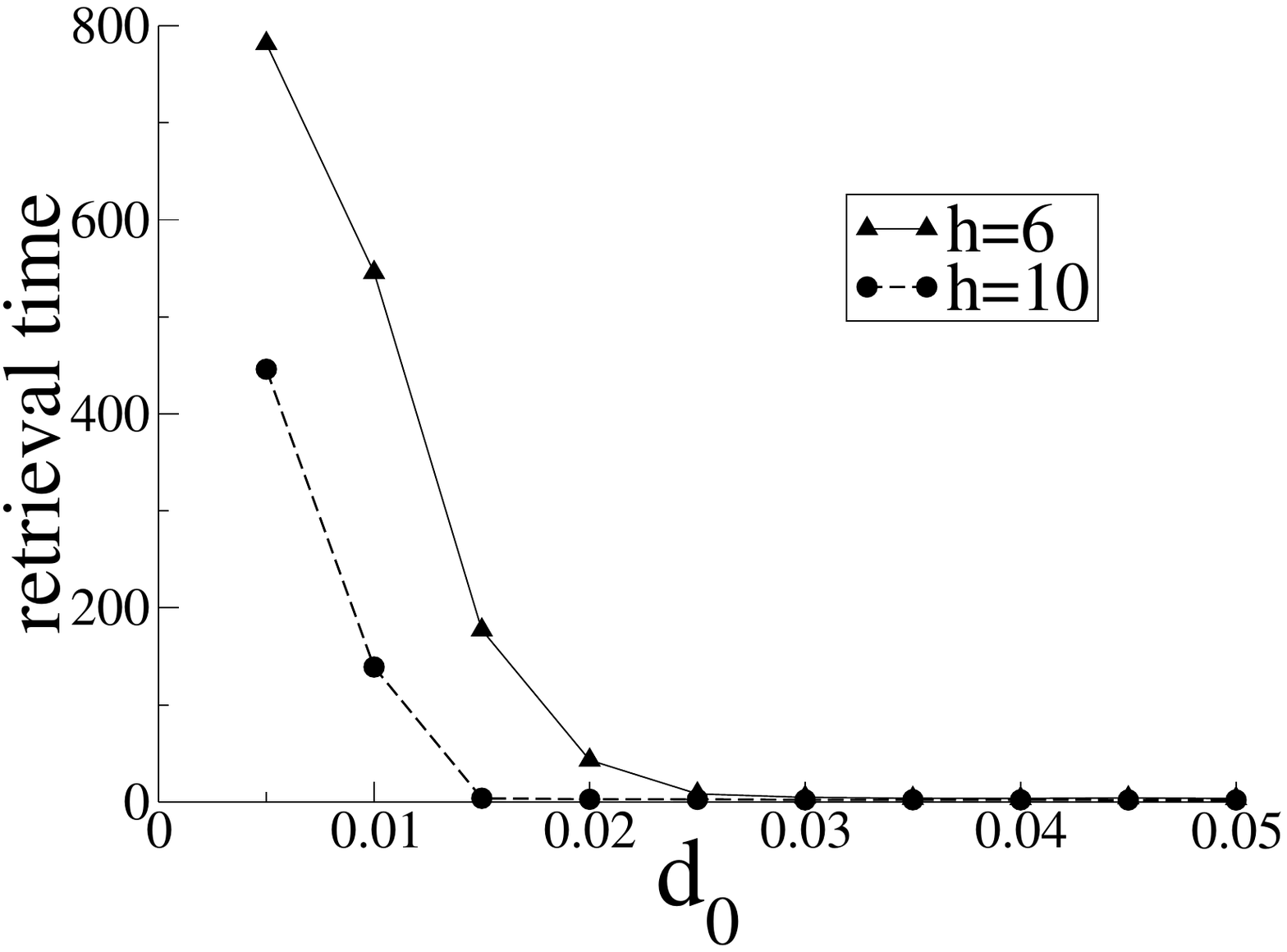}

\caption{Average retrieval time in Monte Carlo simulations as a function of $h$ (top) and $d_0$ (bottom). Each point is averaged over 10 simulations. $N=1000$, $T=0.006$, $\alpha=0.01$, time unit = 1 round of $20N$ steps. In the top panel, $d_0=0.05$.}

\label{fig:retrieval2}

\end{figure}

As expected, the time taken for retrieval is a decreasing function of $h$ and $d_0$ (Fig.~\ref{fig:retrieval2}). It does not depend significantly on the initial conditions of the network.

\subsubsection{Retrieval of one environment}

\label{sec:global}

In order to stabilize one particular map, say of index $\ell$, we ran simulations in which we increased the contribution $J^{\ell}$ to the total synaptic matrix $J$. This artificial modification does not correspond to any physiological mechanism \emph{per se} but could mimic the effect of a 'context dependence' \cite{Smith06}. The synaptic matrix is modified as followed:
\begin{equation}
J_{ij} \to  J_{ij} + h_J\cdot J_{ij}^{\ell}\ ,
\end{equation}
where $h_J>0$ and $J_{ij}^\ell=J_{\pi^\ell(i)\pi^\ell(j)}^0$ (Section \ref{sec:prelim}).

\begin{figure}

\centering

\includegraphics[width=\columnwidth]{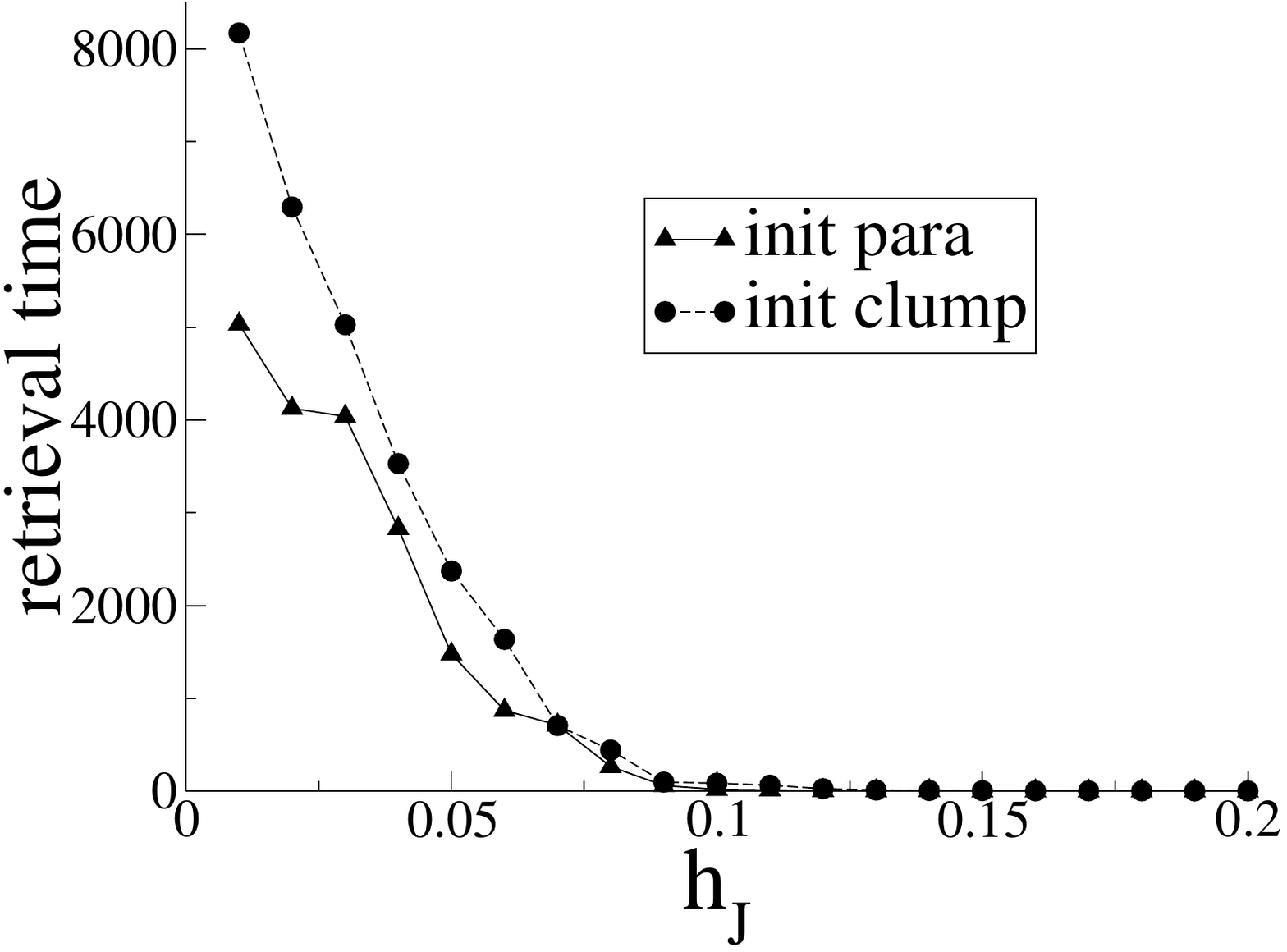}

\caption{Average retrieval time in Monte Carlo simulations as a function of $h_J$, with two different initial conditions: a system in the paramagnetic phase (triangles) or in the clump phase in another environment (circles). Each point is averaged over 100 simulations. $N=1000$, $T=0.006$, $\alpha=0.01$, time unit = 1 round of $20N$ steps.}

\label{fig:globalinput}

\end{figure}

As expected, the time taken for retrieval is a decreasing function of $h_J$ (Fig.~\ref{fig:globalinput}). Note that the retrieval is almost immediate as soon as the additional weight on the environment exceeds 10\%. Interestingly, the retrieval is slightly slowed down if the initial state of the system is a clump in another environment, rather than a paramagnetic configuration. The global input is then in competition with the barriers opposing transitions between environments.

\section{Effects of out-of-equilibrium mechanisms on clump motion}

\label{sec:add}

\subsection{Adaptation}

\label{adapt}

An important biophysical process, which can be incorporated into the model, is spike frequency adaptation. The membrane voltage of frequently active neurons is hyperpolarized by potassium currents, and their firing rates decay to submaximal levels. Adaptation has been observed in hippocampal pyramidal cells \cite{HippocampusBook}. This neural fatigue phenomenon has been proposed as a mechanism to make the clump, otherwise stationary, diffuse in the environment in the absence of external input (mental exploration)\cite{Hopfield10}.

We introduce a mechanism for adaptation in the simulations, to see if it enhances the diffusion process as expected. At the cell level, adaptation can be modelled as an auto-inhibitory current that relaxes with a time constant $\tau_\text{adapt}$ \cite{Hopfield10}. This auto-inhibition was taken into account in the simulations by adding a local field on the spins whose value depends on the spin's past activity. The system is now out of equilibrium, but the time constant $\tau_\text{adapt}$ is chosen to be large compared to thermalization times so that the fields vary slowly. More precisely, we add
\begin{equation}
h_i(t)\equiv h_{\text{adapt}}\sum\limits_{\tau\geq0}e^{-\tau/\tau_\text{adapt}}\,\sigma_i(t-\tau)\ ,
\end{equation}
where $h_{\text{adapt}}$ measures the intensity of the neural fatigue.

\begin{figure}

\centering

 \includegraphics[width=\columnwidth]{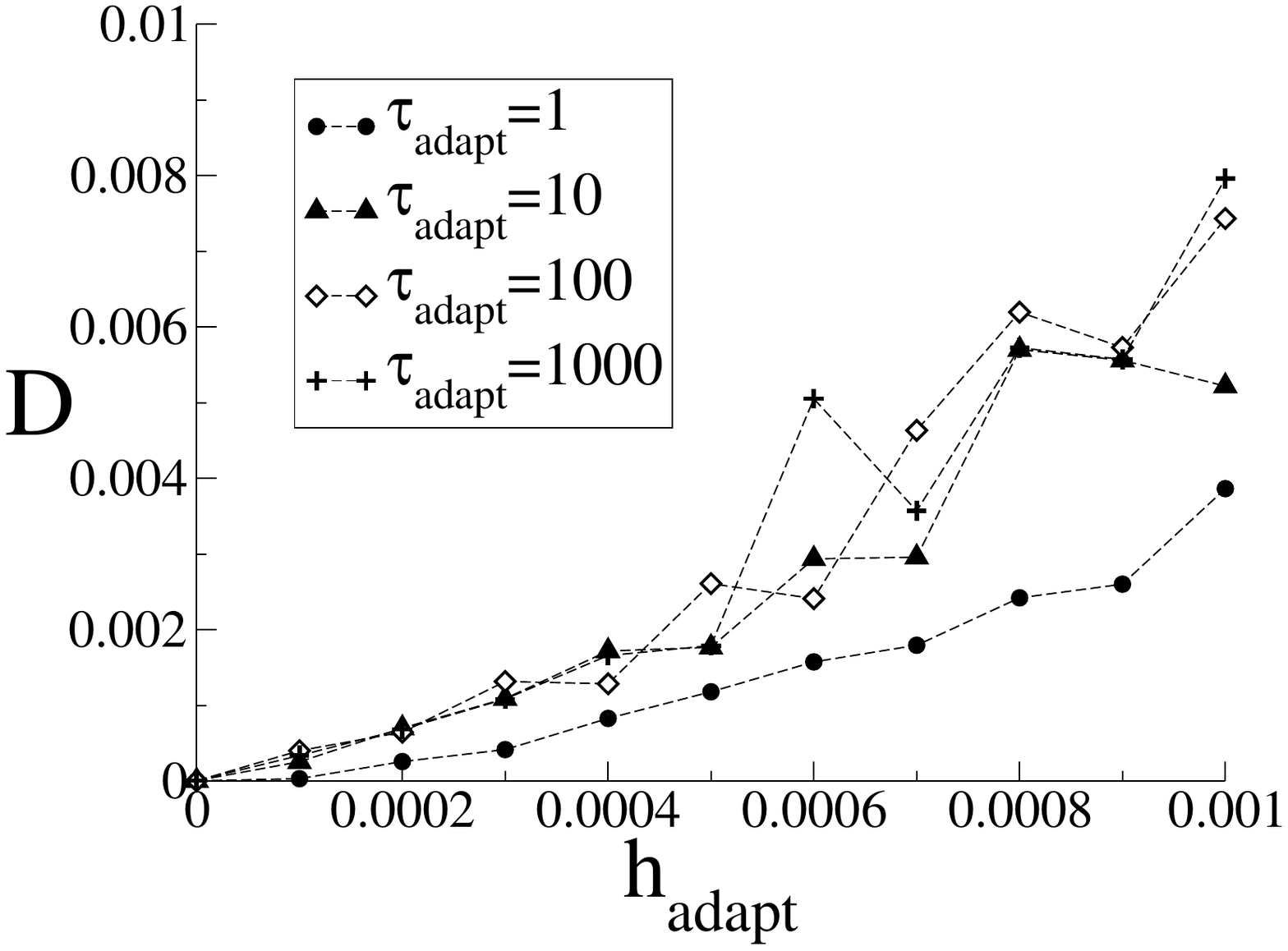}

 \includegraphics[width=\columnwidth]{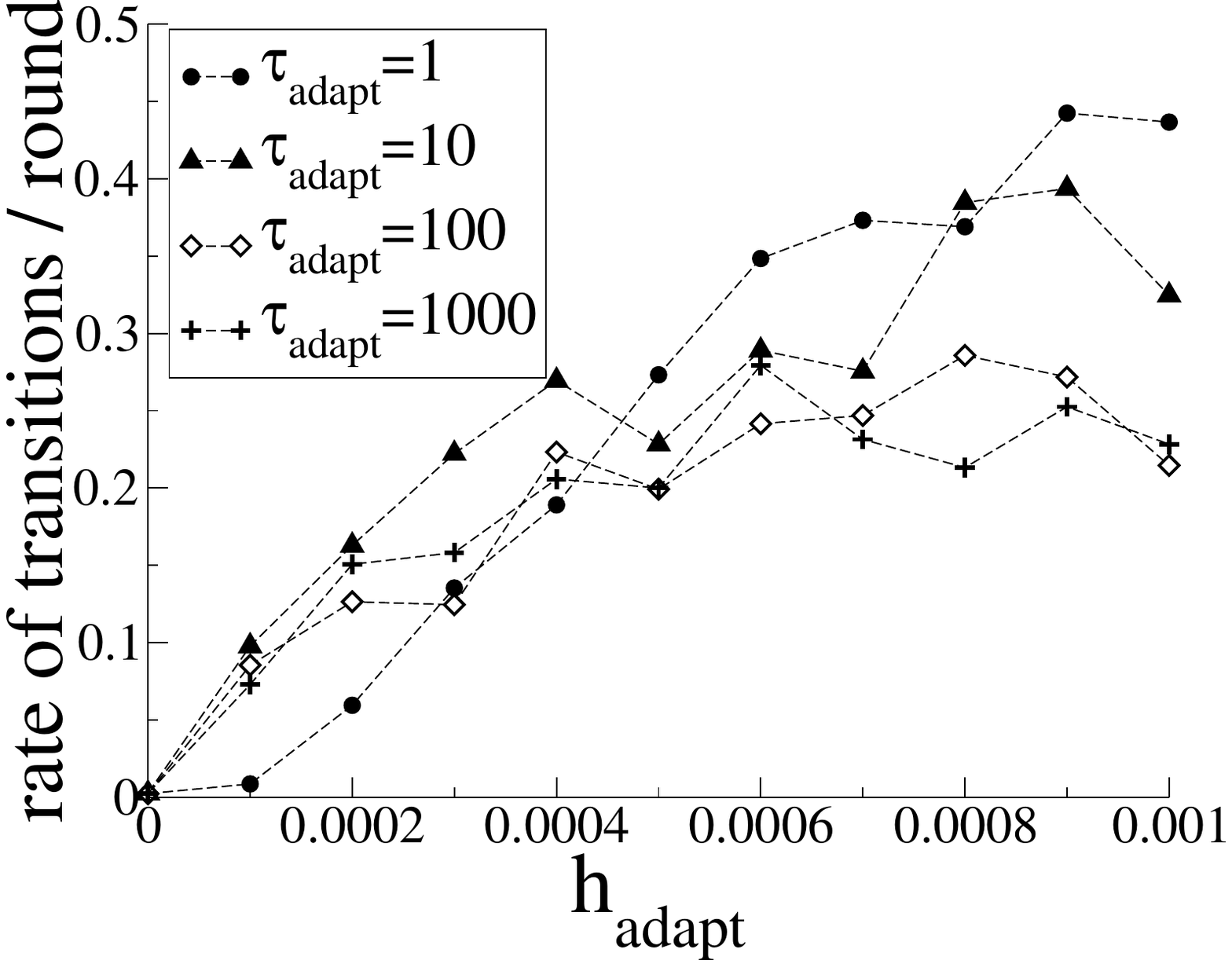}

\caption{Adaptation: $D$ (top) and frequency of transitions (bottom) as a function of the adaptation's intensity $h_{\text{adapt}}$ measured in a Monte Carlo in dimension 1 with $N=333$ spins, $\alpha=0.003$, $T=0.004$, and various values of $\tau_\text{adapt}$ (time unit: 1 round of $20N$ steps). The clump is not stable for stronger $h_{\text{adapt}}$. Depending on the frequency of transitions, simulation durations range from $\sim 10$ to 1000 rounds; each point is averaged over 100 simulations. The estimated error on $D$ varies between $5\cdot10^{-5}$ to a few $10^{-3}$ when $h_{\text{adapt}}$ increases from 0 to 0.001. }

 \label{fig:adapt}

\end{figure}

We ran simulations with various time constants ${\tau_\text{adapt}}$ and intensities $h_{\text{adapt}}$. We used $D$ defined in Section~\ref{sec:numsim} as a measure of the clump square displacement per unit of time. Note that $D$ does not correspond {\em strictly speaking} to a diffusion coefficient any more. As expected intuitively, we observe that increasing $h_{\text{adapt}}$ facilitates the motion of the clump (Fig.~\ref{fig:adapt}, top), but also tends to destabilize it. Transitions to other environments are more frequent (Fig.~\ref{fig:adapt}, bottom) as $h_{\text{adapt}}$ increases, and if $h_{\text{adapt}}$ is too large, the clump breaks apart.

These results support a recent work by Hopfield \cite{Hopfield10}, according to which adaptation (and not the sole neural noise) could be the neural mechanism by which a bump of activity dynamically explores a continuous attractor manifold in the absence of visual or self-motion input. Such a spontaneous motion at the level of the neural activity, taking place without the animal's physically moving, appears useful in the realization of mental exploration tasks such as future trajectories planning or past trajectories remembering. These results also reveal the increasing occurrence of transitions between environments when out-of-equilibrium mechanisms are added to the model and stress the importance of this phenomenon in competition with clump motion within one map.

\subsection{Fluctuations in the global inhibition}

\label{varact}

In our model the effect of inhibitory cells is modeled as a constant activity level $f$ of pyramidal cells. However, in hippocampal recordings in rodents this level varies periodically across time, a phenomenon called theta rythm \cite{Vanderwolf69}. These oscillations play a role in the position coding through the phase precession phenomenon \cite{OKeefe93,Jensen00} and have been proposed as a possible mechanism for resetting of the path integrator \cite{Samsonovich97,Stella11}. Here we address the issue of the effect of theta waves on the diffusing behavior of the clump. We know that changing $f$ quantitatively changes the stability domain of the clump phase and correspondingly moves the $N_c$ contour lines. As a consequence, varying $f$ at a given $(\alpha,T)$ will have the effect of varying the diffusion constant, but in any case this constant remains quite low in the whole stability domain of the clump. So we do not expect the variations of $f$ to improve dramatically the diffusion process.

\begin{figure}

\centering

 \includegraphics[width=\columnwidth]{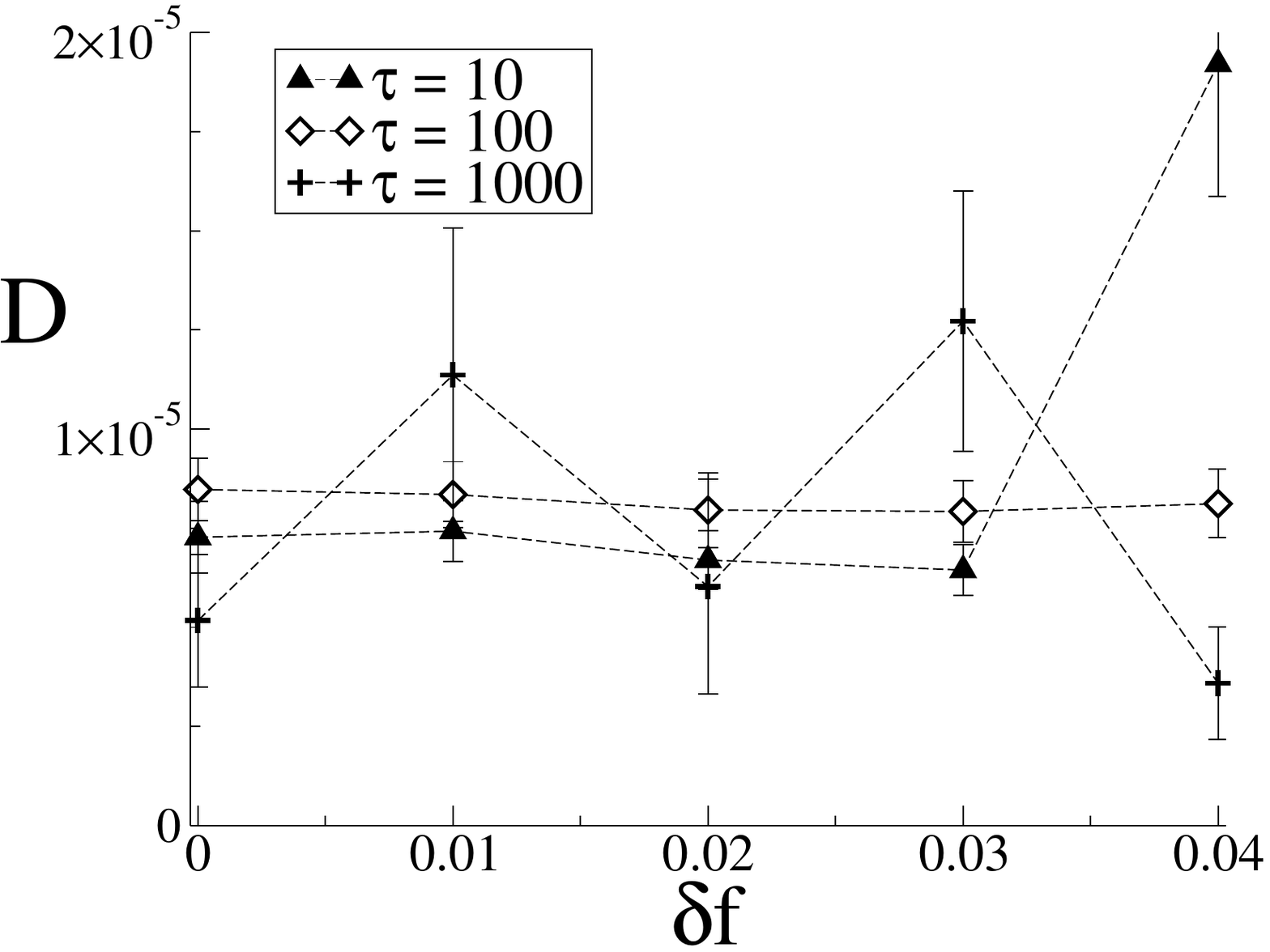}

\caption{Simulations of the case $f(t)=f+\delta f\sin(t/\tau)$: $D$ as a function of $\delta f$ measured in a Monte Carlo in dimension 1 with $N=1000$ spins, $\alpha=0.003$, $T=0.005$, $w=0.05$, and various values of $\tau$ (time unit: $100N$ steps). The clump is not stable for stronger $\delta f$. For $\tau=10$ and $\tau=100$, each point is averaged over 100 simulations of length varying from a few tens of rounds to 1000 rounds depending on the frequency of transitions. For $\tau=1000$, longer simulations were necessary in order to cover several periods of $f(t)$; each point is thus averaged over 10 simulations of duration up to 25000 rounds.}

 \label{fig:varact}

\end{figure}

We simulated the network at a given $(\alpha,T)$ and activity level $f(t)=f+\delta f\sin(t/\tau)$ where $\delta f$ is chosen small enough so that the clump phase remains stable at this $(\alpha,T)$ and $\tau$ is large compared to the simulation unit time. As expected, there is no significant improvement of diffusion, see Fig.~\ref{fig:varact}.

\subsection{Asymmetric synapses}
\label{sec:asym}

In the Hopfield model \cite{Hopfield82}, couplings are given by Hebb's rule and are therefore symmetric. Our synaptic matrix (\ref{rule2}) also follows a Hebbian prescription. Working with symmetric couplings ensures the existence of an equilibrium Gibbs measure over configurations \cite{Amit89}, allowing us to use statistical mechanics tools in this framework. Nevertheless, in biological neural networks asymmetric synaptic plasticity exists \cite{Levy83}. In one-dimensional environments for instance, where most place fields are directional \cite{McNaughton83}, asymmetric learning may take place. In addition, in certain models of the hippocampus, asymmetric synapses have been proposed to play a critical role in some observed phenomena such as phase precession \cite{Tsodyks96}. Attractor neural networks with asymmetric synapses and their storage capacity have been formally studied by \cite{Roudi04}.

\begin{figure}

\centering

 \includegraphics[width=\columnwidth]{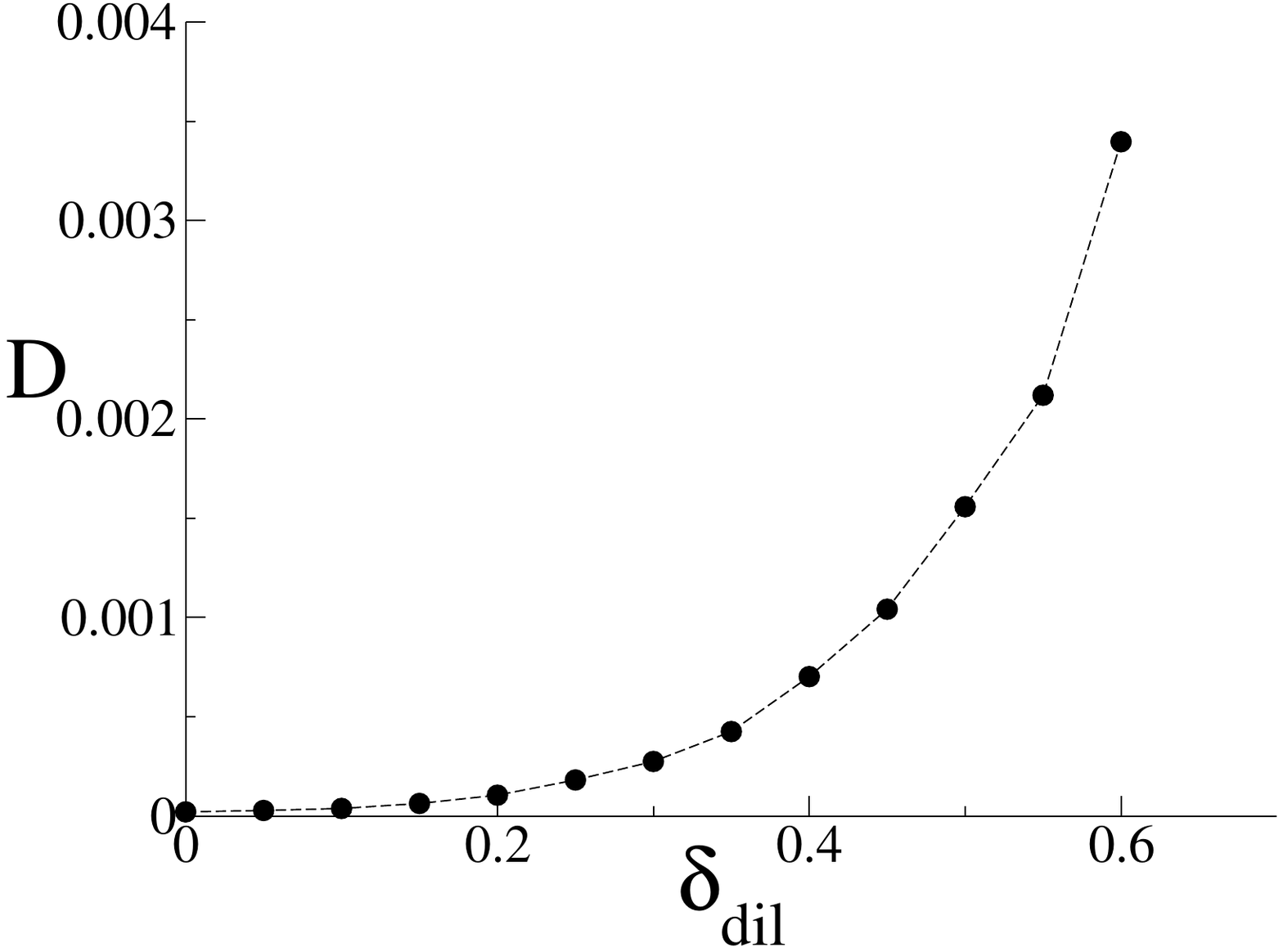}

 \includegraphics[width=6cm,angle=90]{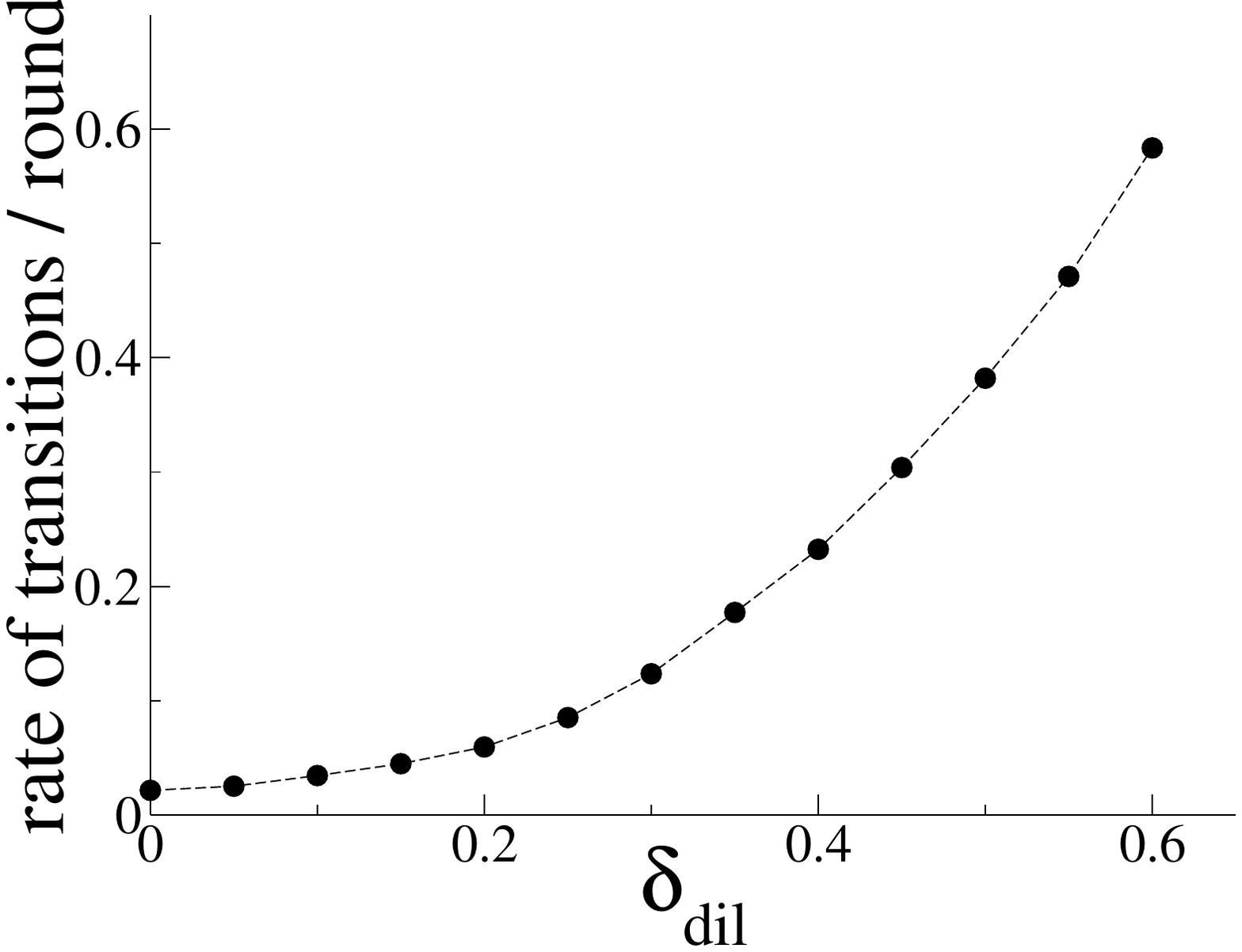}

\caption{Effect of asymmetric random dilution of synapses on $D$ (top) and on the frequency of transitions to other environments (bottom): Monte Carlo simulations in dimension 1 with $N=333$ spins, $\alpha=0.003$, $T=0.005$. The clump is not stable for stronger dilution. Depending on the frequency of transitions, the simulations length varies between 1000 rounds and a few rounds. Each point is averaged over 1000 simulations. The estimated error on $D$ varies between $5\cdot10^{-6}$ to $10^{-4}$ when $\delta_\text{dil}$ increases from 0 to 0.6.}

 \label{fig:asym1d}

\end{figure}

To study the effect of asymmetric synapses on the dynamics of our model we randomly remove a fraction of the couplings $J_{ij}$ \cite{Amit89,Roudi04}. More precisely, if $\delta_\text{dil}$ denotes the dilution fraction, for each $i<j$ we choose
\begin{eqnarray}
\left\{
\begin{array}{ll}
 J_{ij}\rightarrow0\\
J_{ji}\text{ unchanged } 
\end{array}
\right.\text{ with probability } \frac{\delta_\text{dil}}{2}\ ,\nonumber\\
\left\{
\begin{array}{ll}
J_{ij}\text{ unchanged }\\
J_{ji} \rightarrow0
\end{array}
\right.\text{ with probability } \frac{\delta_\text{dil}}{2}\ ,\nonumber\\
J_{ij},J_{ji} \text{ unchanged }
\text{ with probability } 1-\delta_\text{dil}\ .
\end{eqnarray}
We measured $D$ defined in Sec.~\ref{sec:numsim}, with the results shown in Fig.~\ref{fig:asym1d}. We observe that the asymmetric dilution of synapses increases $D$. Nevertheless, because of the concomitant destabilization of the clump, the enhancement of $D$ is here again in competition with more frequent transitions to other environments.

\section{Conclusion}

\subsection{Summary of results}

In this work we have presented analytical and numerical results on the dynamics of a model for hippocampal place cells. Under certain conditions of noise and load, the activity is spatially localized in one of the stored environments (clump phase) \cite{Monasson13}. Here, we have focused on the motion of such a clump across space within one environment, under the influence of neural noise and of quenched disorder due to the other maps contributing to the couplings. In other words, we have studied the dynamics of an attractor neural network storing spatial maps within one of its attractors, with or without external input.

We have first addressed the issue of the macroscopic description of the clump. At equilibrium, the clump shape is described by the average density profile $\rho(x)$. Here we have analytically shown, in the single-environment case, that a macroscopic description of its dynamical evolution within one map was also possible. More precisely the microscopic dynamics of the individual neurons produces an emergent, collective macroscopic motion of diffusion for the clump. The clump therefore acquires the status of a quasi-particle, with very weak fluctuations (for large sizes $N$) of shape, while moving in space. It is legitimate to say that the position of the center of the clump plays the role of a collective coordinate for the neural configurations. In their model of the hippocampus, Samsonovich \& McNaughton \cite{Samsonovich97} had already described the evolution of the clump by a collective coordinate that emerged from the microscopic dynamics in simulations, but the equivalence between both levels of description was not formally justified. Here, we have analytically demonstrated its soundness. We have, in addition, obtained an exact expression for the diffusion coefficient of the clump and its effective mobility as a function of the detailed dynamical rates of the single neurons used in the Monte Carlo simulations. 

We have also considered the dynamical properties of the model in the presence of the quenched disorder caused by multiple-environment storage in the synapses. In this case, the free-energy landscape probed by the clump moving through space is rough. As soon as the number of units exceeds a few hundreds or even tens, the diffusion of the clump appears to be severely hindered by the free energy barriers, especially in 1 dimension. This effect, predicted by the analytical study of the statistics of the free-energy landscape, is corroborated by Monte Carlo simulations. It is found to be very robust to changes in the parameters $f$, $w$, $c$.  Therefore, noise alone is not enough for an efficient motion of the clump, and additional mechanisms must be taken into account. This point had already been underlined by Hopfield in a recent model for mental exploration in the hippocampus \cite{Hopfield10}. It is also related to the clustering effect predicted by Tsodyks \& Sejnowski \cite{TsodyksSejnowski95}, who numerically observed that the presence of disorder in connections tends to make stable bumps collapse into positions corresponding to `places where the synaptic interaction between neurons is strongest', \emph{i.e.} local minima of the energy. Interestingly, in the 2-dimensional case, the possibility of trajectories bypassing the free energy barriers leads to a larger coefficient constant than in 1d. This effect is of particular relevance for biological cognitive maps, often thought to be two-dimensional. Moreover,  the cross-talk between environments also causes transitions from one map to the other, in competition with motion within one map.

We have then investigated the effect of a force on the network, and have showed that a force could, indeed, help the clump overcome free-energy barriers and move across space. This set-up allowed us to exhibit the by-passing of barriers in two dimensions. 

The motion of the clump can also be enhanced by out-of-equilibrium mechanisms. We have modified the model in order to incorporate spike-frequency adaptation,  asymmetry in the synapses, and temporal fluctuations in the level of inhibition. For all mechanisms but the latter, motion is found to be facilitated.

\subsection{Biological relevance}

In order to perform exact, analytical calculations, and to reach a more controlled and accurate understanding of the phenomena at work than with simulations, we intentionally discard many biological features, of various degrees of importance, in our modeling.

We assume first that the learning process is complete (synapses are frozen) and perfect (the $J_{ij}^\ell$ perfectly reflect the topology of the environments, without distortion). In addition each new environment contributes additively to the synapses (Hebb's law). The separation of the learning and the retrieval processes is a common assumption. Quenched distortions in the synapses could be incorporated in the study, {\em e.g.} by making the matrix $J^0$ random rather than perfectly regular on a grid. We expect quenched distortions to have similar effects to the quenched interference noise coming from multiple map storage. Hebb's rule is also a common assumption; it has been shown, in the context of Hopfield's model, that the attractor dynamics is qualitatively robust against the choice of alternative, non-additive rules \cite{Amit89}. We discussed the case of asymmetric synapses in Sec.~\ref{sec:asym}.

Another simplification of the present model is to assume that synaptic interactions code for the topology of the environments, {\em i.e.} spatial information only. We discard any additional 'dimension', such as context dependence \cite{Smith06}, as is the case in most models of place cells. Relaxing this assumption in a meaningful way is a tantalizing task in the absence of a clear experimental guidance. We have also assumed non-directional place fields, in contrast with experimental observations (mostly in one dimension). Directionality could easily be incorporated in the model, and we do not expect it to have a significant effect on most of our results.

The effect of interneurons is modeled through a spatially-homogenous inhibition, which maintains the global level of activity (fraction of active neurons) constant. We ignore spatial inhomogeneities in the inhibitory network, as well as fluctuations in the activity level, such as the theta rhythm. In section \ref{varact} we have relaxed the latter hypothesis in simulations, and have observed that a varying level of activity had no significant effect on the motility of the clump. Nevertheless fluctuations in the activity could have consequences on other phenomena, such as the transitions between maps \cite{Stella11}.

Modeling neurons through binary units is also a big simplification. Realistic conductance-based models would be necessary to describe the dynamics of neurons in a accurate way from the biological point of view. However, such detailed models are intractable in the case of large networks. A majority of works on continuous attractor neural networks make use of rate models \cite{TsodyksSejnowski95,BattagliaTreves98,Tsodyks99,Fung12}. Here, we choose to use binary units, as discussed in our previous study \cite{Monasson13}. The use of binary units allows us to incorporate the noise in the neural response at the time-scale of a spike, while rate variables usually represent the activity of neurons averaged over time, or over a population of neurons. In this respect, the binary description can be considered as more microscopic than rate-based models. Indeed, the rate-based macroscopic description naturally emerges  in our calculation through the order parameters $\rho(x)$ and $\mu(x)$, see also Section II.C in \cite{Monasson13}. Our study therefore offers a microscopic basis for rate-based equations and for the properties of continuous attractors, see for instance the detailed description of the collective motion of the clump from the microscopic dynamical rules  of individual neurons.

A drastic simplification in the present work is the absence of any input. Inputs, be they sensorial or the result of path-integration, are indeed believed to be very important in biologically plausible situations. Yet, our work aims at studying the attractor dynamics. In this context, it is important to understand the spontaneous evolution of the network before taking any external input into consideration. Moreover, the precise form of the inputs to hippocampus, their timing and their intensities are poorly known, which makes their effect on the hippocampal activity hard to model from a quantitative point of view. 

Despite the restrictions listed above we expect that some of our results are quite general, and would hold for more biologically-oriented models. The effect of disorder on the motion of the collective clump within one map is a very robust feature of our model. In one dimension, the motion is drastically hindered, regardless of the parameter values. In two dimensions, this pinning effect is softened by the possibility of by-passing the barriers. We expect that, in higher dimensions, the motion of the clump would be even easier. This behavior, reminiscent of localization phenomena in condensed matter, is likely to remain true even for more realistic models from a biological point of view. A precise coding of position would therefore not be possible, in low dimensions, unless the clump is driven out of free-energy minima by strong enough inputs.

What clearly arises from this study is that, as a result of crosstalk, attractor manifolds coding for different maps are far from being flat, in contradistinction with the usual picture of continuous attractor neural network. Hence, distances in the space of hippocampal neural activities are distorted compared to the 'true' distances in the real space. This finding is consistent with the assumption that the metric system of the brain is encoded in another region, while the hippocampus could serve as an associative system linking together places and other elements of memory.

\subsection{Possible extensions}

Our study could be extended along various directions, some of which are listed below.

An interesting feature of the model is the by-passing of barriers by the clump in two-dimensional maps. To be more quantitative, we could imagine running drift simulations on a strip, that is, a two-dimensional environment with periodic boundary conditions along the $x$-axis and a finite size along the $y$-axis. This would allow us to quantify the minimal 'degree of two-dimensionality' for the motion of the clump, {\em i.e.} the minimal y-width above which the clump can move around the barriers. We expect this width to be of the order of $l_b$.

Our study of biologically-motivated mechanisms possibly enhancing the motility of the clump is not exhaustive. For instance, synapse dynamics, that is, the short-term depression and/or facilitation of synapses, is another candidate. Its effect on the dynamics of a bump of activity in continuous attractor neural networks (in the absence of thermal and quenched disorder) has recently been studied by Fung et al \cite{Fung12}, who showed  that short-term depression increases the motility of the clump. 

In addition, it would be interesting to investigate further the issue of the response to inputs. How the hippocampus integrates the information conveyed by brain areas upstream CA3 is still not fully understood, in spite of a wealth of experimental results during the past ten years (notably the discovery of grid cells \cite{Fyhn04,Hafting05}). The hippocampus is not isolated but a part of a system of interacting regions \cite{Moser08}. The comprehension of the perforant pathway and mossy fibers inputs is a pivotal point. More generally, in the context of attractor network theory, reaching a deep understanding of the effect of these input sources of information on the attractor dynamics would be very important.

Last of all, a striking general result of our study is that diffusion is always in competition with transitions to other environments, whose main features were reported in Section \ref{transit}. All the mechanisms we added to the model in order to make the clump move also increased the probability of these transitions. Two possible (and not mutually exclusive) explanations can be proposed. First, when the clump moves, it explores more positions in space and, thus, has a larger probability to find a 'favorable' position for transitions, that is, a position where the energy barrier opposing a transition is not too large. Secondly, mechanisms enhancing the diffusion of the clump in one environment also tend to destabilize it, which makes transitions to another environment more likely. The study of these transitions is therefore a key issue, not only for the full understanding of the dynamics of our model, but also for the interpretation of experimental results, where manipulations of the visual cues resulted in abrupt swaps of the neural activity \cite{Wills05,Jezek11}. This question will be addressed in a forthcoming publication.

\vskip .3cm
\noindent {\bf Acknowledgements.}
We are indebted to J.~Hopfield for very fruitful discussions, in particular on the emergence of the clump as a collective coordinate of the dynamics. We are grateful to F. Stella for useful discussions. The work of S.R. is supported by a grant from D\'{e}l\'{e}gation G\'{e}n\'{e}rale de l'Armement.

\appendix

\section{Reminder on the free-energy calculation}

\label{app:F}

In \cite{Monasson13} we computed the average free-energy of the system over random remappings. To do so, we used the replica method under the replica-symmetric assumption. In this Appendix we remind the main results of this calculation.\\

The average partition function of the replicated system is 
\begin{equation}
\overline{Z_J^n}=\int\prod\limits_{a<b}\mathrm{d}q^{ab}\mathrm{d}r^{ab}\mathcal{D}\rho^a(x)\mathcal{D}\mu^a(x)\mathrm{d}\lambda^a e^{-N\beta \mathcal{F}_n}\ ,
\end{equation}
where  $a,b$ are the replica indices,  $q^{ab}=\frac1N\sum\limits_i\sigma_i^a\sigma_i^b$ are the overlaps between replicas, the $r^{ab}$ are parameters conjugated to the $q^{ab}$, and
\begin{eqnarray}
\label{eq:Fn}
 \mathcal{F}_n&=&\alpha\beta\sum\limits_{a<b}r^{ab}q^{ab}+\alpha T\sum\limits_{\lambda\neq0}\mathrm{Tr}\ln[\mathbf{Id}_n-\beta\lambda(\mathbf{q}-f^2\mathbf{1}_n)]\nonumber\\
&&-\sum\limits_a\lambda_a\Big(\int\mathrm{d}x\rho^a(x)-f\Big)+\sum\limits_a\int\mathrm{d}x\rho^a(x)\mu^a(x)\nonumber\\
&&-\frac12\sum\limits_a\int\mathrm{d}x\mathrm{d}y\rho^a(x)J_w(x-y)\rho^a(y)\nonumber\\
&&-T\int\mathrm{d}x\ln\left[\sum\limits_{\{\sigma^a\}}e^{\alpha\beta^2\sum\limits_{a<b}\sigma^a\sigma^br^{ab}+\beta\sum\limits_a\mu^a(x)\sigma^a}\right]\ .
\end{eqnarray}
In (\ref{eq:Fn}), $\alpha\equiv\frac LN$; $\mathbf{q}$, $\mathbf{Id}_n$ and $\mathbf{1}_n$ denote respectively the overlap matrix, the $n$-dimensional identity matrix and the $n$-dimensional matrix whose all entries are equal to one. The sum $\sum\limits_{\lambda\neq0}$ runs over all the nonzero eigenvalues of the matrix $J^0$.\\

Within replica symmetric Ansatz we assume
\begin{equation}\forall\ a\neq b,\ \forall\ x,\ \left\{
\begin{array}{l}
r^{ab}=r\\
q^{ab}=q\\
\rho^a(x)=\rho(x)\\
\mu^a(x)=\mu(x)\\
\lambda^a=\lambda
\end{array}
\right.
\end{equation}
Finally, taking the $n\rightarrow0$ limit, we get
\begin{equation}
\overline{Z_J^n}\sim\int\prod\mathrm{d}q\,\mathrm{d}r\,\mathcal{D}\rho(x)\mathcal{D}\mu(x)\mathrm{d}\lambda e^{-N\beta \mathcal{F}}\ ,
\end{equation}
where
\begin{eqnarray}\label{calf}
{\cal F} &=& \frac {\alpha \beta}2 r(f-q) -\frac{\alpha}{\beta} \psi(q,\beta)-\lambda\left(\int\mathrm{d} x\rho(x)-f\right) \nonumber \\
&-& \frac 12 \int dx dy \rho(x)J_w(x-y)\rho(y) +\int dx \mu(x)\rho(x) \nonumber \\
&-&T\int dx\ Dz \log \bigg( 1 + \exp\big[ \beta \big(
z\sqrt{\alpha r} + \mu(x)\big)\big]\bigg)\ .
\end{eqnarray}
$Dz\equiv\exp(-z^2/2)/\sqrt{2\pi}$ is the Gaussian measure, 
\begin{eqnarray}
\label{eq:psi1d}
\psi^{\text{1D}}(q,\beta)\equiv\sum\limits_{k\geq1}&\bigg[\frac{\beta(q-f^2)\sin(k\pi w)}{k\pi-\beta(f-q)\sin(k\pi w)}\nonumber\\
&-\log\big(1-\frac{\beta(f-q)\sin(k\pi w)}{k\pi}\big)\bigg]
\end{eqnarray}
in 1 dimension, and 
\begin{eqnarray}
\label{eq:psi2d}
\psi^{\text{2D}}(q,\beta)\equiv2\sum\limits_{\underset{\neq(0,0)}{(k_1,k_2)}}&\bigg[\frac{\beta(q-f^2)}{\phi(k_1,k_2)-\beta(f-q)}\nonumber\\
&-\log\big(1-\frac{\beta(f-q)}{\phi(k_1,k_2)}\big)\bigg]
\end{eqnarray}
with
\begin{equation}
\phi(k_1,k_2)\equiv\frac{k_1k_2\pi^2}{\sin(k_1\pi\sqrt{w})\sin(k_2\pi\sqrt{w})}
\end{equation}
in 2 dimensions. The fixed-activity constraint is imposed through the parameter $\lambda$. When $N\rightarrow\infty$ the integral is calculated through the saddle-point method. $r,\ q,\ \rho(x)$ and $\mu(x)$ are found by writing the saddle-point equations
\begin{eqnarray}
&\frac{\partial\mathcal{F}}{\partial q}=\frac{\partial\mathcal{F}}{\partial r}=\frac{\partial\mathcal{F}}{\partial \mu(x)}=\frac{\partial\mathcal{F}}{\partial \lambda}=0\nonumber\ ,\\
&\int\mathrm{d}x\rho(x)=f\ ,
\end{eqnarray} 
which give
\begin{eqnarray}
r&=&2T^2(q-f^2)\varphi(q,T)\nonumber\ ,\\
q&=&\int\mathrm{d}x\int\mathrm{D}u[1+e^{-\beta u\sqrt{\alpha r}-\beta\mu(x)}]^{-2}\nonumber\ ,\\
\rho(x)&=&\int\mathrm{D}u[1+e^{-\beta u\sqrt{\alpha r}-\beta\mu(x)}]^{-1}\nonumber\ ,\\
\mu(x)&=&\int\mathrm{d}yJ_w(x-y)\rho(y)+\lambda\nonumber\ ,\\
f&=&\int\mathrm{d}x\rho(x)\ ,
\end{eqnarray}
where $\varphi(q,T)$ is defined by Eq.~(\ref{eq:varphi1d}) in dimension 1 and Eq.~(\ref{eq:varphi2d}) in dimension 2.

\section{Spatial correlations of free-energy fluctuations}
\label{app:spatialcorr}

\begin{figure}

\centering

 \includegraphics[width=\columnwidth]{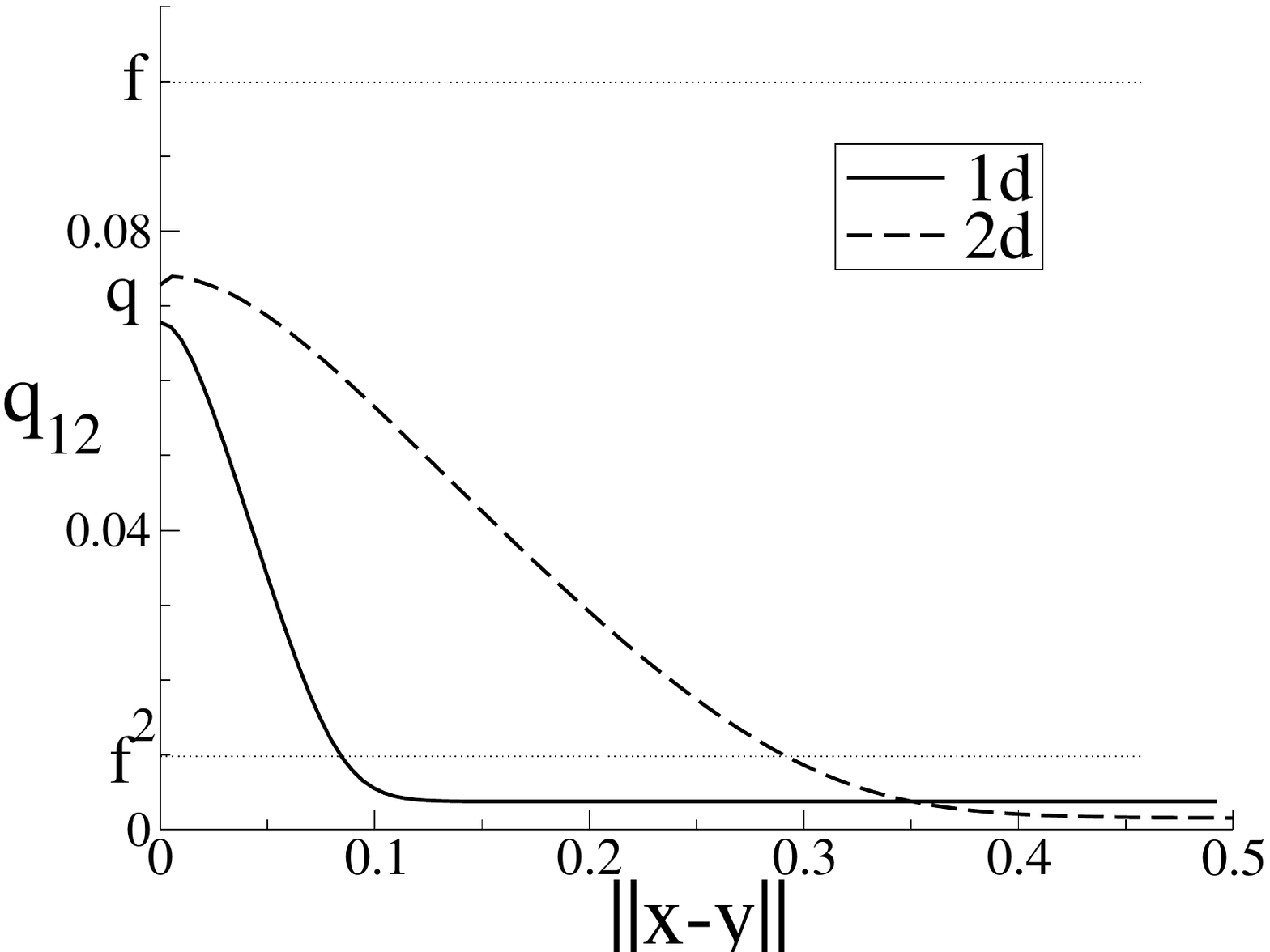}

\caption{Overlap $q_{12}$ between two groups of replicas centered respectively on positions $x$ and $y$, in dimension 1 with ${T=0.006}$, ${\alpha=0.01}$ (full line) and in dimension 2 with ${T=0.004}$, ${\alpha=0.002}$ (dashed line). Dotted lines indicate ${q_{12}=f}$ and ${q_{12}=f^2}$. }

 \label{fig:q12}

\end{figure}

We consider $\frac n2$ copies of the system with a clump centered in $x$ and $\frac n2$ other copies with a clump centered in $y$. In order to lighten notations, we take $y=0$ (the problem is invariant by translation). Under this condition we have
\begin{equation}
\begin{array}{l}
\forall\ a< b\leq\frac n2,\forall x',\ \left\{ \begin{array}{l} r^{ab}=r^{ba}=r_1,\\ q^{ab}=q^{ba}=q_1,\\ \rho^a(x')=\rho_1(x'),\\ \mu^a(x')=\mu_1(x'),\\ \lambda^a=\lambda_1\ ,\end{array}\right.\\
\\
\\
 \forall\ \frac n2<a< b,\ \forall x',\ \left\{ \begin{array}{l} r^{ab}=r^{ba}=r_2,\\ q^{ab}=q^{ba}=q_2,\\ \rho^a(x')=\rho_2(x'),\\ \mu^a(x')=\mu_2(x'),\\ \lambda^a=\lambda_2\ ,\end{array}\right.\\
\\
\\
\forall\ a \leq\frac n2<b,\ r^{ab}=r^{ba}=r_{12},\ q^{ab}=q^{ba}=q_{12}\ .\\
    \end{array}
\end{equation}
(The dependence of $q_{12}$ and $r_{12}$ on $|x|$ will be omitted to lighten notations.) By symmetry, $\ {r_1=r_2=r},\ {q_1=q_2=q},\ {\lambda_1=\lambda_2=\lambda}$, ${\rho_1(x'-x)=\rho_2(x')=\rho(x')}$ and $\mu_1(x'-x)=\mu_2(x')=\mu(x')$. Replacing in (\ref{eq:Fn}) and taking the small $n$ limit, (\ref{eq:Fn}) becomes 
\begin{equation}
\mathcal{F}_n\underset{n\rightarrow0}\sim n\mathcal{F}_0+n^2\mathcal{F}_1+\mathcal{O}(n^3)\ , 
\end{equation}
where $\mathcal{F}_0=\mathcal{F}$ given by (\ref{calf})  and
\begin{eqnarray}
\label{eq:calF1}
\mathcal{F}_1&=&\frac{\alpha\beta}{4}(rq+r_{12}q_{12})\nonumber\\
&&-\frac{\alpha}{2\beta}\left(\big(\frac{q+q_{12}}{2}-f^2\big)^2+\big(\frac{q-q_{12}}{2}\big)^2\right)\varphi(q,T)\nonumber\\
&&+\frac{1}{4\beta}\int\mathrm{d}x'\Bigg[\left(\int\mathrm{D}u\log \left(1+e^{\beta\sqrt{\alpha r}u+\beta\mu(x'-x)}\right)\right)^2\nonumber\\
&&-\int\mathrm{D}u\log^2 \left(1+e^{\beta\sqrt{\alpha r}u+\beta\mu(x')}\right)\nonumber\\
&&+\bigg(\int\mathrm{D}u\log \left(1+e^{\beta\sqrt{\alpha r}u+\beta\mu(x')}\right)\nonumber\\
&&\ \ \cdot\int\mathrm{D}v\log \left(1+e^{\beta\sqrt{\alpha r}v+\beta\mu(x'-x)}\right)\bigg)\nonumber\\
&&-\int\mathrm{D}u\mathrm{D}v\bigg(\kappa(u,v)\log \left(1+e^{\beta\sqrt{\alpha (r-r_{12})}u+\beta\mu(x')}\right)\nonumber\\
&&\ \ \cdot\log \left(1+e^{\beta\sqrt{\alpha (r+r_{12})}v+\beta\mu(x'-x)}\right)\bigg)\Bigg]\nonumber\ ,
\end{eqnarray}
where $\varphi(q,T)$ is given by (\ref{eq:varphi1d}) in dimension 1 and (\ref{eq:varphi2d}) in dimension 2; $\kappa(u,v)$ is given by (\ref{kappa}).

From Eq.~(\ref{eq:VW}) we have
\begin{equation}
 \mathcal{F}_1=-\frac\beta4(V+W(x,y))\ .
\end{equation}
Combining Eqs.~(\ref{eq:calF1}) and (\ref{eq:V}) we obtain expression (\ref{eq:W}) for $W(x,y)$.
Parameters $r_{12}$ and $q_{12}$ are found by writing the saddle-point equations
\begin{equation}
\frac{\partial\mathcal{F}_1}{\partial q_{12}}=\frac{\partial\mathcal{F}_1}{\partial r_{12}}=0\ ,
\end{equation} 
which give
\begin{eqnarray}
r_{12}&=&2T^2(q_{12}-f^2)\varphi(q,T)\nonumber\ ,\\
q_{12}&=&\int\mathrm{d}x'\int\mathrm{D}u\mathrm{D}v\ \kappa(u,v)\nonumber\\
&&\cdot[1+e^{-\beta u\sqrt{\alpha (r-r_{12})}-\beta\mu(x')}]^{-1}\nonumber\\
&&\cdot[1+e^{-\beta v\sqrt{\alpha (r+r_{12})}-\beta\mu(x'-x)}]^{-1}\ .
\end{eqnarray}
The overlap $q_{12}$ as a function of $|x|$ is shown in Fig.~\ref{fig:q12}. When the distance between the two clump centers increases, $q_{12}$ decreases from $q$ (for $x=y$) to a saturation value lower than $f^2$, on a typical distance roughly equal to the width of the clump. More precisely, in 1d ${\frac{\int\mathrm{d}u\,uq(u)}{\int\mathrm{d}u\,q(u)}=0.113}$ and ${\frac{\int\mathrm{d}u\,u\rho(u)}{\int\mathrm{d}u\,\rho(u)}=0.082}$; in 2d ${\frac{\int\mathrm{d}u\,uq(u)}{\int\mathrm{d}u\,q(u)}=0.125}$ and ${\frac{\int\mathrm{d}u\,u\rho(u)}{\int\mathrm{d}u\,\rho(u)}=0.097}$.

\bibliography{bibliodynamics}
\bibliographystyle{apsrev4-1}

\end{document}